\documentclass[epj,nopacs]{svjour}
\usepackage{epsfig}
\usepackage{amssymb}
\usepackage{amsmath}
\usepackage{graphicx,color}
\usepackage{widetext}
\usepackage{rotating}

\usepackage{url}

\usepackage{ifpdf}
\ifpdf
\DeclareGraphicsExtensions{.pdf, .jpg, .tif}
\usepackage[%
  pdftitle={Spin density matrix elements in exclusive omega electroproduction},%
  pdfauthor={The HERMES Collaboration},%
  pdfsubject={HERMES exclusive omega},%
  pdfstartview=FitH,%
  bookmarks=true,%
  bookmarksopen=true,%
  breaklinks=true,%
  colorlinks=true,%
  linkcolor=blue,anchorcolor=blue,%
  citecolor=blue,filecolor=blue,%
  menucolor=blue,pagecolor=blue,%
  urlcolor=blue]{hyperref}
\else
\DeclareGraphicsExtensions{.eps, .jpg}
\usepackage[%
  breaklinks=true,%
  colorlinks=true,%
  linkcolor=blue,anchorcolor=blue,%
  citecolor=blue,filecolor=blue,%
  menucolor=blue,pagecolor=blue,%
  urlcolor=blue]{hyperref}
\fi

\begin{document}
\hugehead

\title{Spin density matrix elements in exclusive $\omega$ electroproduction on $^1$H and $^2$H targets at 27.5 GeV beam energy}

\author{ 
The HERMES Collaboration \medskip \\
A.~Airapetian$^{13,16}$,
N.~Akopov$^{27}$,
Z.~Akopov$^{6}$,
W.~Augustyniak$^{26}$,
A.~Avetissian$^{27}$,
H.P.~Blok$^{18,25}$,  
A.~Borissov$^{6}$,
V.~Bryzgalov$^{20}$,
M.~Capiluppi$^{10}$,
G.P.~Capitani$^{11}$,
E.~Cisbani$^{22}$,
G.~Ciullo$^{10}$,
M.~Contalbrigo$^{10}$,
P.F.~Dalpiaz$^{10}$,
W.~Deconinck$^{6}$,
R.~De~Leo$^{2}$,
E.~De~Sanctis$^{11}$,
M.~Diefenthaler$^{15,9}$,
P.~Di~Nezza$^{11}$,
M.~D\"uren$^{13}$,
M.~Ehrenfried$^{13}$,
G.~Elbakian$^{27}$,
F.~Ellinghaus$^{5}$,
E.~Etzelm\"uller$^{13}$,
R.~Fabbri$^{7}$,
L.~Felawka$^{23}$,
S.~Frullani$^{22}$,
D.~Gabbert$^{7}$,
G.~Gapienko$^{20}$,
V.~Gapienko$^{20}$,
F.~Garibaldi$^{22}$,
G.~Gavrilov$^{19,6,23}$,
V.~Gharibyan$^{27}$,
M.~Hartig$^{6}$,
D.~Hasch$^{11}$,
Y.~Holler$^{6}$,
I.~Hristova$^{7}$,
A.~Ivanilov$^{20}$,
H.E.~Jackson$^{1}$,
S.~Joosten$^{15,12}$,
R.~Kaiser$^{14}$,
G.~Karyan$^{27}$,
T.~Keri$^{13}$,
E.~Kinney$^{5}$,
A.~Kisselev$^{19}$,
V.~Korotkov$^{20}$,
V.~Kozlov$^{17}$,
P.~Kravchenko$^{19}$,
V.G.~Krivokhijine$^{8}$,
L.~Lagamba$^{2}$,
L.~Lapik\'as$^{18}$,
I.~Lehmann$^{14}$,
P.~Lenisa$^{10}$,
W.~Lorenzon$^{16}$,
B.-Q.~Ma$^{3}$,
D.~Mahon$^{14}$,
S.I.~Manaenkov$^{19}$,
Y.~Mao$^{3}$,
B.~Marianski$^{26}$,
H.~Marukyan$^{27}$,
A.~Movsisyan$^{10,27}$,
M.~Murray$^{14}$,
Y.~Naryshkin$^{19}$,
A.~Nass$^{9}$,
W.-D.~Nowak$^{7}$,
L.L.~Pappalardo$^{10}$,
R.~Perez-Benito$^{13}$,
A.~Petrosyan$^{27}$,
P.E.~Reimer$^{1}$,
A.R.~Reolon$^{11}$,
C.~Riedl$^{15,7}$,
K.~Rith$^{9}$,
A.~Rostomyan$^{6}$,
D.~Ryckbosch$^{12}$,
A.~Sch\"afer$^{21}$,
G.~Schnell$^{4,12}$,
K.P.~Sch\"uler$^{6}$,
B.~Seitz$^{14}$,
T.-A.~Shibata$^{24}$,
M.~Stahl$^{13}$,
M.~Stancari$^{10}$,
M.~Statera$^{10}$,
E.~Steffens$^{9}$,
J.J.M.~Steijger$^{18}$,
S.~Taroian$^{27}$,
A.~Terkulov$^{17}$,
R.~Truty$^{15}$,
A.~Trzcinski$^{26}$,
M.~Tytgat$^{12}$,
Y.~Van~Haarlem$^{12}$,
C.~Van~Hulse$^{4,12}$,
V.~Vikhrov$^{19}$,
I.~Vilardi$^{2}$,
S.~Wang$^{3}$,
S.~Yaschenko$^{6,9}$,
S.~Yen$^{23}$,
D.~Zeiler$^{9}$,
B.~Zihlmann$^{6}$,
P.~Zupranski$^{26}$
}

\institute{ 
$^1$Physics Division, Argonne National Laboratory, Argonne, Illinois 60439-4843, USA\\
$^2$Istituto Nazionale di Fisica Nucleare, Sezione di Bari, 70124 Bari, Italy\\
$^3$School of Physics, Peking University, Beijing 100871, China\\
$^4$Department of Theoretical Physics, University of the Basque Country UPV/EHU, 48080 Bilbao, Spain and IKERBASQUE, Basque Foundation for Science, 48013 Bilbao, Spain\\
$^5$Nuclear Physics Laboratory, University of Colorado, Boulder, Colorado 80309-0390, USA\\
$^6$DESY, 22603 Hamburg, Germany\\
$^7$DESY, 15738 Zeuthen, Germany\\
$^8$Joint Institute for Nuclear Research, 141980 Dubna, Russia\\
$^9$Physikalisches Institut, Universit\"at Erlangen-N\"urnberg, 91058 Erlangen, Germany\\
$^{10}$Istituto Nazionale di Fisica Nucleare, Sezione di Ferrara and Dipartimento di Fisica e Scienze della Terra, Universit\`a di Ferrara, 44122 Ferrara, Italy\\
$^{11}$Istituto Nazionale di Fisica Nucleare, Laboratori Nazionali di Frascati, 00044 Frascati, Italy\\
$^{12}$Department of Physics and Astronomy, Ghent University, 9000 Gent, Belgium\\
$^{13}$II. Physikalisches Institut, Justus-Liebig Universit\"at Gie{\ss}en, 35392 Gie{\ss}en, Germany\\
$^{14}$SUPA, School of Physics and Astronomy, University of Glasgow, Glasgow G12 8QQ, United Kingdom\\
$^{15}$Department of Physics, University of Illinois, Urbana, Illinois 61801-3080, USA\\
$^{16}$Randall Laboratory of Physics, University of Michigan, Ann Arbor, Michigan 48109-1040, USA \\
$^{17}$Lebedev Physical Institute, 117924 Moscow, Russia\\
$^{18}$National Institute for Subatomic Physics (Nikhef), 1009 DB Amsterdam, The Netherlands\\
$^{19}$B.P. Konstantinov Petersburg Nuclear Physics Institute, Gatchina, 188300 Leningrad Region, Russia\\
$^{20}$Institute for High Energy Physics, Protvino, 142281 Moscow Region, Russia\\
$^{21}$Institut f\"ur Theoretische Physik, Universit\"at Regensburg, 93040 Regensburg, Germany\\
$^{22}$Istituto Nazionale di Fisica Nucleare, Sezione di Roma, Gruppo Collegato Sanit\`a and Istituto Superiore di Sanit\`a, 00161 Roma, Italy\\
$^{23}$TRIUMF, Vancouver, British Columbia V6T 2A3, Canada\\
$^{24}$Department of Physics, Tokyo Institute of Technology, Tokyo 152, Japan\\
$^{25}$Department of Physics and Astronomy, VU University, 1081 HV Amsterdam, The Netherlands\\
$^{26}$National Centre for Nuclear Research, 00-689 Warsaw, Poland\\
$^{27}$Yerevan Physics Institute, 375036 Yerevan, Armenia\\
} 

%\date{Compiled: \today  / Received:  / Revised version:}
\date{DESY Report 14-116 / Compiled: \today\  / Version: 6.0 (final, Erratum incorporated)}

\titlerunning{omega SDMEs}
\authorrunning{The HERMES Collaboration}

\abstract{
Exclusive electroproduction of $\omega$ mesons on unpolarized hydrogen and deuterium targets is studied in the kinematic region of $Q^2>$ 1.0 GeV$^2$, 3.0 GeV $< W <$ 6.3 GeV, and $-t'< $ 0.2 GeV$^2$. Results on the angular distribution of the $\omega$ meson, including its decay products, are presented. The data were accumulated with the HERMES forward spectrometer during the 1996-2007 running period using the 27.6 GeV longitudinally polarized electron or positron beam of HERA. The determination of the virtual-photon longitudinal-to-transverse cross-section ratio reveals that a considerable part of the cross section arises from transversely polarized photons. Spin density matrix elements are presented in projections of $Q^2$ or  $-t'$. Violation of $s$-channel helicity conservation is observed for some of these elements. A sizable contribution from unnatural-parity-exchange amplitudes is found and the phase shift between those amplitudes that describe transverse $\omega$ production by longitudinal and transverse virtual photons, $\gamma^{*}_{L} \to \omega_{T}$ and $\gamma^{*}_{T} \to \omega_{T}$, is determined for the first time. A hierarchy of helicity amplitudes is established, which mainly means that the unnatural-parity-exchange amplitude describing the $\gamma^*_T \to \omega_T$ transition dominates over the two natural-parity-exchange amplitudes describing the $\gamma^*_L \to \omega_L$ and  $\gamma^*_T \to \omega_T$ transitions, with the latter two being of similar magnitude. Good agreement is found between the HERMES proton data and results of a pQCD-inspired phenomenological model that includes pion-pole contributions, which are of unnatural parity.
}

\maketitle

% pseudo line numbers
%\setvruler[11.1pt][0][1][3][1][20pt][20pt][-9pt]

%%%%%%%%%%%%%%%%%%%%%%%%%%%%%%%%%%%%%%%%%
\section{Introduction}
\label{intro}

Exclusive electroproduction of vector mesons on nucleons offers a rich source of
information on the mechanisms that produce these mesons, see
e.g., Refs.~\cite{Fran,diehl1}. This process can be considered to consist of 
three   subprocesses: i) the incident lepton emits a virtual photon 
$\gamma^*$, which  dissociates into a $q\bar{q}$ pair; ii) this  pair 
interacts strongly with the nucleon; iii) from the scattered $q\bar{q}$ 
pair the observed vector meson is formed.

In Regge phenomenology, the interaction of the $q\bar{q}$ pair with the 
nucleon proceeds through the exchange of a pomeron or (a combination of) the
exchanges of other regge\-ons (e.g., $\rho$, $\omega$, $\pi$, ...).
If the quantum numbers of the particle lying on the Regge trajectory are 
$J^P=0^+,\;1^-$, ..., the process is denoted  Natural  Parity Exchange 
(NPE). Alternatively, the case of   $J^P=0^- ,\;1^+$, ... is denoted 
Unnatural Parity Exchange (UPE). 
In perturbative quantum chromodynamics (pQCD), the interaction of the $q\bar{q}$ pair with 
the nucleon can proceed via two-gluon exchange or quark-antiquark 
exchange, where the former corresponds to the exchange of a pomeron and 
the latter to the exchange of a (combination of) reggeon(s).

Spin density matrix elements (SDMEs) describe the  final spin states of the
produced  vector meson. In this work, SDME values will be determined and
discussed in the formalism that was developed in Ref.~\cite{Schill} for the 
case of an unpolarized or
longitudinally polarized beam and an unpolarized target. For completeness,
we also present SDME values in the more general formalism of Ref.~\cite{Diehl}.
The SDMEs can be expressed in terms of helicity amplitudes that describe 
the transitions from the initial helicity states of virtual photon and 
incoming nucleon to the final helicity states of the produced vector meson and the 
outgoing nucleon. The values of SDMEs will be used to establish a 
hierarchy of helicity amplitudes, to test the hypothesis of $s$-channel 
helicity conservation, to investigate UPE contributions, and to determine the 
longitudinal-to-transverse cross-section ratio.

In the framework of pQCD, the nucleon structure can also be studied 
through hard exclusive meson production as the process amplitude contains 
Generalized  Parton Distributions (GPDs)~\cite{gpd1,gpd2,gpd3}. For 
longitudinal virtual photons, this amplitude is proven to factorize 
rigorously into a perturbatively calculable hard-scattering part and two soft parts (collinear
factorization)~\cite{Radyushkin:1996ru,Collins:1996fb}. The soft parts of the 
convolution contain GPDs and a meson distribution amplitude. At leading twist, the chiral-even GPDs $H^f$ and 
$E^f$ are sufficient to describe exclusive vector-meson production on a 
spin-$1/2$ target such as a proton or a neutron, where $f$ denotes a quark of flavor $f$ or a gluon. 
These GPDs are of special interest 
as they are related to the total angular momentum carried by quarks or gluons in the 
nucleon~\cite{Ji:1996e}.

Although there is no such rigorous proof for transverse virtual photons, 
phenomenological models use the modified perturbative approach~\cite{Botts:1989kf} instead, 
which takes into account parton transverse momenta. 
The latter are included at subleading twist in the subprocess $\gamma^* f \rightarrow {\cal M} f$, 
where ${\cal M}$ denotes the meson, while the partons are still emitted and reabsorbed by 
the nucleon collinear to the nucleon momentum. By using this approach, 
the pQCD-inspired phenomenological ``GK model" can   
describe existing data on cross sections, SDMEs and spin asymmetries in 
exclusive vector-meson production for values of Bjorken-$x$, $x_{B}$, below about 
0.2~\cite{Goloskokov:2005sd,Goloskokov:2007nt,Goloskokov:2013mba}. It 
can also describe exclusive leptoproduction of pseudoscalar mesons by 
including the full contribution  to the electromagnetic form factor from the 
pion, in contrast to earlier studies at leading-twist, which took into account only 
the relatively small perturbative contribution to this 
form factor (see Ref.~\cite{Goloskokov:2009ia} and references therein). 
The GK model also applies successfully to the description of deeply virtual Compton scattering~\cite{k_mo_sa}.
The results of the most recent variant of the GK model, in which the 
unnatural-parity contributions due to pion exchange are included to describe 
exclusive $\omega$ leptoproduction~\cite{thcal},
will be  compared in this paper to the HERMES proton data in terms of SDMEs and 
certain combinations of them. 

Early  papers on exclusive $\omega$  electroproduction are summarized 
in Ref.~\cite{Bauer}, which particularly contains results on SDMEs obtained at DESY
for 0.3 GeV$^2$ $< Q^{2} <$ 1.4 GeV$^2$  and 0.3 GeV $< W <$ 2.8 GeV.
The symbol $Q^{2}$ represents the
negative square of the virtual-photon four-momentum and
$W$ is the invariant mass of the photon-nucleon system.
Recently, SDMEs in exclusive $\omega $ electroproduction  were 
studied for 1.6 GeV$^2$ $< Q^{2}<$ 5.2 GeV$^2$ by CLAS~\cite{clas} and 
it was  found that the exchange of the pion Regge
trajectory dominates exclusive $\omega$ production, even for Q$^2$ values 
as large as 5 GeV$^2$.

\section {Formalism}

\subsection {Spin density matrix elements}

The $\omega$ meson is produced in the following reaction: 
\begin{eqnarray} 
e + p \to e + p +  \omega,
\label{omprod}
\end{eqnarray}
with a branching ratio $Br = 89.1 \%$ for the $\omega$ decay: 
\begin{eqnarray} 
\omega \to \pi^+ + \pi^- + \pi^0,~\pi^0 \to2\gamma.
\label{omdecay}
\end{eqnarray}
The angular distribution of  the three final-state pions     
depends on SDMEs. 
The first subprocess of vector-meson production, the emission of a
virtual photon
($ e \rightarrow e+ \gamma^{*}$), is
described by the photon spin density matrix~\cite{Schill}, 
\begin{equation}
\varrho^{U+L}_{\lambda_{\gamma} \lambda '_{\gamma }}
 = \varrho^{U}_{\lambda_{\gamma} \lambda '_{\gamma }} +
P_{b}\varrho^{L}_{\lambda_{\gamma} \lambda '_{\gamma }},
\label{phspden}
\end{equation}
where U and L denote unpolarized and longitudinally polarized beam,
respectively, and $P_{b}$ is the value of the beam polarization.
 The photon spin density matrix can be calculated in quantum electrodynamics.

The vector-meson spin density matrix
$\rho_{\lambda_{V}\lambda_{V}^{'}}$ is expressed  through  helicity amplitudes
{$F_{\lambda _{V} \lambda '_{N}\lambda _{\gamma} \lambda_{N}}$}.
 These  amplitudes describe the transition of a
virtual photon with helicity $\lambda _{\gamma}$ to a vector meson with  
helicity $ \lambda _{V}$, while  $\lambda_{N}$  and $\lambda '_{N}$ are the
helicities of the nucleon in the initial and final states, respectively.
Helicity amplitudes depend on $W$, $Q^{2}$, and $t'=t - t_{min}$, where
$t$ is the Mandelstam variable  and $-t_{min}$ represents the
smallest kinematically allowed value of $-t$ at fixed virtual-photon energy and
$Q^{2}$. The quantity $\sqrt{-t'} $ is approximately equal to the transverse
momentum of the vector meson with respect to the direction of the virtual
photon in the  $\gamma^{*}N$  centre-of-mass (CM) system. 
In this system,  the spin density matrix of the vector meson is given by the
von Neumann equation~\cite{Schill}, 
\begin{align}
\rho_{\lambda_{V} \lambda '_{V}}= \frac{1}{2 \mathcal{N} }
  \sum_{\lambda_{\gamma}
\lambda '_{\gamma}\lambda_N \lambda '_N}
   F_{\lambda_{V}\lambda '_N\lambda_{\gamma}\lambda _N}
 \varrho^{U+L}_{\lambda_{\gamma} \lambda '_{\gamma }}   
  F_{\lambda '_{V} \lambda '_N\lambda '_{\gamma}\lambda
 _N}^{*} \, ,
\end{align}
where $\mathcal{N}$ is a normalization factor, see Refs.~\cite{Schill,DC-24}.

After the decomposition of 
$\varrho^{U+L}_{\lambda_{\gamma} \lambda'_{\gamma}}$ 
into the standard  set of $3\times 3$ Hermitian matrices $ \Sigma^{\alpha}$, 
the vector-meson spin density matrix is expressed in 
terms of a set of nine matrices $\rho^{\alpha}_{\lambda_V\lambda'_V}$  related to various 
photon polarization states: transversely polarized photon ($\alpha$=0,...,3), 
longitudinally   polarized photon ($\alpha$=4), and  terms describing  
their interference  ($\alpha$=5,...,8)~\cite{Schill}. 
When contributions of transverse and longitudinal photons cannot be separated, 
the SDMEs are customarily defined as 
\begin{align}
r^{04}_{\lambda_{V}\lambda '_{V}} &= (\rho^{0}_{\lambda_{V}\lambda '_{V}}
+ \epsilon R \rho^{4}_{\lambda_{V}\lambda '_{V}})( 1 + \epsilon R )^{-1},   
\nonumber\\
r^{\alpha}_{\lambda_{V}\lambda'_{V}} &=
\begin{cases}
{  \rho^{\alpha}_{\lambda_{V}\lambda'_{V}}}{(  1 + \epsilon R )^{-1}},
\; \alpha = 1,2,3,\\
{ \sqrt{R} \rho^{\alpha}_{\lambda_{V}\lambda '_{V}}}
{(1 + \epsilon R )^{-1}}, \; \alpha = 5,6,7,8.
\end{cases}
  \hspace*{0.25cm}
\label{rmatr}
\end{align}
The quantity $R= d\sigma_{L}/ d\sigma_{T}$ is the longitudinal-to-trans\-verse
virtual-photon differential cross-section ratio and $\epsilon$ is the
ratio of fluxes of longitudinal and transverse virtual photons. \\

\subsection{Helicity amplitudes} \label{hel_ampli}

 A helicity amplitude can be  decomposed into a sum of a NPE amplitude T
 and a UPE amplitude U, 
\begin{equation}
F_{\lambda_{V} \lambda '_{N} \lambda_{\gamma}  \lambda_{N} } =
T_{\lambda_{V} \lambda '_{N} \lambda_{\gamma}  \lambda_{N} }+
U_{\lambda_{V}\lambda '_{N} \lambda_{\gamma}  \lambda_{N}},
\label{nu}
\end{equation} 
for details see Refs.~\cite{Schill,DC-24}. 
The relations between the
amplitudes $F$, $T$, and $U$ are the following~\cite{Schill}:
\begin{align}
T_{\lambda_V \lambda'_N \lambda_{\gamma} \lambda_N}=\frac{1}{2}[
F_{\lambda_V \lambda'_N \lambda_{\gamma} \lambda_N}\nonumber ~~~~~~~~~~~~~~~~~~~~~~~~~~~\\
+(-1)^{\lambda_V-\lambda_{\gamma}}F_{-\lambda_V \lambda'_N -\lambda_{\gamma}\lambda_N}],
\label{fnat}\\
U_{\lambda_V \lambda'_N \lambda_{\gamma} \lambda_N}=\frac{1}{2}[
F_{\lambda_V \lambda'_N \lambda_{\gamma} \lambda_N} \nonumber~~~~~~~~~~~~~~~~~~~~~~~~~~~ \\
-(-1)^{\lambda_V-\lambda_{\gamma}}F_{-\lambda_V \lambda'_N -\lambda_{\gamma} \lambda_N}].
\label{funnnat}
\end{align}
The asymptotic behaviour of amplitudes $F$
at small $-t'$~\cite{Diehl},
\begin{equation}
F_{\lambda_V \lambda'_N \lambda_{\gamma} \lambda_N} \propto \Bigl (\frac{\sqrt{-t'}}{M}\Bigr
)^{|(\lambda_V-\lambda'_N)-(\lambda_{\gamma}-\lambda_N)|},
\label{asytpr}
\end{equation}
follows from  angular-momentum conservation. Equations \eqref{fnat}-\eqref{asytpr} show that the double-helicity-flip amplitudes with 
$|\lambda _V-\lambda_{\gamma}|=2$ are suppressed at least by a factor of $\sqrt{-t'}/M$, and the contributions of these double-helicity-flip amplitudes to
the SDMEs are suppressed by $-t'/M^2$. Therefore they will be neglected throughout the paper.

For an unpolarized target, there exists no interference between NPE and UPE
amplitudes and there is no linear contribution from
nucleon-helicity-flip amplitudes to SDMEs.
For brevity, the following notations will be used:
\begin{equation}
\widetilde{\sum}T_{\lambda_V \lambda_{\gamma}} T^*_{\lambda'_V
\lambda'_{\gamma}}\equiv
 \frac{1}{2} \sum_{\lambda_N \lambda'_N}
T_{\lambda_V \lambda'_N\lambda_{\gamma}\lambda_N}
T^*_{\lambda'_V \lambda'_N\lambda'_{\gamma}\lambda_N}.
\label{tilde-sum}
\end{equation}
Using the symmetry properties~\cite{Schill,DC-24} of the amplitudes
$T$, Eq.~(\ref{tilde-sum}) can be rewritten  as 
\begin{align}
\widetilde{\sum}T_{\lambda_V \lambda_{\gamma}} T^*_{\lambda'_V
\lambda'_{\gamma}}=~~~~~~~~~~~~~~~~~~~~~~~~~~~~~~~~~~~~~~~~~~~~\nonumber\\ 
T_{\lambda_V \frac{1}{2}\lambda_{\gamma}\frac{1}{2}}
T^*_{\lambda'_V \frac{1}{2}\lambda'_{\gamma}\frac{1}{2}}+
T_{\lambda_V -\frac{1}{2}\lambda_{\gamma}\frac{1}{2}}
T^*_{\lambda'_V -\frac{1}{2}\lambda'_{\gamma}\frac{1}{2}}.\label{sum-two}
\end{align}
Here, the first and second product on the right-hand side gives the contribution of 
NPE amplitudes without and with nucleon-helicity flip, respectively. 
Analogous relations hold for  UPE amplitudes.
An additional abbreviated notation in the text will be the omission of the nucleon-helicity indices when discussing the amplitudes  with
$\lambda_N=\lambda'_N$, i.e.,
\begin{align}
T_{\lambda_V \lambda_{\gamma}}&\equiv
T_{\lambda_V \frac{1}{2}\lambda_{\gamma}\frac{1}{2}}=
T_{\lambda_V -\frac{1}{2}\lambda_{\gamma}-\frac{1}{2}} \nonumber\\  
 U_{\lambda_V \lambda_{\gamma}}&\equiv
U_{\lambda_V \frac{1}{2}\lambda_{\gamma}\frac{1}{2}}=
-U_{\lambda_V -\frac{1}{2}\lambda_{\gamma}-\frac{1}{2}}.
\label{abrr}
\end{align}
The dominance of diagonal $\gamma^{*} \to V$ transitions 
($\lambda_{V}=\lambda_{\gamma}$) is called $s$-channel helicity conservation (SCHC).

\subsection{Angular distribution}

 The SDMEs in exclusive electroproduction of $\omega$ mesons are determined using the process in Eq.~(\ref{omprod}). 
They are fitted as parameters of $\mathcal{W}^{U+L}(\Phi,\phi,\cos\Theta)$, which is the three-di\-men\-sio\-nal angular distribution, 
to the corresponding experimental distribution of the three pions originating from the
$\omega$-meson decay.
The angular distribution $\mathcal{W}^{U+L}(\Phi,\phi,\cos\Theta)$ is decomposed
into $\mathcal{W}^{U}$ and $\mathcal{W}^{L}$, see Eq.~\eqref{eqang1}, 
which are the respective
distributions for unpolarized and longitudinally polarized beams. 
From the fit, 15 ``unpolarized'' SDMEs (see Eq.~\eqref{eqang2}) are extracted and
additionally  8 ``polarized" SDMEs (see Eq.~\eqref{eqang3}) from  data collected  with a longitudinally polarized
beam.

%%%%%%%%%%%%%%%%%%%%%
\begin{widetext}
\begin{align}
\mathcal{W}^{U+L}(\Phi,\phi,\cos{\Theta}) 
	=&  \ \mathcal{W}^{U}(\Phi,\phi,\cos{\Theta}) + P_{b}\mathcal{W}^{L}(\Phi,\phi,\cos{\Theta}), \label{eqang1}\\
	  \mathcal{W}^{U}(\Phi,\phi,\cos{\Theta})  
	=& \ \frac{3} {8 \pi^{2}} \Bigg[
         \frac{1}{2} (1 - r^{04}_{00}) + \frac{1}{2} (3 r^{04}_{00}-1) \cos^2{\Theta}
	- \sqrt{2} \mathrm{Re} \{ r^{04}_{10} \} \sin 2\Theta 
 	\cos \phi - r^{04}_{1-1}  \sin ^{2} \Theta \cos 2 \phi \hspace*{1.0cm}
	\nonumber \\ 
	&- \epsilon \cos 2 \Phi \Big( r^{1}_{11} \sin ^{2} \Theta  + r^{1}_{00} \cos^{2}{\Theta}
  	 - \sqrt{ 2}  \mathrm{Re} \{r^{1}_{10}\} \sin 2  \Theta  \cos  \phi
    	 - r^{1}_{1-1} \sin ^{2} \Theta \cos 2 \phi   \Big)   \nonumber  \\
	&- \epsilon \sin 2 \Phi \Big( \sqrt{2} \mathrm{Im} \{r^{2}_{10}\} \sin 2 \Theta \sin \phi +
        	 \mathrm{Im} \{ r^{2}_{1-1} \} \sin ^{2} \Theta \sin 2 \phi  \Big)  \nonumber \\
	&+ \sqrt{ 2 \epsilon (1+ \epsilon)}  \cos \Phi
	 \Big(  r^{5}_{11} \sin ^2 {\Theta} +
 	 r^{5}_{00} \cos ^{2} \Theta - \sqrt{2} \mathrm{Re} \{r^{5}_{10}\} \sin 2 \Theta \cos \phi -
 	 r^{5}_{1-1} \sin ^{2} \Theta \cos 2 \phi  \Big)  \nonumber \\
	&+ \sqrt{ 2 \epsilon (1+ \epsilon)}  \sin \Phi
	 \Big( \sqrt{ 2} \mathrm{Im} \{ r^{6}_{10} \} \sin 2 \Theta \sin \phi
	 + \mathrm{Im} \{r^{6}_{1-1} \} \sin ^{2} \Theta \sin 2 \phi \Big) \Bigg],
	\label{eqang2} \\
\mathcal{W}^{L}(\Phi,\phi,\cos \Theta)  
	=& \ \frac{3}{8 \pi^{2}}  \Bigg[
  	 \sqrt{ 1 - \epsilon ^{2} }  \Big(  \sqrt{ 2}  \mathrm{Im} \{ r^{3}_{10} \}
	 \sin 2 \Theta \sin \phi +
   	 \mathrm{Im} \{ r^{3}_{1-1}\} \sin ^{2} \Theta \sin 2 \phi  \Big)  \nonumber  \\
	&+ \sqrt{ 2 \epsilon (1 - \epsilon)} \cos \Phi
	  \Big( \sqrt{2} \mathrm{Im} \{r^{7}_{10}\} \sin 2 \Theta \sin \phi
	 +  \mathrm{Im} \{ r^{7}_{1-1} \}  \sin ^{2} \Theta \sin 2 \phi   \Big)  \nonumber \\
	&+ \sqrt{ 2 \epsilon (1 - \epsilon)} \sin \Phi
	 \Big( r^{8}_{11} \sin ^{2} \Theta + r^{8}_{00} \cos ^{2}
	 \Theta -  \sqrt{2} \mathrm{Re}\{ r^{8}_{10}\} \sin 2 \Theta \cos \phi 
	 - r^{8}_{1-1} \sin ^{2} \Theta \cos 2\phi \Big)  \Bigg]. 
	\label{eqang3}
\end{align}
\end{widetext}
%%%%%%%%%%%%%%%%%%%%

\begin{figure}[t]
\includegraphics[width=.5\textwidth]{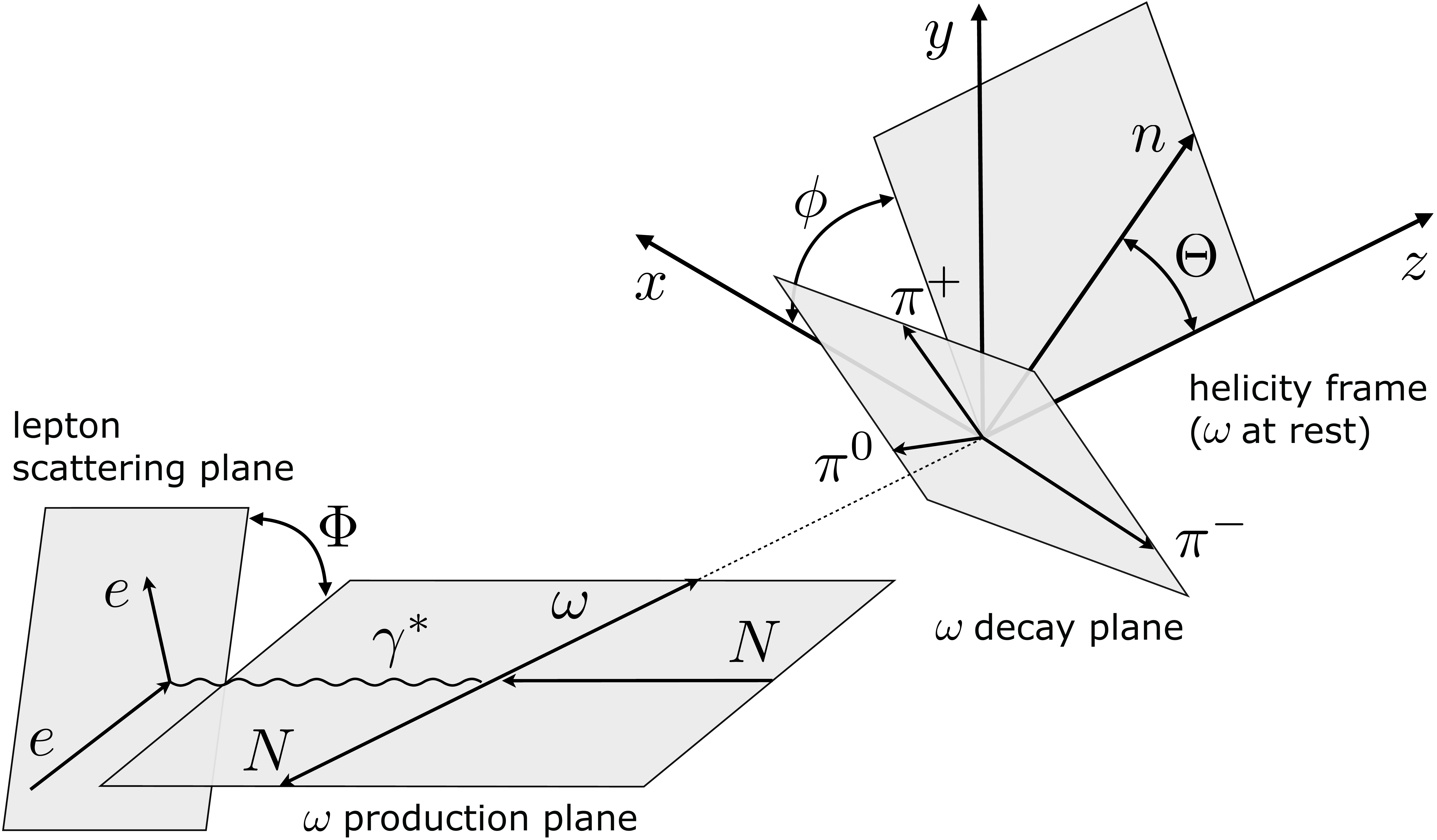}
\vspace{0.5cm}   
\caption{
Definition of angles in the process
$e N \to e N  \omega$, where  $\omega 
 \to \pi^+ \pi^- \pi^0$. 
Here, $\Phi$ is the angle between the $\omega$ production plane
and the lepton scattering plane  in the
center-of-mass system
of the virtual photon and the target nucleon.
The variables $\Theta$ and $\phi$ 
are respectively the polar and azimuthal
angles of the unit vector normal to the decay plane in the
$\omega$-meson rest frame.}
\label{defang}
\end{figure}

Definitions of angles and reference frames are shown in Fig.~\ref{defang}. The directions of the axes of the hadronic CM system
 and of the $\omega$-meson rest frame follow the directions of the axes of the helicity frame~\cite{Schill,DC-24,joos}.

The angle  $\Phi$ between the $\omega $ production  
and the lepton scattering plane in the hadronic CM system 
is given by 
\begin{align}
\cos \Phi &= \frac{ (\vec{q} \times \vec{v}) \cdot (\vec{k} \times \vec{k}')}
{ | \vec{q} \times \vec{v} | \cdot |\vec{k} \times \vec{k}'| } ,  \\ 
\sin \Phi &= 
 \frac{ [ (\vec{q} \times \vec{v} )\times (\vec{k} \times \vec{k}' )] \cdot \vec{q} }
 { |\vec{q} \times \vec{v}| \cdot |\vec{k} \times \vec{k}'| \cdot |\vec{q}|
 }.
\label{phicap-def}
\end{align}   
Here  $\vec{k}$, $\vec{k'}$, $\vec{q}=\vec{k}-\vec{k'}$, and  $\vec{v}$  
are the three-momenta of the incoming and outgoing
leptons, virtual photon, and $\omega $ meson respectively.

The unit vector normal  to the decay plane in the $\omega$ rest frame is defined by 
\begin{equation}
\vec n = \frac{\vec{p}_{\pi^+} \times \vec{p}_{\pi^-}}{|\vec{p}_{\pi^+} \times\vec{p}_{\pi^-}| },
\label{normal}
\end{equation}
where $\vec{p}_{\pi^+}$ and  $\vec{p}_{\pi^-}$ are the three-momenta of the
positive and  negative  decay pions in the $\omega$ rest frame.

The polar angle  $\Theta$ of the unit vector  $\vec{n}$ in the 
$\omega$-meson rest frame, with the $z$-axis aligned opposite to the outgoing nucleon
momentum $\vec{p}'$ and the $y$-axis directed along
$\vec{p}' \times \vec{q}$, is defined by 
\begin{align} 
\cos \Theta  &= -\frac{ \vec{p}' \cdot \vec n }{| \vec{p}' |},
\label{theta-def}
\end{align} 
while the azimuthal angle $\phi$ of the unit vector $\vec n$ is given
by
\begin{align} 
\cos \phi &= 
\frac{ (\vec{q} \times \vec{p}' )\cdot (\vec{p}' \times \vec n) }
{ | \vec{q} \times \vec{p}'| \cdot |\vec{p}' \times \vec{n} | },  \\ 
 \sin \phi &=  
 - \frac{[ (\vec{q} \times \vec{p}' )\times \vec{p}' ] \cdot ( \vec n  \times \vec{p}' ) } 
{ | (\vec{q} \times \vec{p}' )\times \vec{p}' | \cdot |\vec n \times \vec{p}' |
 } .
\label{phismall-def}
\end{align}
 
\section {Data analysis}

\subsection{HERMES experiment}

The data analyzed in this paper were accumulated with the HERMES spectrometer during the 
running period of 1996 to 2007 using the 27.6 GeV longitudinally polarized electron or positron beam
of HERA, and gaseous hydrogen or deuterium targets. 
The HERMES forward spectrometer, which is described in detail in Ref.~\cite{identif}, was built of two 
identical halves situated above and below the lepton beam pipe. 
It consisted of a dipole magnet in conjunction with tracking and particle identification detectors. 
Particles were accepted when their polar angles were in the range $\pm 170$~mrad 
in the horizontal direction and $\pm(40-140)$~mrad in the vertical direction.  
The spectrometer permitted a precise measurement of charged-particle momenta, with a resolution of $1.5\%$. 
A separation of leptons was achieved with an average efficiency of $98\%$ and a hadron 
contamination below $1\%$. 
 
\subsection{Selection of exclusively produced $\omega$ mesons}

%%%%%%%%%FIG
\begin{figure}[t!]
\centering
\includegraphics[width=.45\textwidth]{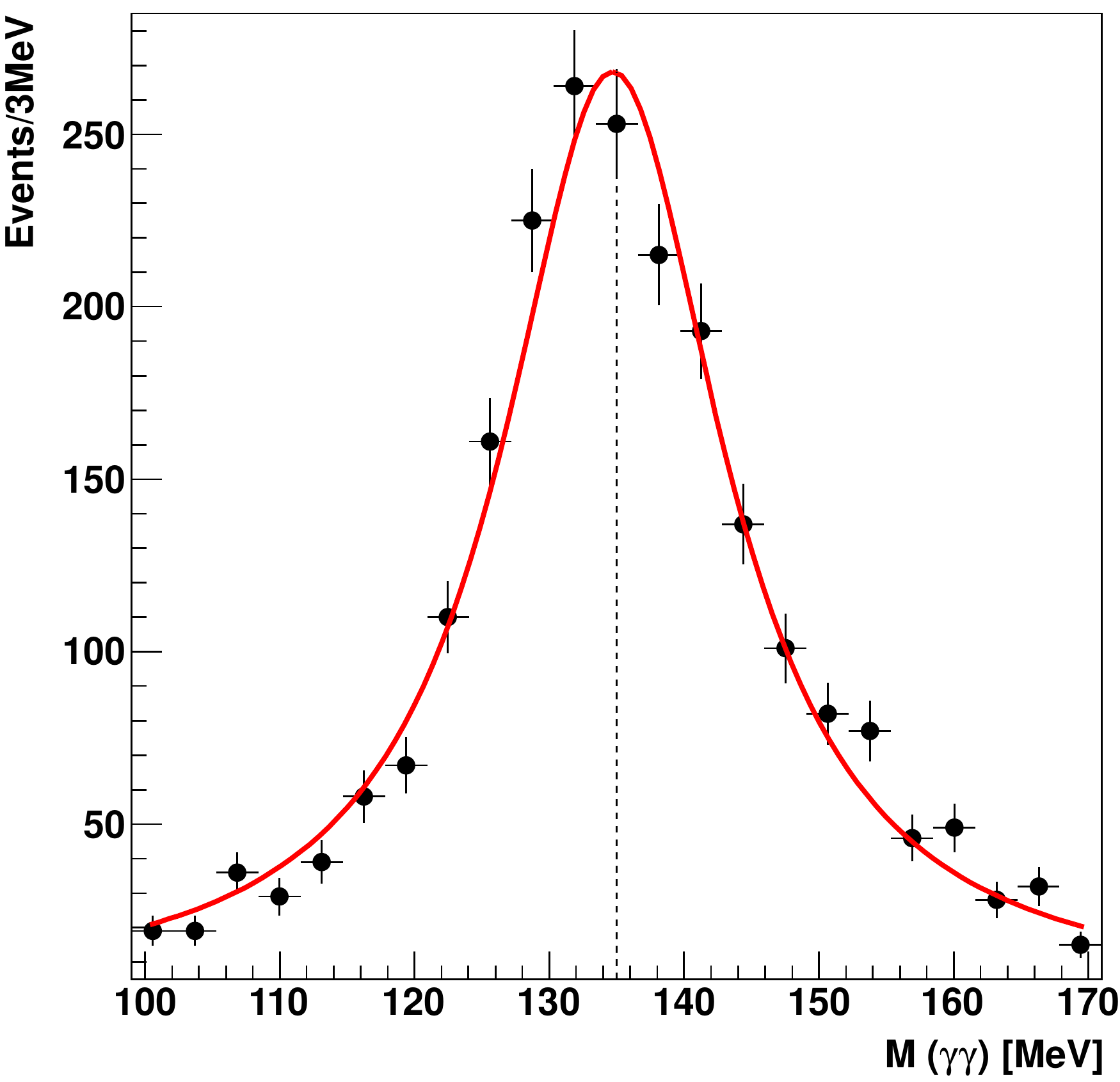}
\caption{Two-photon invariant mass distribution after application of
all criteria to select exclusively produced $\omega$  mesons. 
The Breit--Wigner fit to the mass distribution is shown as a continuous line and the dashed
line indicates the PDG value of the $\pi^{0}$ mass.} 
\label{pi0}
\end{figure}

The following requirements were applied to select exclusively produced
$\omega$ mesons from  reaction (\ref{omprod}):\\
i) Exactly two oppositely charged hadrons, which are assumed to be pions, and one lepton with the same charge as the beam lepton are identified through the analysis of the combined
responses of the  four particle-identification detectors~\cite{identif}.\\
ii) A $\pi^{0}$ meson that is reconstructed from two calorimeter
clusters as explained in Ref.~\cite{arny} is selected requiring the two-photon invariant mass to be in the interval $0.11$~GeV $ < M({\gamma\gamma})<$ 0.16 GeV. The distribution of $M ({\gamma\gamma})$ is shown in Fig.~\ref{pi0}.
This distribution  is centered at $m_{\pi^{0}}=134.69\pm19.94$~MeV, which agrees well with 
the PDG~\cite{pdg} value of the $\pi^{0}$ mass.\\
iii) The three-pion invariant mass is required to obey 0.71 GeV$\le M(\pi^+ \pi^- \pi^0) \le$ 0.87~GeV.
\\
iv) The kinematic requirements  for exclusive production of $\omega$ mesons
are the following:\\
a) The scattered-lepton momentum lies above $3.5$~GeV.\\
b) The constraint $-t' <$ 0.2 GeV$^{2}$ is used. \\
c) For exclusive production the missing energy $ \Delta E $ must vanish. 
Here, the missing energy is calculated both for proton and deuteron as $ \Delta E = \frac{ M^{2}_{X} -M^{2}_{p}}{2 M_{p}}$, with $M_{p}$ being
the proton mass and $ M^{2}_{X}=({p} + {q}- {p}_{\pi^+} - {p}_{\pi^-} -
{p}_{\pi^0})^{2}$  the missing mass squared, where  ${p}$, ${q}$, ${p}_{\pi^+}$, $
{p}_{\pi^-}$, and  ${p}_{\pi^0}$ are the four-momenta of target nucleon, virtual photon, and each of the three
 pions respectively. In this analysis, taking into account the spectrometer resolution, the  missing energy has to
lie in the interval $-1.0$ GeV $<\Delta E < 0.8$ GeV, which is referred to as
``exclusive region" in the following.\\
d) The requirement $ Q^2> $ 1.0 GeV$^2$  is applied  in order to
 facilitate the application of pQCD.\\ 
e) The requirement $W > 3.0$~GeV is applied in order to be outside of the resonance region, while
   an upper cut of $W < 6.3$~GeV is applied in order to define a clean kinematic phase space.

%%%%%%%%%FIG
\begin{figure}[t!]
\centering
\includegraphics[width=.45\textwidth]{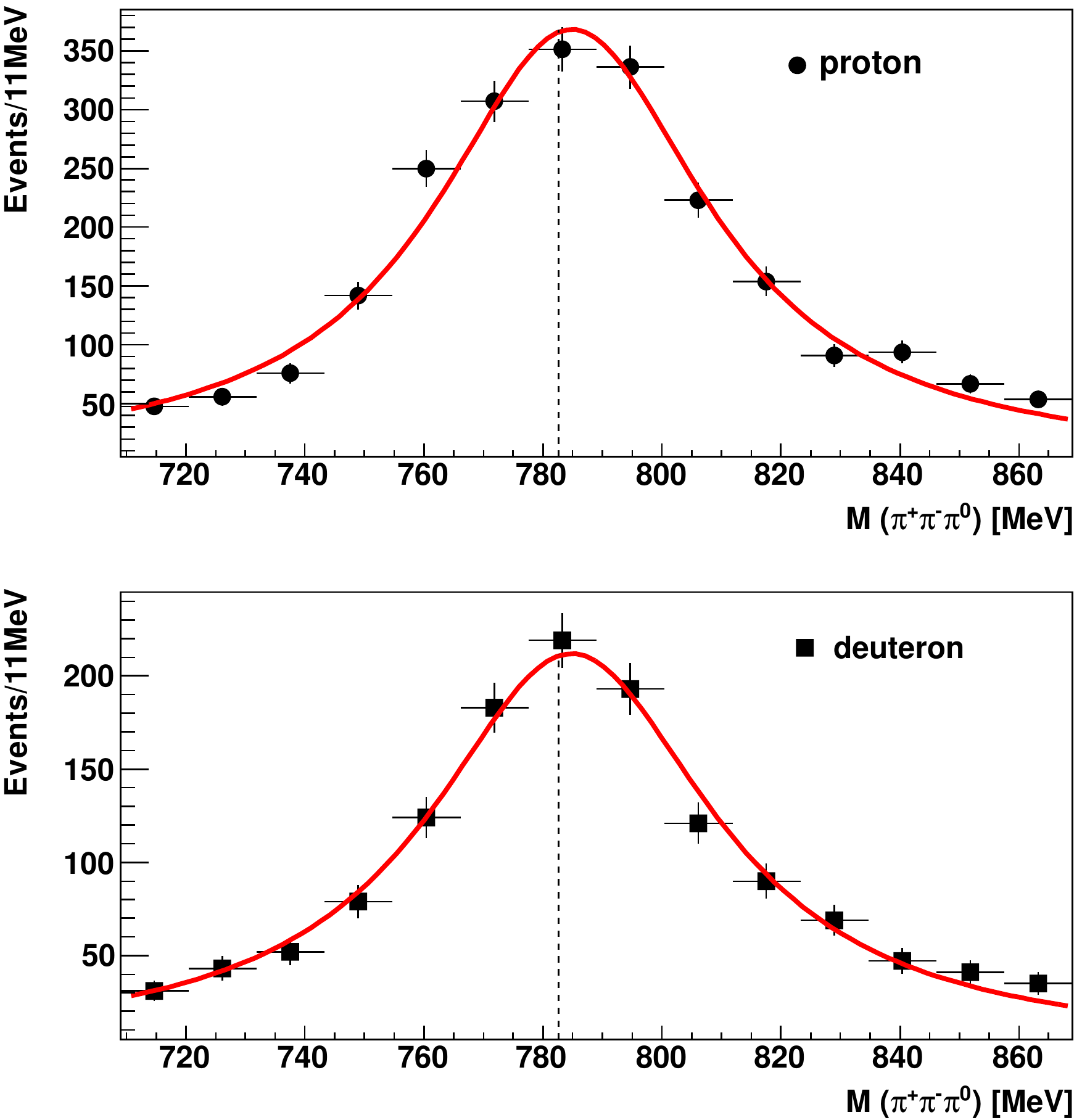}
\caption{Breit-Wigner fit (solid line) of  $ \pi^+ \pi^- \pi^0$
 invariant mass distributions after application of
all criteria to select $\omega$ mesons produced  exclusively from 
 proton (top) and from deuteron (bottom). The dashed line represents the PDG
 value  of the $\omega$ mass.}
\label{omega} 
\end{figure} 
   
\begin{figure}[hbtc!]
\centering
\includegraphics[width=.24\textwidth]{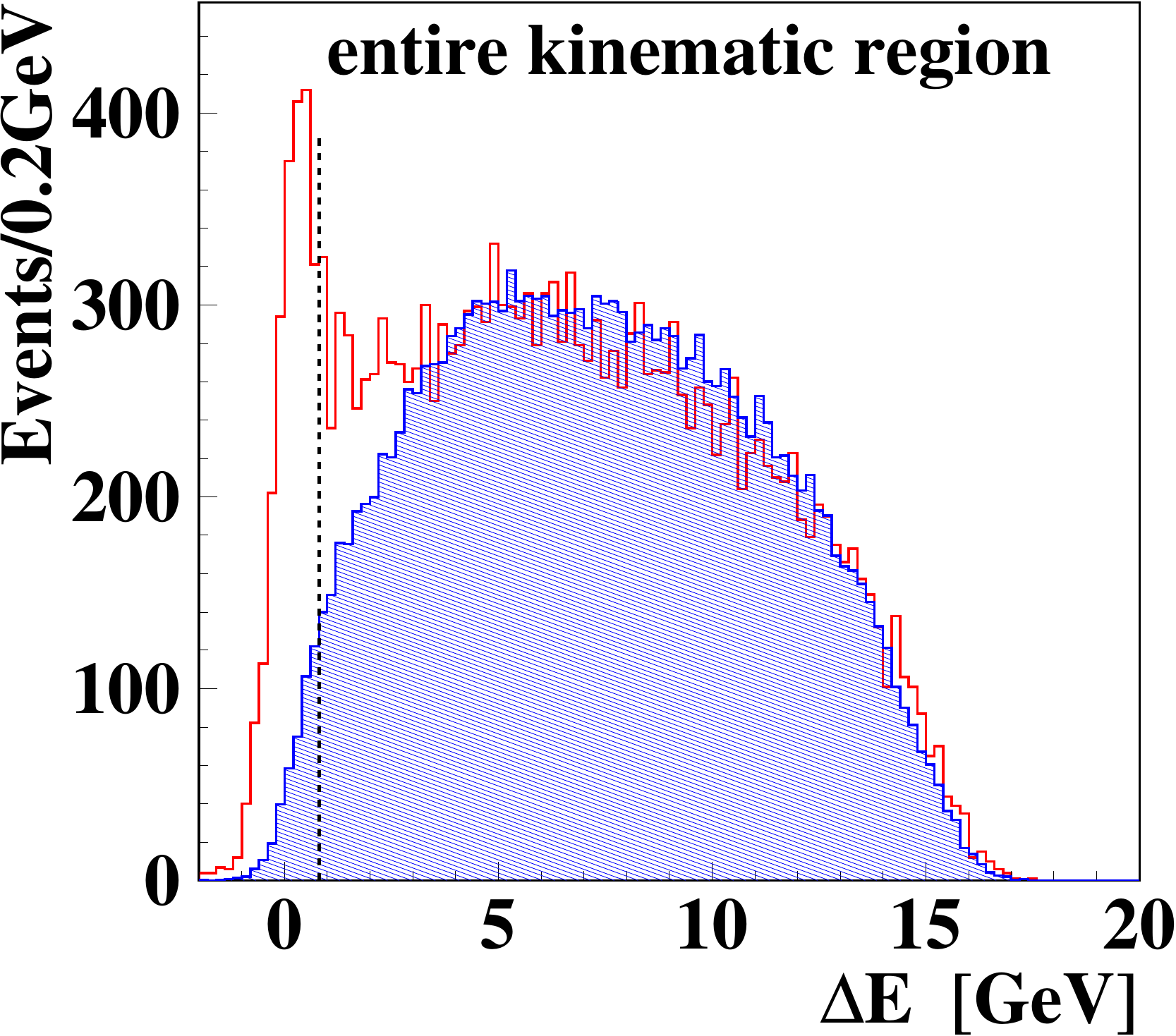}   
\includegraphics[width=.24\textwidth]{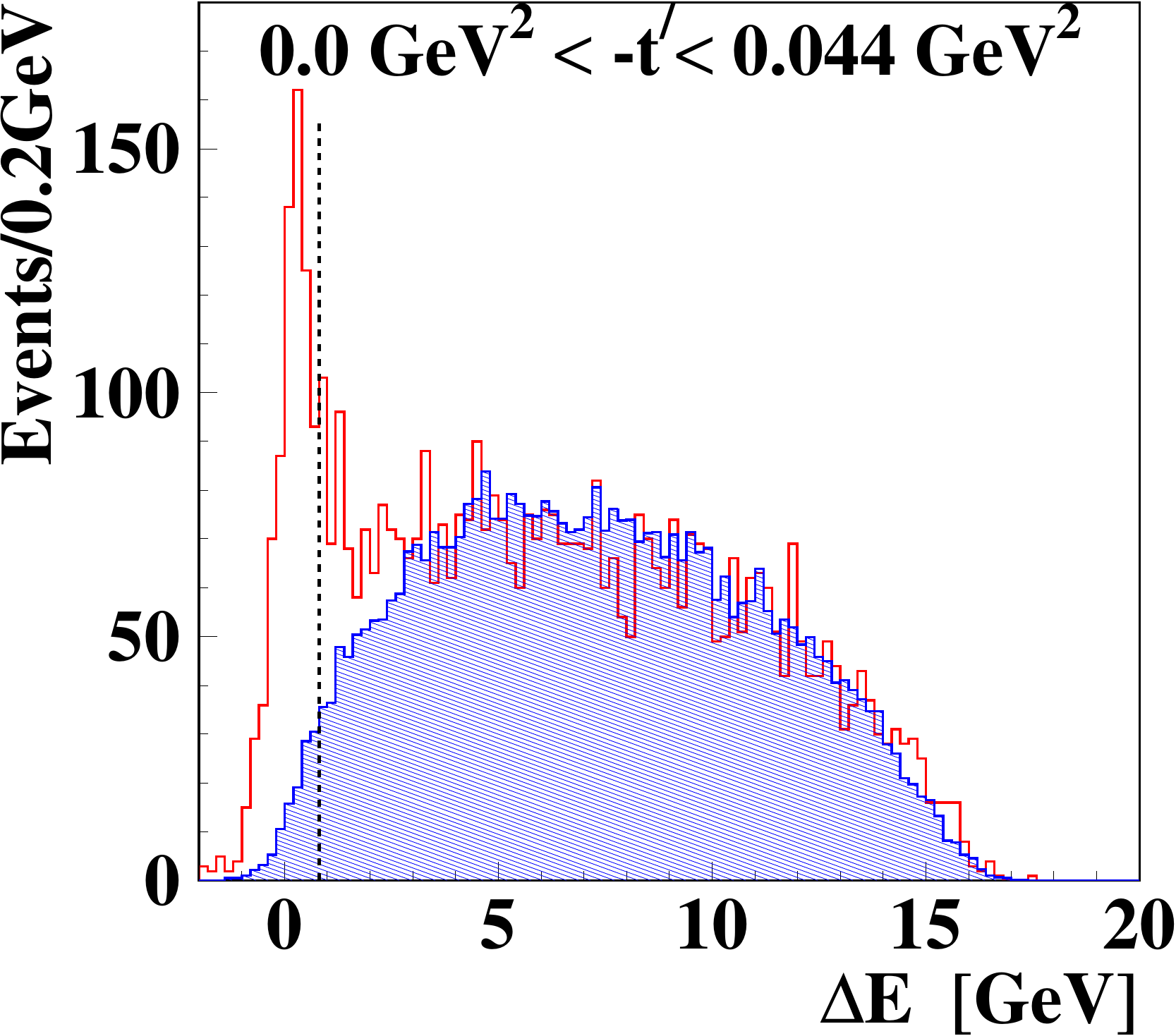}\\
\includegraphics[width=.24\textwidth]{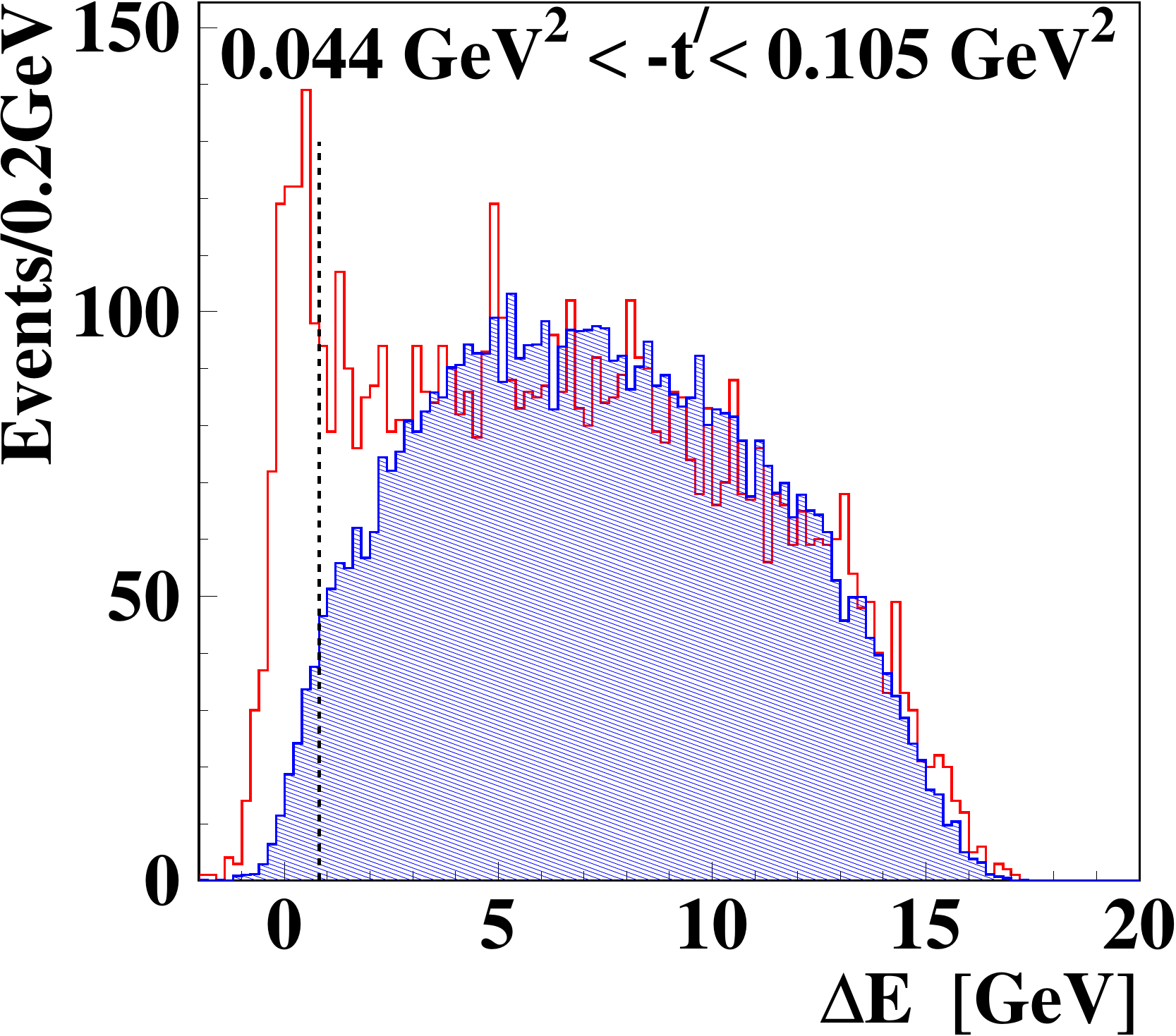}
\includegraphics[width=.24\textwidth]{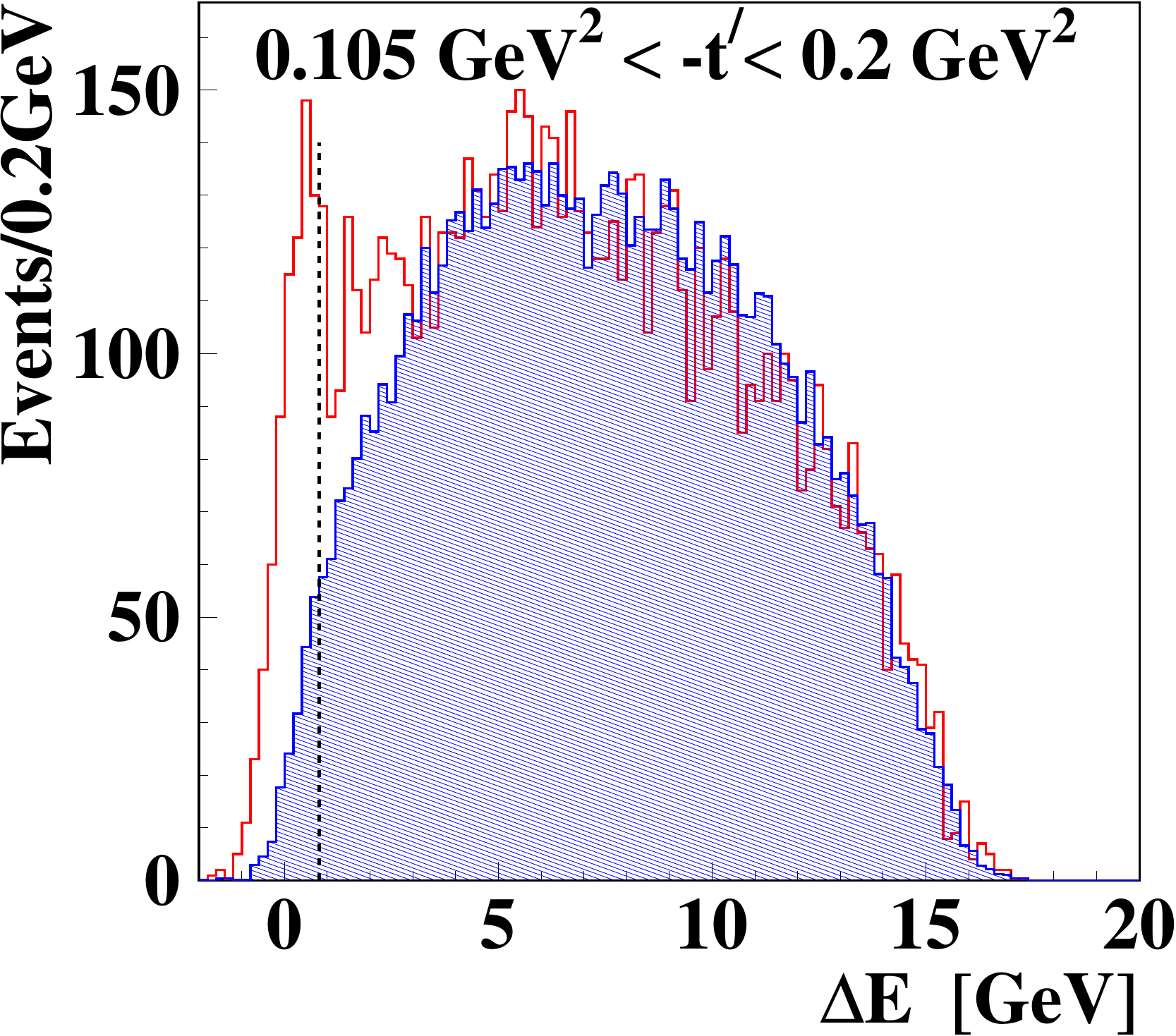}  
\caption{
The $\Delta E$  distributions of  $\omega$ mesons produced in the entire
kinematic region and in three kinematic bins in $-t'$ are
compared with SIDIS $\Delta E$  distributions from PYTHIA (shaded area).
The vertical dashed line denotes the upper limit of the exclusive region.}
\label{deltae}
\end{figure}

   After application of all these constraints, the proton sample contains 2260
and the deuteron  sample 1332 events of exclusively produced  $\omega$ mesons. 
These data samples are referred to in the following  as data in the ``entire
kinematic region".
The invariant-mass distributions  for exclusively produced $\omega$ mesons 
are shown in Fig.~\ref{omega}. Note the reasonable 
agreement of the fit result, $m_{\omega} = 784.8\pm55.8$~MeV for proton data and $m_{\omega} = 784.6\pm58.2$~MeV for deuteron data, 
with the PDG~\cite{pdg} value  of the  $\omega$ mass.
The distributions of missing energy $\Delta$E, 
shown in Fig.~\ref{deltae}, exhibit clearly visible exclusive peaks. The  shaded
histograms
represent semi-inclusive deep-inelastic scattering (SIDIS)
background obtained from a PYTHIA~\cite{pythia} Monte Carlo  simulation that is normalized to data
in the region $2~\text{GeV} < \Delta E <  20~\text{GeV}$. 
The simulation is  used to determine the fraction of background under the exclusive
peak, which is calculated as the ratio of number of background events to the total number of events.
It amounts to  about 20\% for the entire
kinematic region and increases  from 16\% to 26\% with increasing $-t'$. 

%%%%%Fig

%%%%%%%%%%%%%%
\subsection{Comparison of data and Monte Carlo events}

\begin{figure}[hbt]
\centering
\includegraphics[width=.48\textwidth]{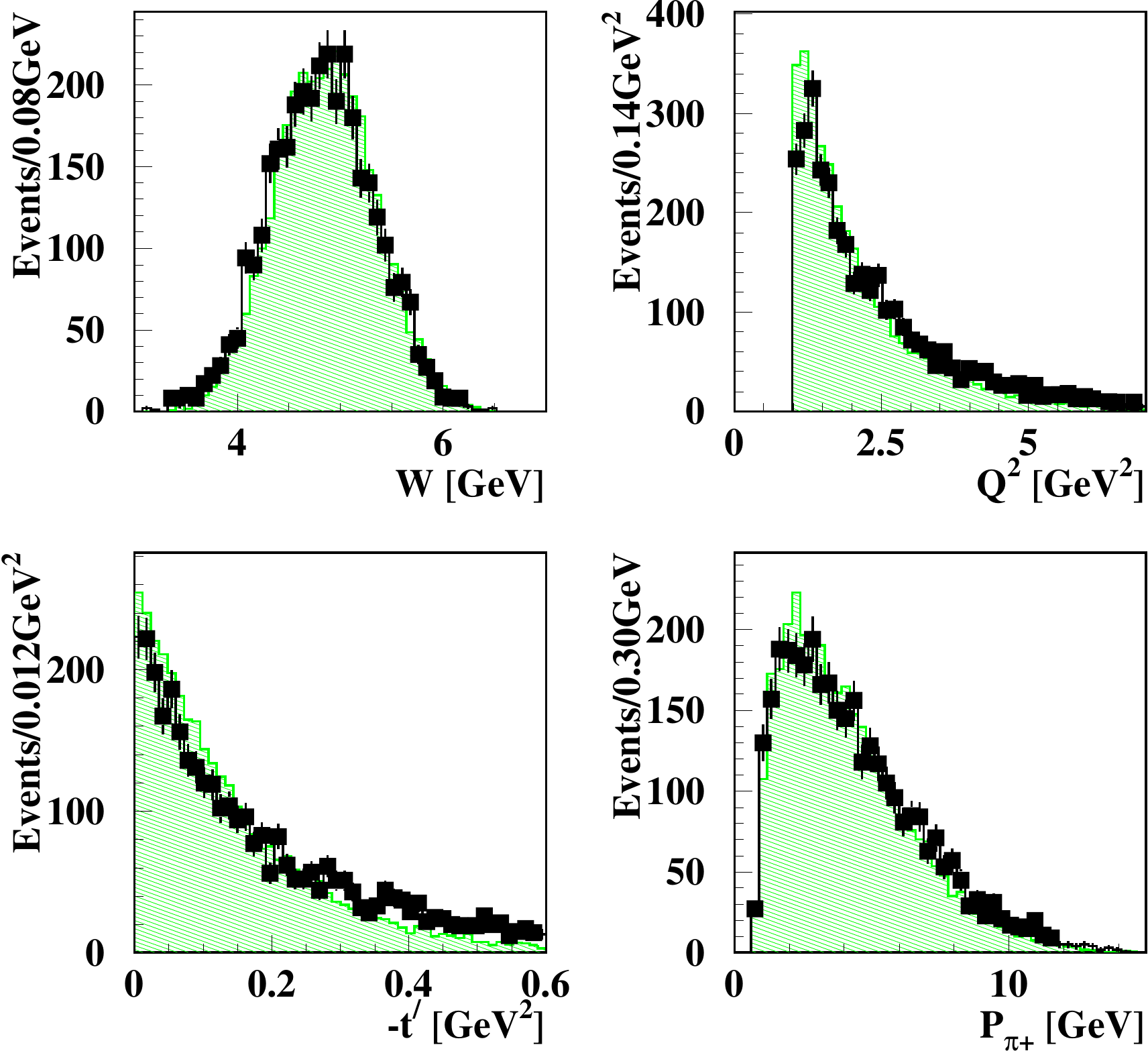} 
\includegraphics[width=.48\textwidth]{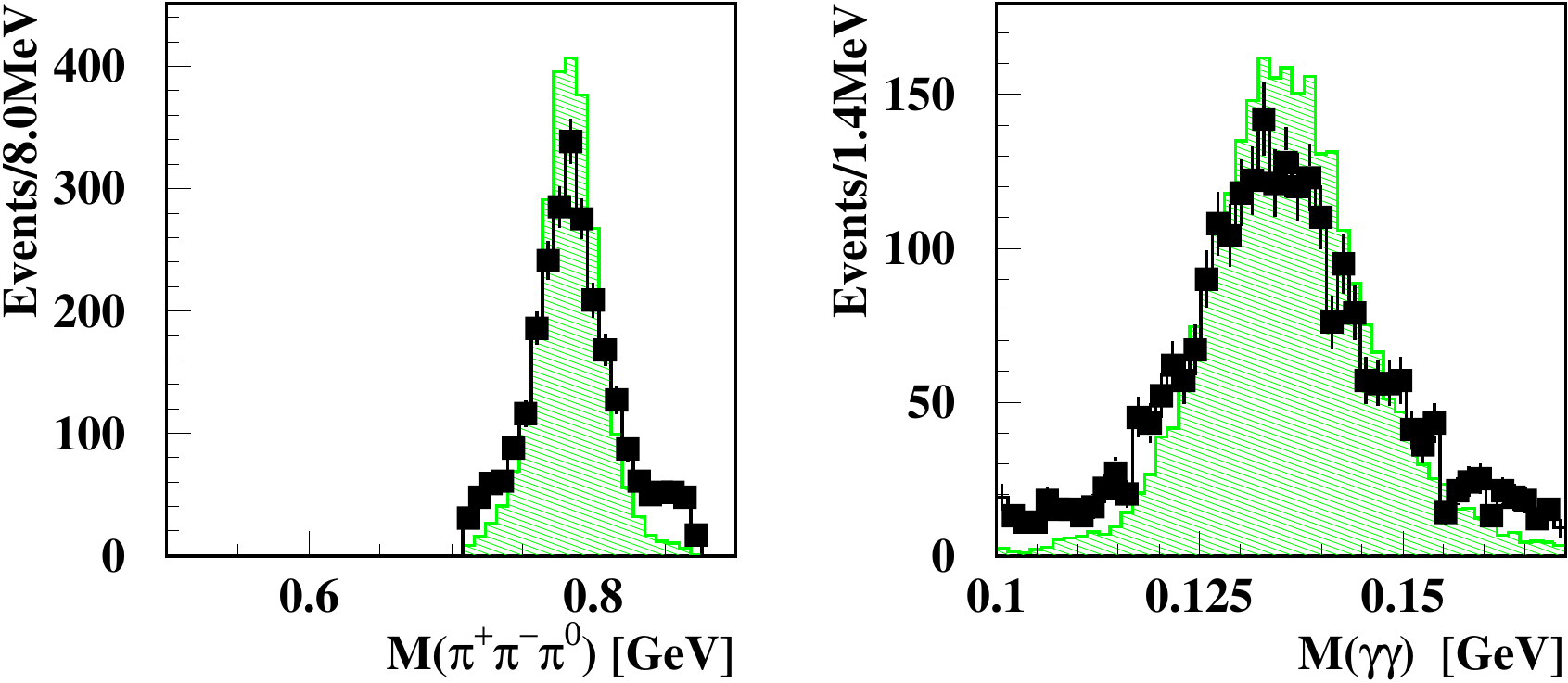} 
\caption{Distributions of several kinematic variables
from experimental  data on exclusive $\omega$-meson leptoproduction (black squares)
in comparison with simulated exclusive events from the  PYTHIA  generator (dashed areas).
Simulated events are normalized to the experimental data.}
\label{comdatpyt}
\end{figure}

 Distributions of experimental data in some kinematic variables are
compared to those simulated by  PYTHIA. The comparison is shown
in Fig.~\ref{comdatpyt} and  mostly demonstrates good agreement
between experimental and simulated data. 

%%%%%Fig0

%%%%%%%%%%%%%%
\section {Extraction of $\omega$ spin density matrix elements}
%%%%%%%%%%%%%%

\subsection{The unbinned  maximum likelihood method}

The SDMEs are extracted from data by fitting  the angular distribution
$\mathcal{W}^{U+L}(\Phi,\phi,\cos{\Theta})$ to the experimental angular
distribution using an unbinned  maximum likelihood method.
The probability distribution function is  
$\mathcal{W}^{U+L}({\cal R};\Phi,\phi,\cos{\Theta})$, where ${\cal R}$
represents the set of 23 SDMEs, i.e., the coefficients of the trigonometric
functions in Eqs.~(\ref{eqang2}, \ref{eqang3}).
The negative log-likelihood function to be minimized reads  
\begin{equation} 
-\ln L({\cal R})=
-\sum_{i=1}^{N}\ln\frac{\mathcal{W}^{U+L}({\cal R};\Phi_{i},\phi_{i},\cos{\Theta_{i}})}{\widetilde{\mathcal
N}({\cal R})},
\label{loglik-def}
\end{equation}
where the normalization factor
\begin{equation}
 \widetilde {\mathcal N}({\cal R})=
\sum_{j=1}^{N_{MC}}\mathcal{W}^{U+L}({\cal R};\Phi_{j},\phi_{j},\cos{\Theta_{j}})
\label{loglik-def1}
\end{equation}
is calculated numerically using events from a PYTHIA Monte Carlo generated
 according to an isotropic three-dimensional  angular distribution and 
passed through the same analytical process as experimental data. The numbers of data  and Monte Carlo events are
 denoted by ${N}$ and  $N_{MC}$, respectively.

%%%%%%%%%%%%%%
\subsection{ Background treatment}

 In order to account for the SIDIS background in the fit, first ``SIDIS-background
SDMEs" are obtained using Eqs. (\ref{loglik-def}, \ref{loglik-def1}) for the PYTHIA SIDIS sample
in the exclusive region. Then, SDMEs corrected for SIDIS background are  obtained
as follows~\cite{phdami}:
\begin{eqnarray} 
-\ln L({\cal R})=\nonumber ~~~~~~~~~~~~~~~~~~~~~~~~~~~~~~~~~~~~~~~~~~~~~~~~~~~\\
-\sum_{i=1}^{N}\ln\Bigl[ \frac{ (1-f_{bg})*\mathcal W^{U+L}({\cal R};\Phi_{i},\phi_{i},\cos\Theta_{i})}
{\widetilde {\mathcal N}({\cal R},\Psi)} \nonumber\\
+\frac{f_{bg}*\mathcal W^{U+L}(\Psi; \Phi_{i},\phi_{i}, \cos \Theta_{i})}
{\widetilde{\mathcal N}({\cal R},\Psi)}\Bigr].~~~
\label{logbac}
\end{eqnarray}
From now on, 
${\cal R}$ denotes the set of  SDMEs corrected for background,
$\Psi$ the set of  the SIDIS-background SDMEs,  
and  $f_{bg}$ is the fraction of SIDIS  background.
The normalization factor reads correspondingly 
\begin{eqnarray}
\widetilde{\mathcal N}({\cal R},\Psi)=
\sum_{j=1}^{N_{MC}}\left[(1-f_{bg})*\mathcal W^{U+L}({\cal R}; \Phi_{j},
\phi_{j},\cos\Theta_{j})\right.\nonumber\\
\left. +f_{bg}*\mathcal W^{U+L}(\Psi; \Phi_{j}, \phi_{j},\cos\Theta_{j})\right].~~~                
\label{logbacnor}
\end{eqnarray}

%%%%%%%%%%%%%%
\subsection{Systematic uncertainties}
 
The total systematic uncertainty on a given extracted SDME $r$ is obtained by adding in
quadrature the uncertainty
 from the background subtraction procedure, $\Delta r_{sys}^{bg}$,  and the
 one  due to the extraction
 method, $\Delta r_{sys}^{MC}$.
The former uncertainty is assigned to be  the difference between the SDME obtained with and without
background correction. This conservative approach also covers the small
uncertainty on the fraction of  SIDIS background, $f_{bg}$.
 The uncertainty $\Delta r_{sys}^{MC}$  is estimated using
the Monte Carlo data that  were generated  with
an angular distribution determined  by the  set  of SDMEs ${\cal R}$.
The statistics of the Monte Carlo data exceed those of the experimental
data by about a factor of six. The generated events were passed through a realistic model of the HERMES apparatus
using GEANT~\cite{geant} and were then reconstructed and analyzed in the same way as experimental data. 
These Monte Carlo data were used to extract the SDME set ${\cal R}^{MC}$. 
 In this way,  effects from  detector acceptance, efficiency, smearing, and misalignment are
 accounted for. 
Two  uncertainties are considered to be responsible for the difference between
input and output value of a given  SDME $r$, 
\begin{equation}
(r - r^{MC})^2 = (\Delta r_{sys}^{MC})^2 + (\Delta r_{stat}^{MC})^2,
\label{sysextr}
\end{equation}
where $\Delta r_{stat}^{MC}$ is the  statistical uncertainty of
$r^{MC}$ as obtained in the fitting procedure that uses MINUIT~\cite{minuit}.
 From Eq.~(\ref{sysextr}), $\Delta r_{sys}^{MC}$ is determined, using the
 convention that
$\Delta r_{sys}^{MC}$ is set to zero if $[(r -
r^{MC})^2- (\Delta r_{stat}^{MC})^2]$ is negative. 

%%%%%%%%%%%%%%
\section{Results}\label{sec-2}
%%%%%%%%%%%%%%

The results on SDMEs in the Schilling-Wolf~\cite{Schill}  representation are given in
Tables~\ref{tab1}-\ref{tab5} in Appendix~\ref{sec:tables} and in the Diehl~\cite{Diehl} representation in 
Table~\ref{tab6} in the same Appendix. The SDMEs for the entire kinematic region are discussed 
in Sect.~\ref{sec_sdme_fullkin}, 
while their dependences on $ Q^2$ and $-t'$ are discussed in Sect.~\ref{sec_sdme_kindep}.

%%%%%%%%%%%%%%
\subsection{SDMEs for the entire kinematic region}\label{sec_sdme_fullkin}

%%%%%FIG06
\begin{figure*}\centering
\includegraphics[width=.75\textwidth]{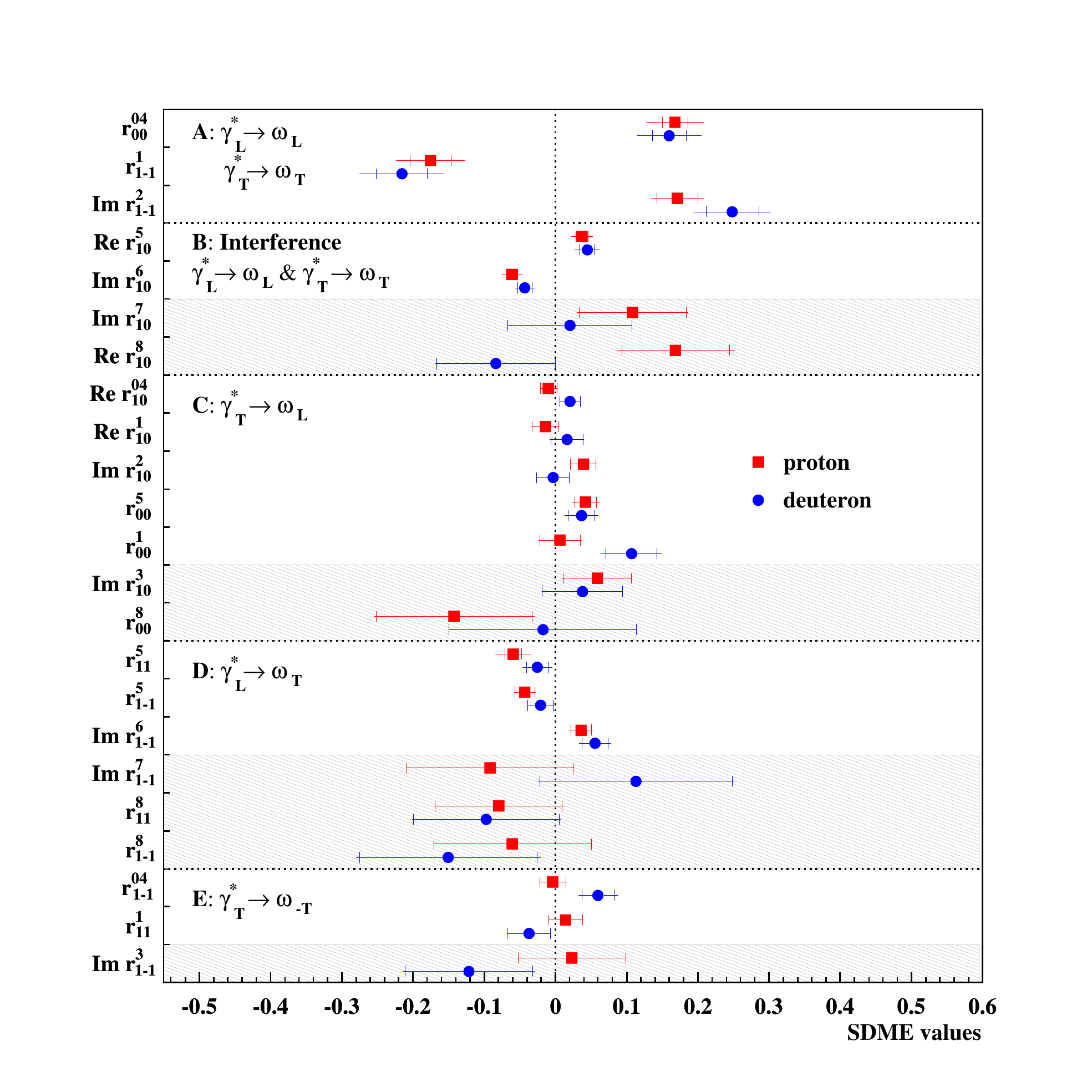}
\caption{The 23 SDMEs for exclusive $\omega$ electroproduction extracted in the entire  HERMES kinematic
region with $ \langle Q^2 \rangle = 2.42$~GeV$^2$, $\langle W\rangle =4.8$~GeV, 
$\langle -t'\rangle = 0.080$~GeV$^2$. Proton data are denoted by squares and deuteron data by circles.
The inner error bars represent the statistical uncertainties, while the
outer ones indicate the statistical and systematic uncertainties added in
quadrature. Unpolarized (polarized) SDMEs are displayed in the unshaded
(shaded) areas.}
\label{sdmescaled}
\end{figure*}

The  SDMEs of the $\omega$ meson  for the entire  kinematic region
($\left< Q^{2} \right>= 2.42$~GeV$^2$, $\left< W \right>= 4.8$~GeV, and
$\left<-t^{\prime} \right> = 0.080$~GeV$^2$) are presented in Fig.~\ref{sdmescaled}.
These SDMEs are divided into five classes corresponding to different
helicity transitions.
The main terms in the expressions of class-A SDMEs correspond to the transitions from
longitudinal virtual photons to  longitudinal vector mesons, $\gamma^*_L \to V_L$,
and from transverse virtual photons to transverse vector mesons, $\gamma^*_T \to V_T$.
The dominant terms of class B correspond to the interference of these two transitions. 
The main terms of class-C, class-D, and  class-E SDMEs are proportional to small
amplitudes describing $\gamma^*_T \to V_L$, $\gamma^*_L \to V_T$, and
$\gamma^*_T \to V_{-T}$ transitions respectively.\\   
\indent The SDMEs for the proton and deuteron data are
found to be  consistent with each other within their quadratically combined total uncertainties, 
with a $\chi^2$ per degrees of freedom of $28/23  \approx 1.2$. 
 In Fig.~\ref{sdmescaled}, the eight polarized SDMEs are presented in  shaded areas.
Their experimental uncertainties are larger
in comparison to those of the unpolarized  SDMEs because the lepton beam polarization is
smaller than unity ($|P_{b}| \approx 40\%$) and in the equation for the  angular distribution they are
multiplied by the small kinematic factor $|P_{b}|\sqrt{1- \epsilon}\approx 0.2 $, 
cf.~Eq.~(\ref{eqang2}) vs.~Eq.~(\ref{eqang3}).

%%%%%%%%%%%%%%
\subsection{Test of the SCHC hypothesis}
\label{sec_schc}

  In the case of SCHC, the seven SDMEs of class A and class B ($r_{00}^{04}$, $r_{1-1}^1$, $\mathrm{Im}
\{r_{1-1}^2\}$, $\mathrm{Re}\{r_{10}^5\}$, $\mathrm{Im}\{r_{10}^{6}\}$,
$\mathrm{Im} \{r_{10}^{7}\}$, $\mathrm{Re}\{r_{10}^{8}\}$) are not restricted to be
zero, but six of them have to obey the following relations~\cite{Schill}:
\begin{eqnarray}
r_{1-1}^1 &=&-\mathrm{Im} \{r_{1-1}^2\},\nonumber\\
\mathrm{Re}\{r_{10}^5\} &=&-\mathrm{Im}\{r_{10}^{6}\},\nonumber\\
\mathrm{Im} \{r_{10}^{7}\} &=& \mathrm{Re} \{r_{10}^{8}\}.\nonumber
\end{eqnarray}
The proton data yield  
\begin{eqnarray}
r^{1}_{1-1} +\mathrm{Im}\{r^{2}_{1-1}\}&=& -0.004 \pm 0.038  \pm0.015,\nonumber\\ 
 \mathrm{Re}\{r^{5}_{10}\} +  \mathrm{Im}\{r^{6}_{10}\}&=& -0.024 \pm  0.013 \pm0.004 ,\nonumber\\
 \mathrm{Im}\{r^{7}_{10}\} -\mathrm{Re}\{r^{8}_{10}\} &=&-0.060 \pm  0.100 \pm0.018,  \nonumber
\end{eqnarray}
and the deuteron data yield 
\begin{eqnarray}
r^{1}_{1-1} +\mathrm{Im}\{r^{2}_{1-1}\}&=& 0.033 \pm  0.049\pm 0.016,\nonumber\\
 \mathrm{Re} \{r^{5}_{10}\} +  \mathrm{Im} \{r^{6}_{10}\}&=& 0.001 \pm 0.016 \pm0.005,\nonumber\\
 \mathrm{Im} \{r^{7}_{10}\} -\mathrm{Re} \{r^{8}_{10}\} &=& 0.104
 \pm0.110\pm0.023. \nonumber
\end{eqnarray}
Here and in the following, the first uncertainty is statistical
and the  second  systematic. In the calculation of the statistical uncertainty, the correlations between the different SDMEs 
are taken into account, see correlation matrices in Tables~\ref{tab8} and \ref{tab9}. 
It can  be concluded that the above SCHC relations are 
fulfilled for class A and B.
The SCHC relations for the class-A SDMEs $r^1_{1-1}$ and
$\mathrm{Im} \{r^2_{1-1}\}$
can be violated only by the quadratic contributions of the double-helicity-flip amplitudes
$T_{1\pm\frac{1}{2}-1 \frac{1}{2}}$ and
$U_{1 \pm\frac{1}{2}-1 \frac{1}{2}}$ with  
$|\lambda_V-\lambda_{\gamma}|=2$.
The observed validity of SCHC means that their possible  contributions are smaller than the
experimental uncertainties. Also for class-B SDMEs,  
 to which the same small double-helicity-flip amplitudes contribute
 linearly, no SCHC violation is observed. 
 In addition, class-B SDMEs  contain the contribution of the 
two small products  $T_{0 \pm\frac{1}{2}1 \frac{1}{2}}
T^*_{1 \pm\frac{1}{2}0 \frac{1}{2}}$
($U_{0 \pm\frac{1}{2}1 \frac{1}{2}}U^*_{1 \pm\frac{1}{2}0 \frac{1}{2}}$).
As the SCHC hypothesis is fulfilled, all these contributions are concluded to be negligibly small
compared to the experimental uncertainties. This validates the assumption made in Sect.~\ref{hel_ampli} that the 
double-helicity-flip amplitudes can be neglected.

All SDMEs of class C to E have to be zero in
the case of SCHC. The class-C SDME $r^5_{00}$ deviates from zero by about
three standard deviations for the proton and two standard deviations for the
deuteron (see Fig.~\ref{sdmescaled}). Since the numerator of the equation for $r_{00}^{5}$~\cite{DC-24}, 
\begin{equation}
r^{5}_{00}=\frac{\mathrm{Re} \left\{T_{0-\frac{1}{2}1 \frac{1}{2}} T^*_{0-\frac{1}{2}0\frac{1}{2}}+
T_{0\frac{1}{2}1 \frac{1}{2}} T^*_{0\frac{1}{2}0 \frac{1}{2}}\right\}}{\cal{N}},  
\label{r500}
\end{equation}
contains two amplitude products, at least one product is nonzero.
 However, without an  amplitude analysis of the presented data it cannot be concluded which contribution to $r^5_{00}$ dominates.
 Both amplitudes $T_{0-\frac{1}{2}1 \frac{1}{2}}$ and
$T_{0\frac{1}{2}1 \frac{1}{2}}$ have to be zero if the SCHC hypothesis
holds.

Figure \ref{sdmescaled} shows  that out of  the six SDMEs of class D three, i.e.,  
 $r_{11}^5$, $r_{1-1}^5$, and  $\mathrm{Im}\{r_{1-1}^6\}$, slightly differ from zero (see Table~\ref{tab1}). 
As will be discussed in Sections~\ref{sec:UPE} and \ref{sec:hierarchy},
 the largest UPE amplitudes in  $\omega$ production are $U_{11}$ and
 $U_{10}$, and 
 $|U_{11}|\gg|U_{10}|$. The main term of the first two SDMEs is proportional
 to $\mathrm{Re}[U_{10}U_{11}^*]$, 
while  $\mathrm{Im}\{r_{1-1}^6\}$  is proportional to $-\mathrm{Re}[U_{10}U_{11}^*]$. 
The calculated linear combination of these three
 SDMEs, $r_{11}^5+r_{1-1}^5-\mathrm{Im}\{r_{1-1}^6\}$, is $-0.14 \pm 0.03 \pm
 0.04$  for the proton  and  $-0.10 \pm 0.03 \pm 0.03$ for the deuteron. 
These values differ from zero by about three standard deviations of the total uncertainty 
for the proton.
This, together with the experimental information on measured class-C and class-D SDMEs, indicates a 
violation of the SCHC hypothesis in exclusive $\omega$ production.

%%%%%%%%%%%%%%
\subsection{Dependences of SDMEs on $Q^{2}$ and $-t'$ and comparison to a phenomenological model}
\label{sec_sdme_kindep}

%%%%%%%%%FIG
\begin{figure*}[hbtc!]
\centering
\includegraphics[width=.8\textwidth]{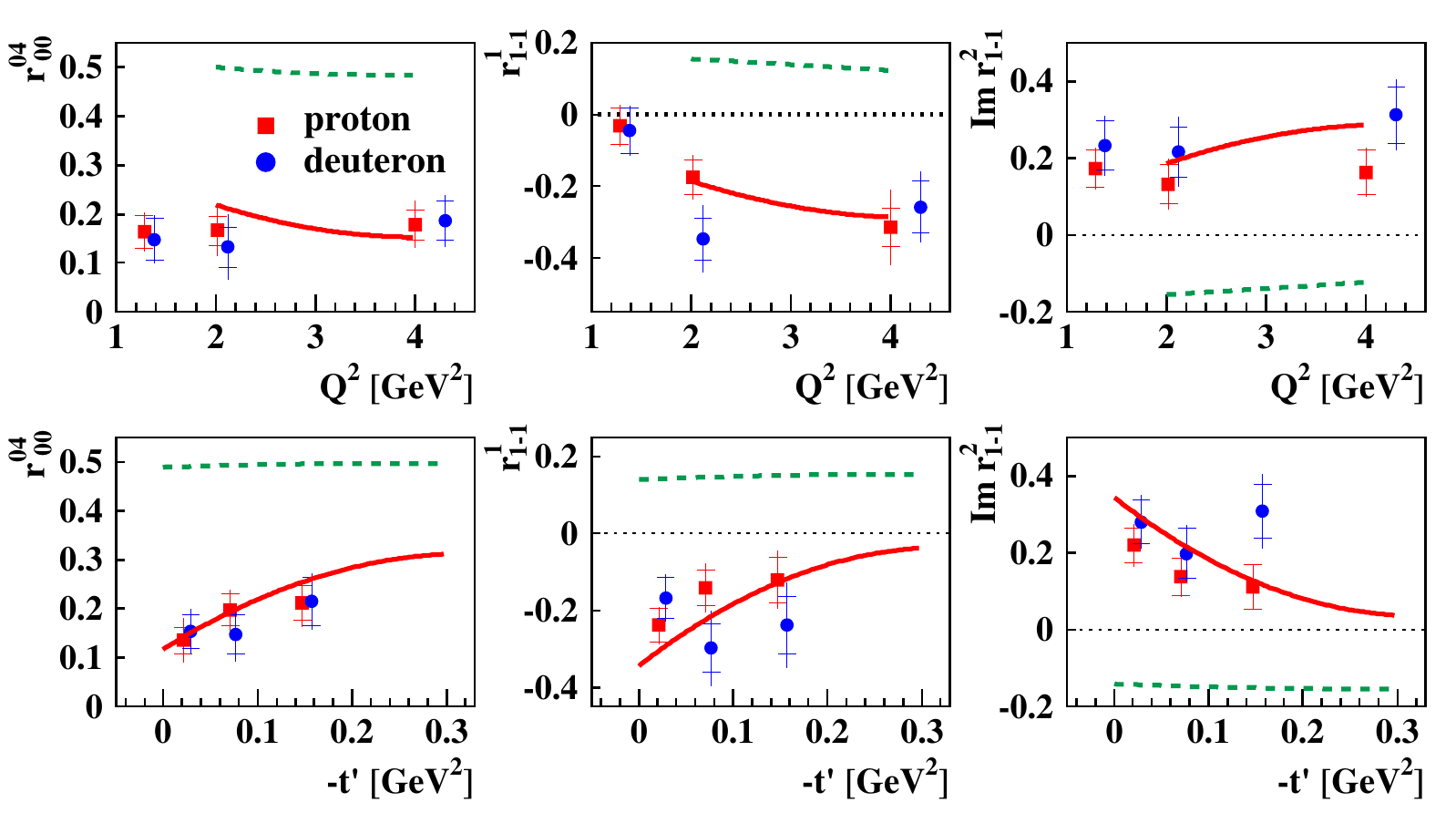}                     
\caption{Q$^{2}$ and $-t'$  dependences of  class-A SDMEs.
Proton data are denoted by squares and deuteron data  by
circles. Data points for deuteron data are slightly shifted horizontally for legibility. 
The representation of the uncertainties follows that of Fig.~\ref{sdmescaled}. The proton data are compared to calculations of a
phenomenological model~\cite{thcal}, where solid (dashed) lines
denote results with (without) pion-pole contributions.
}
\label{Aq2t}   
\end{figure*}

In the following sections, kinematic dependences of the
measured SDMEs and certain combinations of them are presented
and interpreted wherever possible. In particular,  the proton
data presented in this paper are compared to the calculations of the phenomenological GK  model described in
Sect.~\ref{intro}.
In each case, model calculations are   shown with and without inclusion of  
the pion-pole contribution. 
 In order to stay in the framework of handbag factorization and to avoid large $1/Q^2$ corrections, model calculations are only shown for $Q^2 > 2$~GeV$^2$, 
which leaves for the $Q^2$ dependence only two data points that can be compared to the model calculation. This paucity of comparable points makes it 
sometimes difficult to draw useful conclusions about the 
data-model comparison.

The kinematic dependences of SDMEs on $Q^{2}$ and $-t'$ are
presented  in
three bins of $Q^{2}$ with  
$\langle Q^2 \rangle = 1.28$ GeV$^2$, 
$\langle Q^2 \rangle = 2.00$ GeV$^2$, 
$\langle Q^2 \rangle = 4.00$ GeV$^2$, 
and  $t'$ with 
$\langle -t' \rangle = 0.021$ GeV$^2$, 
$\langle -t' \rangle = 0.072$ GeV$^2$,
$\langle -t' \rangle = 0.137$ GeV$^2$.
Table~\ref{tab7} shows the average value of $Q^2$ and $-t'$ for bins in $-t'$ and $Q^2$, respectively.

%%%%%%%%%FIG
\begin{figure*}[hbtc!]
\centering
\includegraphics[width=.8\textwidth]{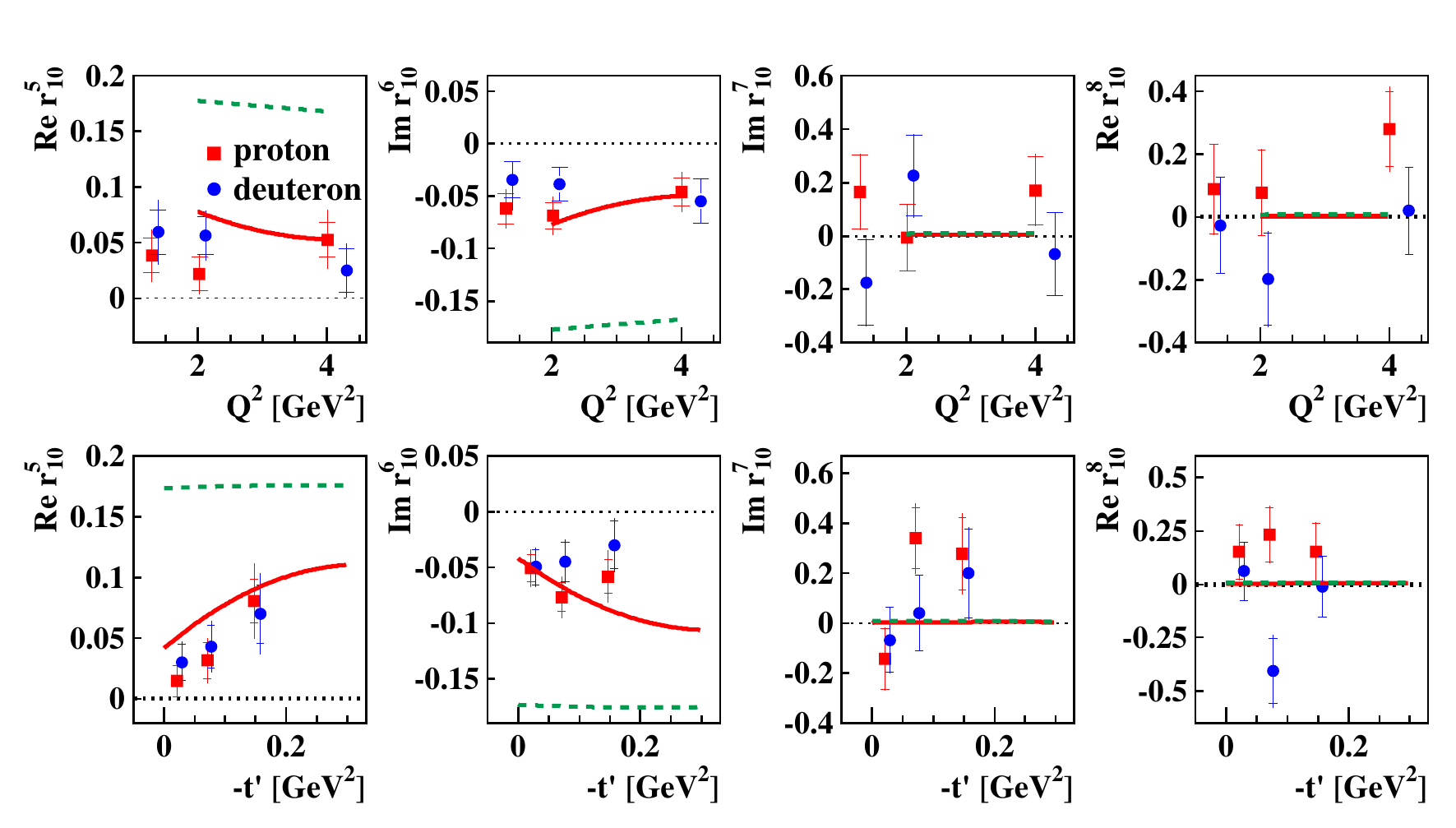}                     
\caption{$Q^{2}$ and $-t'$  dependences of  class-B SDMEs.
Otherwise as for Fig.~\ref{Aq2t}.}
\label{Bq2t}   
\end{figure*}

\begin{figure*}[hbtc!]
\centering
\includegraphics[width=.8\textwidth]{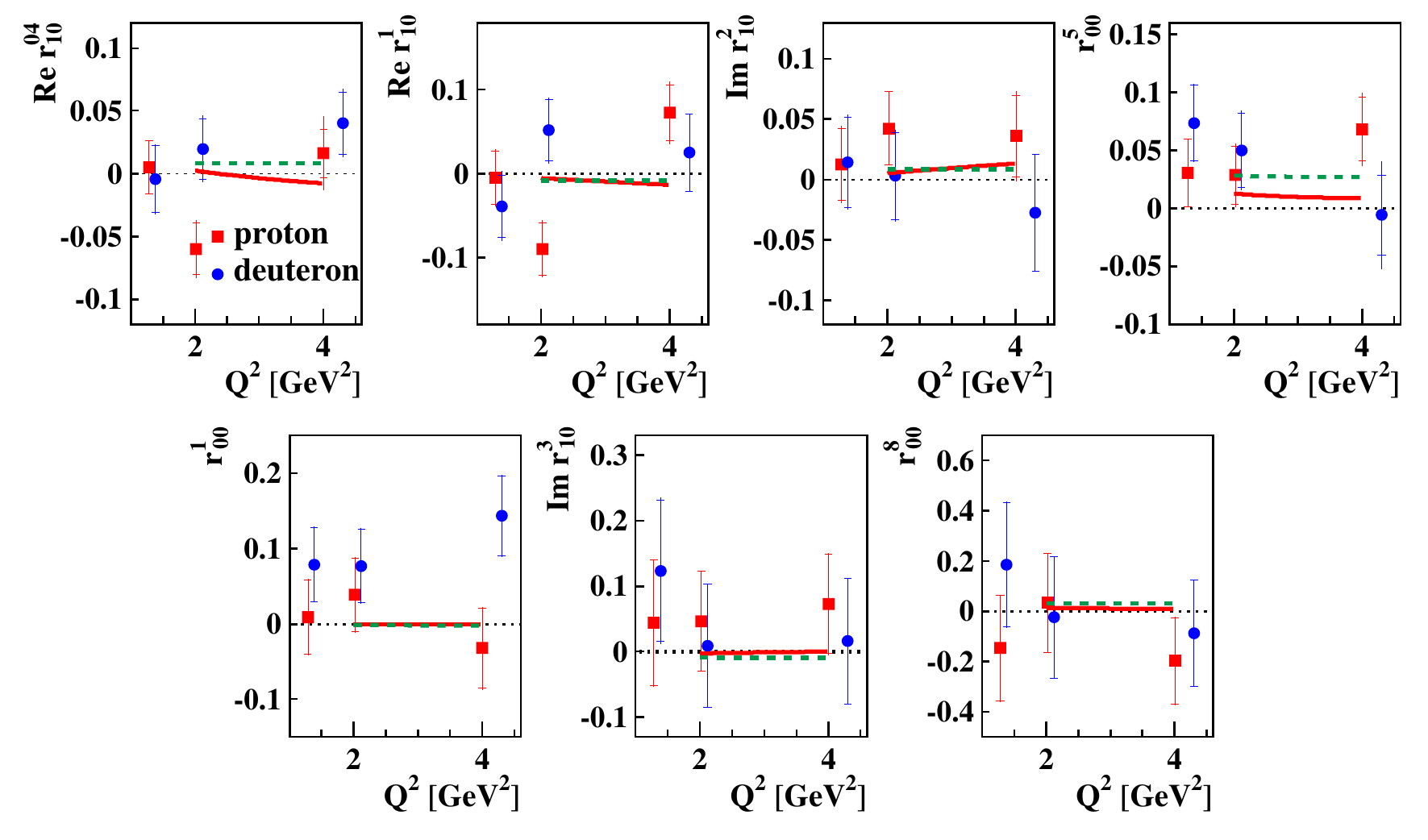}                     
\includegraphics[width=.8\textwidth]{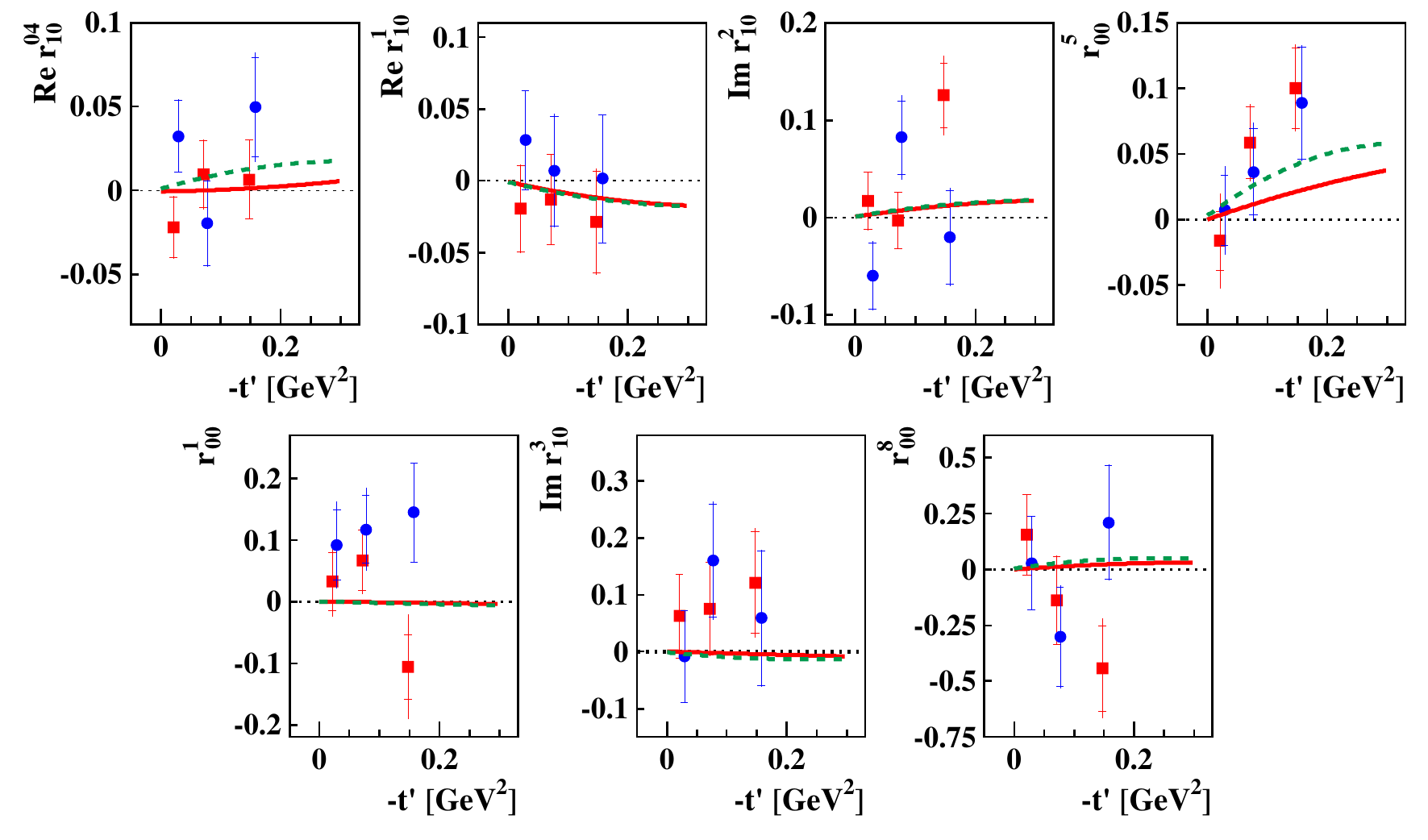}                                                                  
\caption{$Q^{2}$ and $-t'$ dependences of  class-C SDMEs. Otherwise as for Fig.~\ref{Aq2t}.}
\label{Cq2t}   
\end{figure*}

\begin{figure*}[hbtc!]
\centering
\includegraphics[width=.8\textwidth]{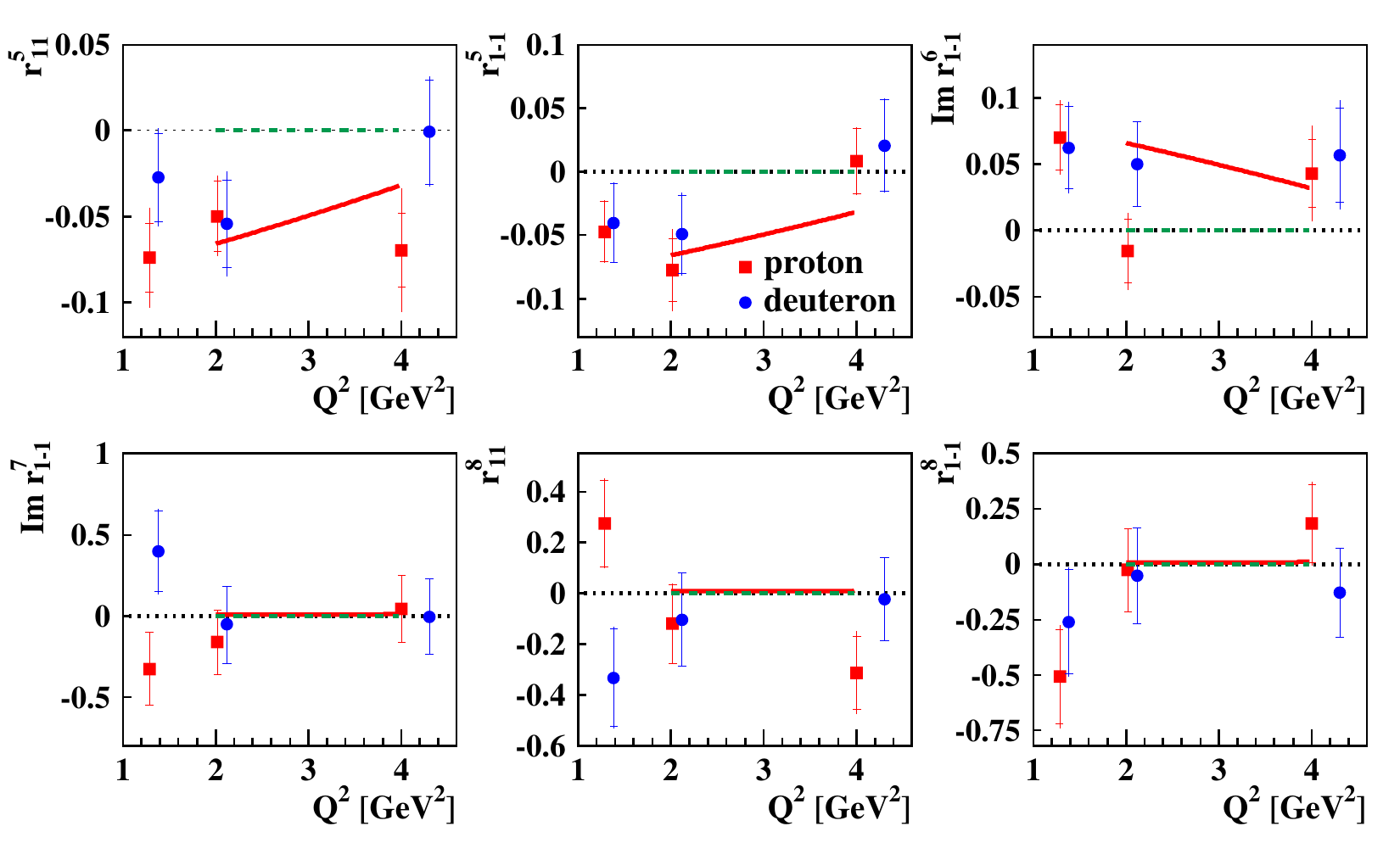}                      
\includegraphics[width=.8\textwidth]{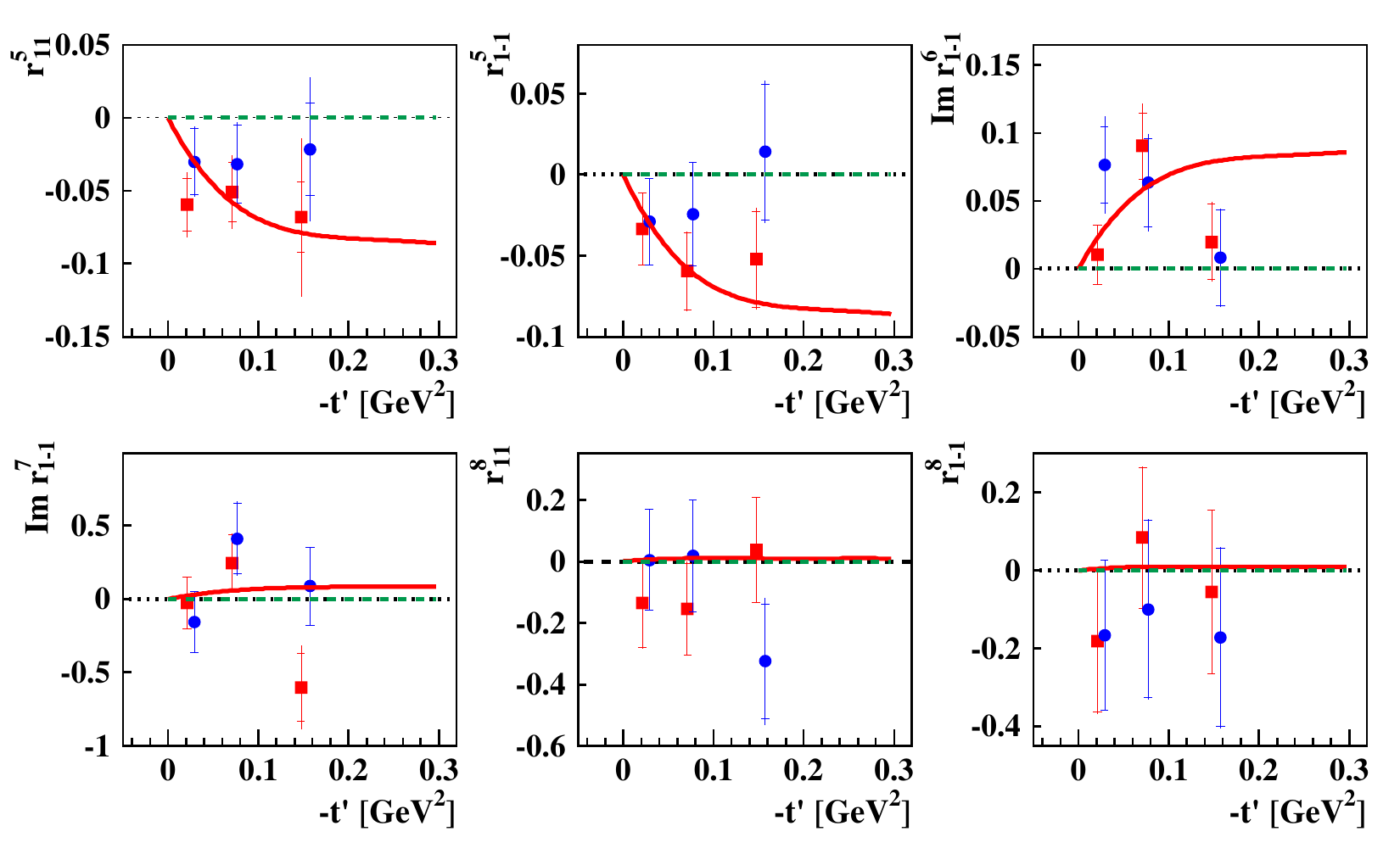}                                                                  
\caption{ $Q^{2}$ and $-t'$  dependences of  class-D SDMEs. Otherwise as for Fig.~\ref{Aq2t}.}
\label{Dq2t}   
\end{figure*}

\begin{figure*}[hbtc!]
\centering
\includegraphics[width=.8\textwidth]{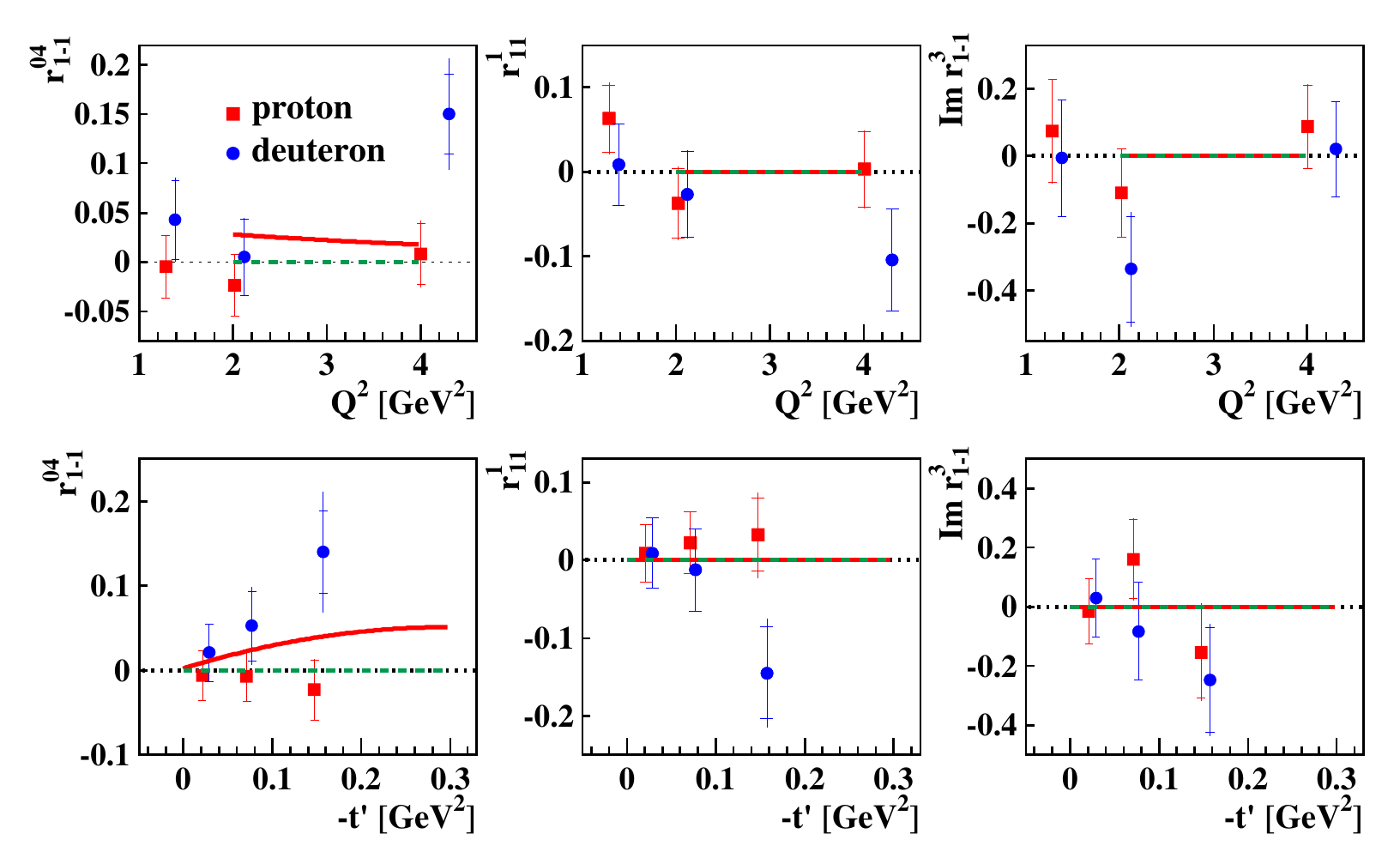}
\caption{$Q^{2}$ and $-t'$  dependences of  class-E SDMEs. Otherwise as for Fig.~\ref{Aq2t}.}
\label{Eq2t}
\end{figure*}

The $Q^2$ and $-t'$ dependences of class-A SDMEs are shown 
and compared to the model calculations in Fig.~\ref{Aq2t}. All three SDMEs clearly show the 
need for the unnatural-parity contribution of the pion pole and the 
measured $-t'$ dependence is well reproduced both in shape and magnitude. 
The same holds for the two unpolarized class-B SDMEs that are shown in 
Fig.~\ref{Bq2t}. For the polarized class-B SDMEs as well as for all class-C 
SDMEs, which are shown in Fig.~\ref{Cq2t}, the pion-pole contribution has  
little or no effect, and the model describes the magnitude of the data 
reasonably well. The class-D and E SDMEs  are shown in Figs.~\ref{Dq2t} and \ref{Eq2t}, 
respectively. These SDMEs are expected to be zero if the pion-pole 
contribution is not included. When comparing the $-t'$ dependences of 
the three unpolarized class-D SDMEs to the model calculation, also here 
the unnatural-parity pion-exchange contribution seems to be required. 
The two unpolarized class-E SDMEs are measured with reasonable precision, and 
agreement with the model calculation can be seen.

Within experimental uncertainties, the SDMEs measured on the proton are seen to be very similar to those measured on the deuteron.
This can be understood by considering the different contributions to exclusive omega production. 
The pion-pole contribution is seen to be substantial~\cite{thcal}.
For the NPE amplitudes, the dominant contribution comes from gluons and sea quarks, which are the same for protons and neutrons, while the valence-quark contribution is different. Thus altogether, only small differences between the proton and deuteron SDMEs are expected for incoherent scattering. As coherence effects are difficult to estimate, one can not exclude that they are of the size of the valence-quark effects. Therefore, the deuteron SDMEs are presently difficult to calculate reliably.

%Within experimental accuracy, no conclusions can be drawn concerning possible differences in the kinematic dependences between proton and deuteron data.

%%%%%%%%%FIG

%%%%%%%%%FIG

%%%%%%%%%FIG

\subsection{UPE in $\omega$-meson production}\label{sec:UPE}

\begin{figure*}\centering
\includegraphics[width=.8\textwidth]{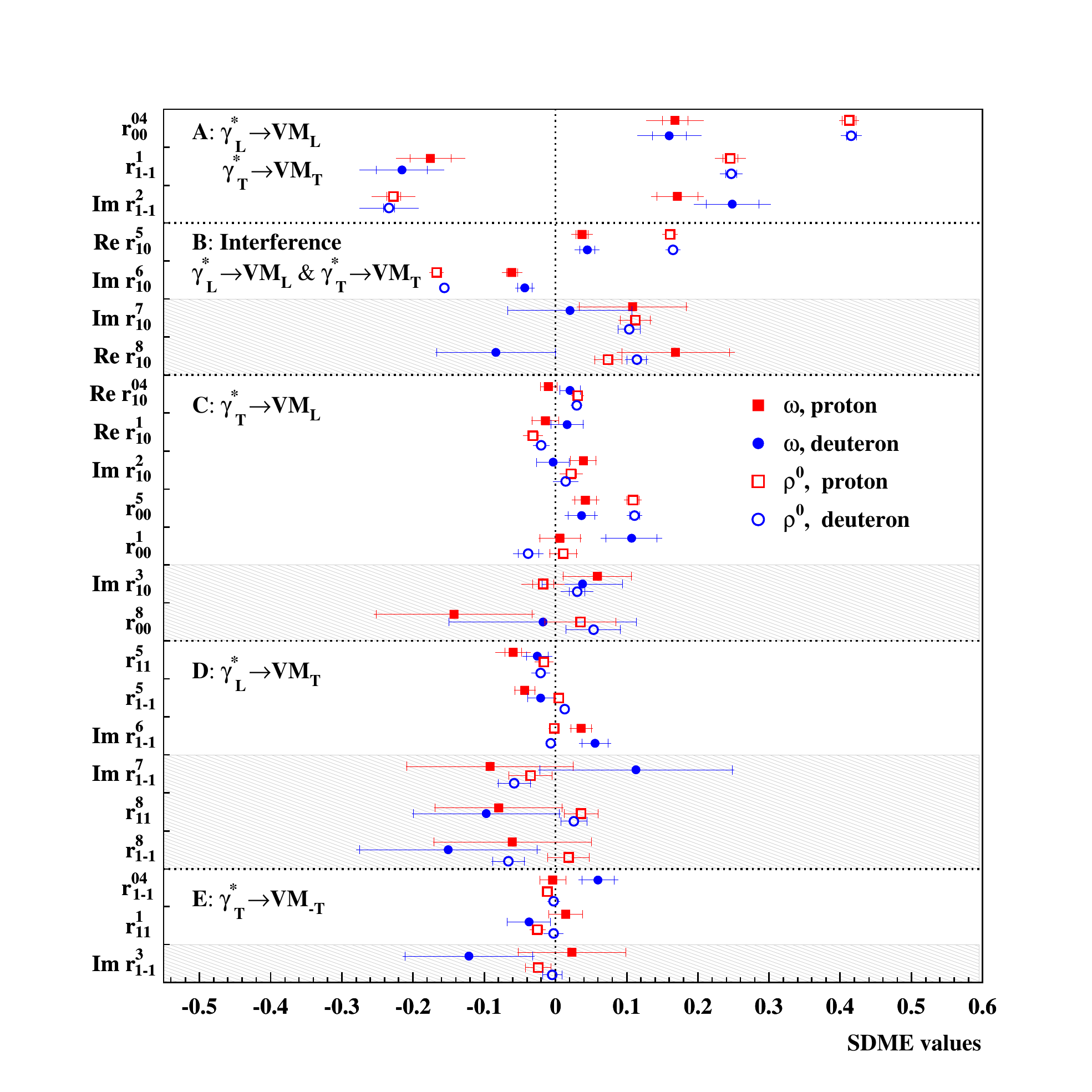}
\caption{Comparison of SDMEs in  exclusive $\omega$  and 
$\rho^{0}$~\cite{DC-24} electroproduction at HERMES for the  entire kinematic region.
The average values of the kinematic variables in exclusive $\rho^{0}$ production are  
$\langle Q^2  \rangle = 1.95$~GeV$^2$, $\langle W\rangle = 4.8$~GeV, and  $\langle -t' \rangle = 0.13$~GeV$^2$.}
\label{sdmeomrho}
\end{figure*}

In Fig.~\ref{sdmeomrho}, the comparison of $\omega$ and $\rho^{0}$~\cite
{DC-24} SDMEs is  shown. One can see that the SDMEs  $r^{1}_{1-1}$ and
$\mathrm{Im} \{r^{2}_{1-1}\} $ of class A have opposite sign for  $\omega$ and
$\rho^{0}$.
The SDME $r^{1}_{1-1}$  is negative for the $\omega$ meson and positive for
$\rho^{0}$, while 
$\mathrm{Im}\{r^{2}_{1-1}\} $ is positive for  $\omega$ and negative for  $\rho^{0}$.
In terms of helicity amplitudes, these two SDMEs are written~\cite{DC-24} as 
\begin{align}
r_{1-1}^1  = \frac {1}{2\mathcal{N}}\widetilde{\sum} & \left( |T_{11}|^2+|T_{1-1}|^2 \right. \nonumber \\
 & \left. -|U_{11}|^2-|U_{1-1}|^2 \right), \label{sdprz} \\
\mathrm{Im} \{r_{1-1}^2 \} =\frac {1}{2\mathcal{N}} \widetilde{\sum} & \left( -|T_{11}|^2+|T_{1-1}|^2 \right. \nonumber\\  
& \left. +|U_{11}|^2 -|U_{1-1}|^2 \right) .\label{sdprz1}
\end{align}
The difference between Eqs.~(\ref{sdprz1}) and (\ref{sdprz}) reads   
\begin{equation}
\mathrm{Im} \{r_{1-1}^2\} - r_{1-1}^1 = \frac
{1}{\mathcal{N}}\widetilde{\sum}(-|T_{11}|^2+|U_{11}|^2). 
\label{difff} 
\end{equation}
For the entire kinematic region, this difference is clearly positive for the
$\omega$ meson, hence
$\widetilde{\sum}|U_{11}|^2 >\widetilde{\sum}|T_{11}|^2$, while for
the $\rho^{0}$ meson $\widetilde{\sum}|T_{11}|^2 > \widetilde{\sum}|U_{11}|^2$~\cite{DC-24}.  
This  suggests a large UPE contribution in exclusive $\omega$-meson production.
Applying Eq.~(\ref{sum-two}) to relation (\ref{difff}), the latter can be 
rewritten as 
\begin{align}
\mathrm{Im}
\{r^2_{1-1}\}-r^1_{1-1}   =\frac{1}{\mathcal{N}}
(&-|T_{1 \frac{1}{2}1 \frac{1}{2}}|^2-|T_{1 -\frac{1}{2}1 \frac{1}{2}}|^2
\nonumber \\
&+|U_{1 \frac{1}{2}1 \frac{1}{2}}|^2+|U_{1 -\frac{1}{2}1\frac{1}{2}}|^2).
\label{appl}
\end{align}
The amplitudes with  nucleon helicity flip,
 $T_{1 -\frac{1}{2}1 \frac{1}{2}}$ and $U_{1 -\frac{1}{2}1 \frac{1}{2}}$, should be
zero at $t'=0$ and are proportional to $\sqrt{-t'}$ at small $t'$ (see Eq.~(\ref{asytpr}) and Ref.~\cite{Diehl}). The small
 contribution of $|T_{1-\frac{1}{2}1\frac{1}{2}}|^2$ will be neglected from now on. 
As it was established above, the UPE contribution is larger than the
NPE one. This means that if the dominant UPE  helicity-flip amplitude is
 $U_{1 -\frac{1}{2}1 \frac{1}{2}}$,  expression (\ref{appl}) would increase proportionally to
 $-t'$. However, the experimental values of
($\mathrm{Im} \{r^2_{1-1}\}-r^1_{1-1}$) (see Tables~\ref{tab3} and \ref{tab5}) do not
demonstrate such an increase; the values for the proton data even decrease
smoothly with $-t'$. Hence  the dominant UPE  amplitude is
$U_{1\frac{1}{2}1\frac{1}{2}}$, and it holds
 $|U_{11}|^2>|T_{11}|^2$.
 
\begin{figure*}\centering
\includegraphics[width=.75\textwidth]{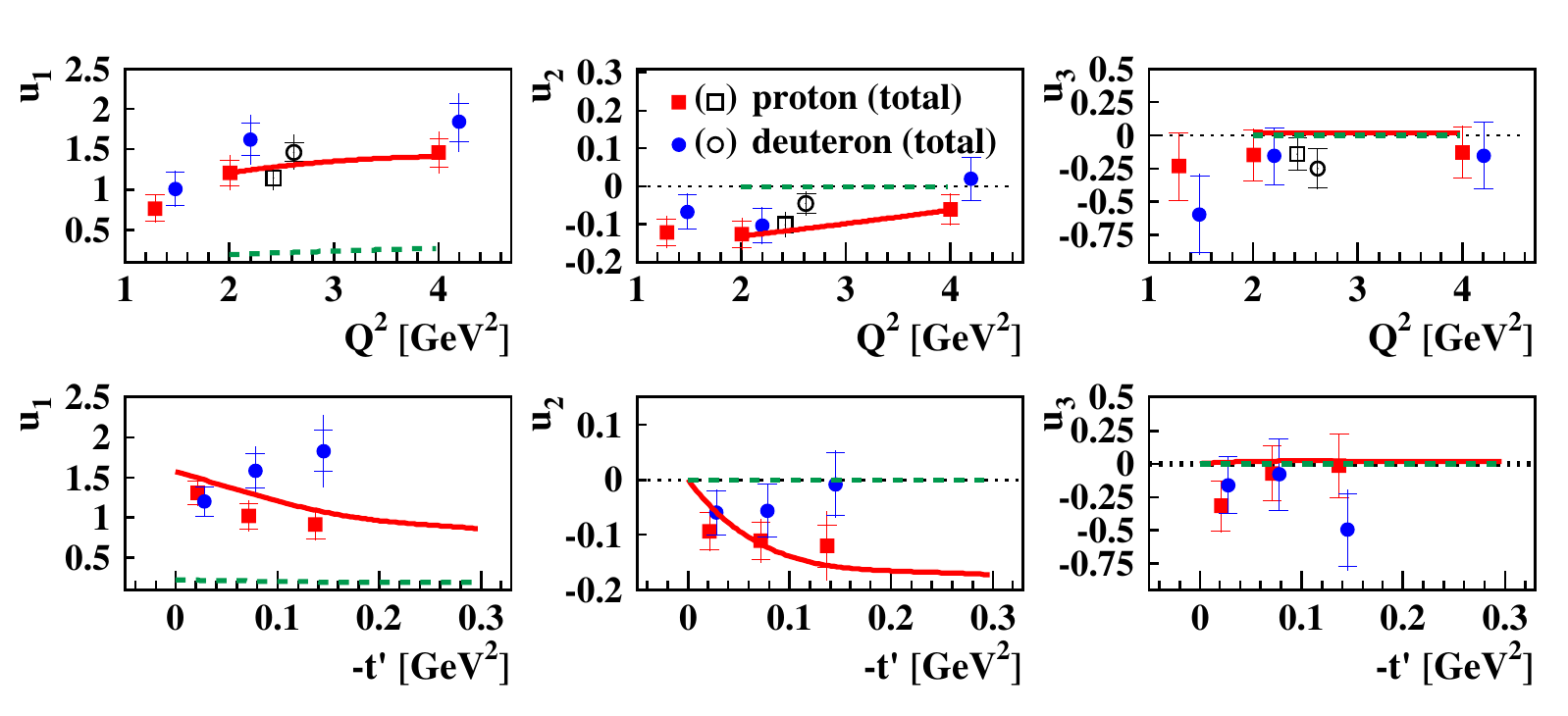} 
\caption {The $Q^{2}$ and $-t'$ dependences of $u_1$, $u_2$, and
  $u_3$. The open symbols represent the values over the entire
  kinematic region. Otherwise as for Fig.~\ref{Aq2t}.}
\label{u1}     
\end{figure*}

 The existence of UPE in  $\omega$ production on
the proton and deuteron can also be tested with  linear combinations
of SDMEs such as
\begin{equation}
u_1=1-r^{04}_{00}+2r^{04}_{1-1}-2r^{1}_{11}-2r^{1}_{1-1},
\label{uu1}
\end{equation}
\begin{equation}
u_2=r^{5}_{11}+r^{5}_{1-1},
\label{uu2}
\end{equation}
\begin{equation}
u_3=r^{8}_{11}+r^{8}_{1-1}.                                
\label{uu3}
\end{equation}
\noindent The quantity $u_1$ can be  expressed  in terms of helicity
amplitudes  as 
\begin{equation}
u_1=\frac{1}{\mathcal{N}} \ \widetilde{\sum}{\left( 4\epsilon|U_{10}|^2+2|U_{11}+U_{-11}|^2\right)}.
\label{u1u}
\end{equation}
A non-zero  result for $u_1$, implying that at least one of the four amplitudes $U_{1\pm\frac{1}{2}0\frac{1}{2} }$ or
($U_{1\pm\frac{1}{2}1\frac{1}{2} }+U_{-1\pm\frac{1}{2}1\frac{1}{2} } $) is nonzero, indicates the existence of UPE contributions.
In the entire kinematic region, $u_1$ is 1.15 $\pm$ 0.09 $\pm$ 0.12 and 1.47 $\pm$ 0.12 $\pm$ 0.18 for 
proton and deuteron data, respectively.
In Fig.~\ref{u1}, the  $Q^{2}$ and $-t'$ dependences of $u_1$ for 
proton and deuteron data are presented. It can be seen that $u_1$ is  larger than unity, which implies the existence of large
contributions from UPE transitions.

The expression for the quantities $u_2$ and $u_3$ in terms of
helicity amplitudes is 
\begin{equation}
u_2 + iu_3
=\frac{\sqrt2}{\mathcal{N}} \ \widetilde{\sum} {(U_{11}+U_{-11})U^{*}_{10}},
\label{u2u3n}
\end{equation}
showing that these quantities are nonzero if at least one of the products
$U^*_{1\frac{1}{2}0\frac{1}{2}}(U_{1\frac{1}{2}1\frac{1}{2}}+U_{-1\frac{1}{2}1\frac{1}{2}})$ 
or $U^*_{1-\frac{1}{2}0\frac{1}{2}}(U_{1-\frac{1}{2}1\frac{1}{2}}+U_{-1-\frac{1}{2}1\frac{1}{2}})$ 
is nonzero.
Therefore $u_2$ and $u_3$ provide  information complementary  to that given by
$u_1$. In Fig.~\ref{u1}, also the quantities $u_2$ and $u_3$ versus
$Q^2$ and $-t'$
are  presented both for proton and deuteron data. As seen from this figure,
there are no clear dependences on $Q^2$
and $-t'$, but $u_2$ for the proton data is definitely nonzero and there is some
evidence that it is also nonzero for the deuteron data. 
Note that $u_2$ and
$u_3$ are compatible with zero in $\rho^0$-meson electroproduction~\cite{DC-24}.

Figure~\ref{u1} also demonstrates good agreement between proton
data and the model 
calculation. It appears that including the pion-pole into the model fully
accounts for the unnatural-parity contribution measured through $u_1$ and
$u_2$, both in $-t'$ shape and magnitude. Conclusions on $u_3$ are 
prevented by the considerable experimental uncertainties.  

\subsection{Phase difference between amplitudes}
  
  %%%%%FIG

Taking the  amplitude without  helicity flip,
$U_{1 \frac{1}{2}1 \frac{1}{2}}$, as the dominant UPE  one, Eq.~(\ref{u2u3n}) can be
simplified as 
\begin{equation}
u_2+iu_3=\frac{\sqrt{2}}{\mathcal{N}}U_{1 \frac{1}{2}1 \frac{1}{2}}
U^*_{1 \frac{1}{2}0 \frac{1}{2}}
\equiv \frac{\sqrt{2}}{\mathcal{N}} U_{11}U^*_{10}.
\label{u2u3simpl}
\end{equation}
The expressions for the phase difference $\delta_U$ between
the UPE amplitudes $U_{11}$ and $U_{10}$ follow immediately from
Eq.~(\ref{u2u3simpl}): 
\begin{equation}
\cos \delta_{U} =u_2/\sqrt{(u_2)^2+(u_3)^2},
\end{equation}
\begin{equation} 
\sin \delta_{U} =u_3/\sqrt{(u_2)^2+(u_3)^2},
\end{equation}
\begin{equation} 
\tan \delta_{U} =u_3/u_2 = \frac{r^{8}_{11}+r^{8}_{1-1}}{r^{5}_{11}+r^{5}_{1-1}}. 
\end{equation}
The phase differences  obtained  for the entire kinematic region
are  $\delta_{U}$ = (-126 $\pm$ 12 $\pm$ 2)  degrees for proton and 
$\delta_{U}$ =  (-100 $\pm$ 61 $\pm$ 3) degrees for deuteron data.

The phase difference $\delta_{N}$ between the NPE amplitudes $T_{11}$ and
 $T_{00}$ can be calculated as follows~\cite{DC-24}:
\begin{equation}
\cos \delta_{N}= \frac{ 2  \sqrt{\epsilon}
(\mathrm{Re}\{r^{5}_{10}\}-\mathrm{Im}\{r^{6}_{10} \})}
{\sqrt{ r^{04}_{00}(1-r^{04}_{00}+r^{1}_{1-1}-\mathrm{Im}\{r^{2}_{1-1}\} )}
}.
\label{eq:cosdelta}
\end{equation}
 The  phase differences obtained for the entire kinematic region are 
$|\delta_{N}|$ = (51 $\pm$ 5 $\pm$ 14) degrees   and $|\delta_{N}|$ = (50 $\pm$ 7 $\pm$ 16) degrees for
proton and deuteron data, respectively.    
Using polarized SDMEs, in principle  also the sign of $\delta_{N}$ can be determined from the following equation: 
\begin{equation}
\sin \delta_{N}= \frac{ 2  \sqrt{\epsilon}
(\mathrm{Re}\{r^{8}_{10}\}+\mathrm{Im}\{r^{7}_{10}\} ) }
{\sqrt{  r^{04}_{00}(1-r^{04}_{00}+r^{1}_{1-1}-\mathrm{Im}\{r^{2}_{1-1} \})
}},
\label{eq:sindelta}
\end{equation}
which is given in Ref.~\cite{DC-24}.
For the present data, the large experimental  uncertainties of the  polarized SDMEs make it impossible to
determine the sign of  $\delta_{N}$.

\subsection{Longitudinal-to-transverse cross-section ratio}

%%%%%%%%%FIG
\begin{figure*}\centering
\includegraphics[width=.75\textwidth]{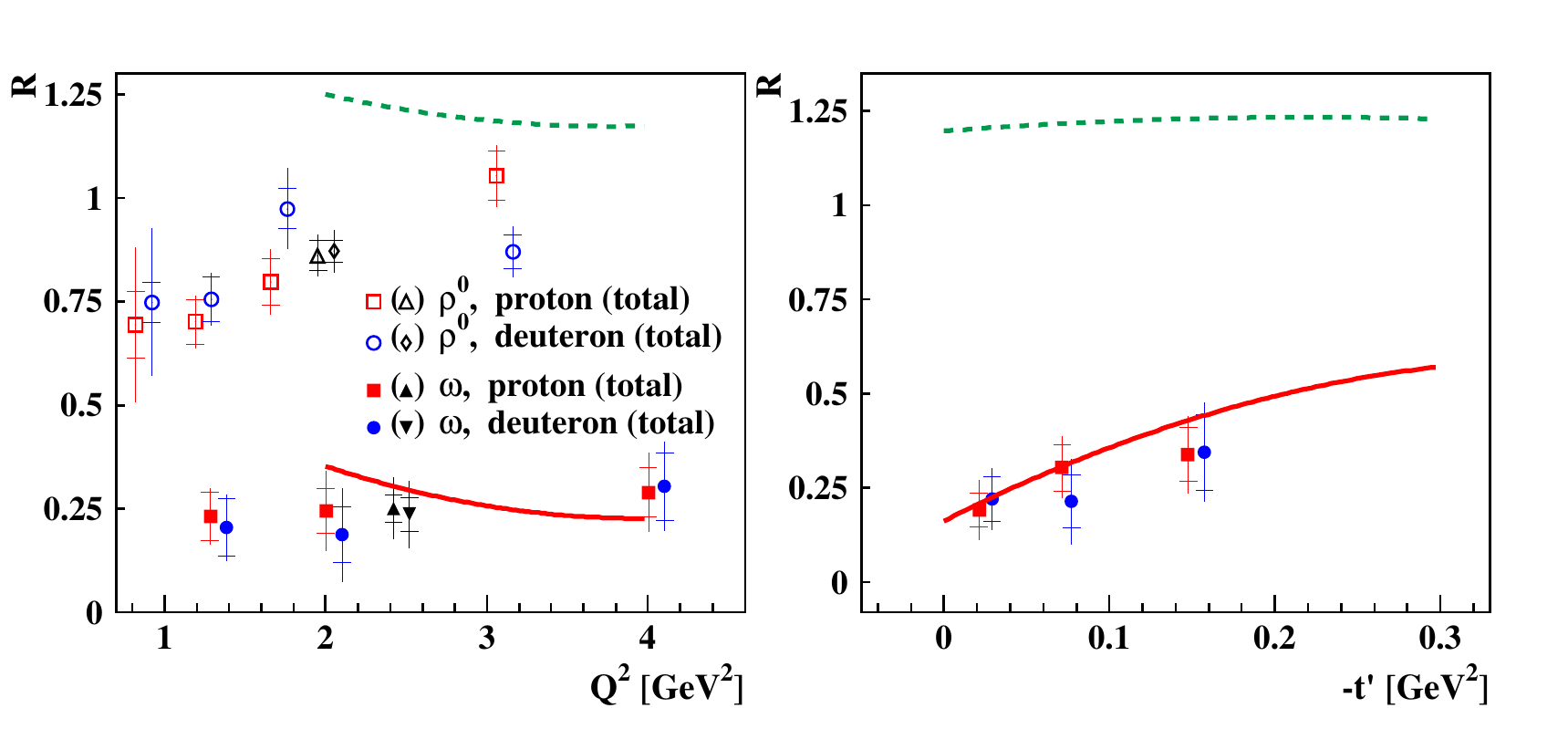}
\caption{The $Q^{2}$ (left) and $-t'$ (right) dependences of the longitudinal-to-transverse
virtual-photon differential cross-section ratio for  exclusive $\omega$ and  $\rho^{0}$
electroproduction at HERMES, where  the $-t'$ bin covers the interval   [0.0-0.2] GeV$^2$ for $\omega$
production and [0.0-0.4] GeV$^2$ for $\rho^{0}$ production~\cite{DC-24}. The symbols that are parenthesized in the legend represent the 
value of $R$ in the entire kinematic region. Otherwise as for Fig.~\ref{Aq2t}.}
\label{R_sigma}
\end{figure*}

Usually, the longitudinal-to-transverse virtual-photon  differential cross-section ratio 
\begin{equation}
R= \frac{d\sigma_{L}(\gamma^{*}_{L} \to V)}{ d\sigma_{T}(\gamma^{*}_{T} \to V)}\nonumber
\end{equation}
 is
experimentally determined from the measured SDME $r^{04}_{00}$ using
the approximated equation~\cite{DC-24}
\begin{equation}
R  \approx
\frac{1}{\epsilon}\frac{r^{04}_{00}}{1-r^{04}_{00}}.
\label{sigto}
\end{equation}
This relation is exact in the case of SCHC.
The $Q^{2}$ dependence of $R$ for the $\omega$ meson is 
shown in the left panel of Fig.~\ref{R_sigma}, where also for comparison the same dependence
for the $\rho^{0}$ meson~\cite{DC-24} is shown. For $\omega$ mesons produced in the entire
kinematic region, it is found that  $R$ =
0.25 $\pm$ 0.03 $\pm$ 0.07 for the proton and $R$ = 0.24 $\pm$ 0.04 $\pm$ 0.07 
for the deuteron data. Compared to the case of exclusive $\rho^{0}$
production,  this ratio is about four
times smaller,  and  for the  $\omega$ meson this ratio is almost independent of Q$^2$.
The $-t'$ dependence of $R$ is shown in the right panel of Fig.~\ref{R_sigma}. 
The comparison of the proton data to the GK model calculations with and without inclusion of the pion-pole 
contribution demonstrates the clear need to include the pion pole.
The data are well described by the model and appear to follow the $-t'$ dependence suggested by the model when the
pion-pole contribution is included. This implies that transverse and longitudinal virtual-photon
cross sections have different $-t'$ dependences. Hence the usual high-energy assumption that their
ratio can be identified with the corresponding ratio of the integrated cross sections does not hold 
in exclusive $\omega$ electroproduction at HERMES kinematics, due to the pion-pole contribution. The GK model appears to fully account 
for the unnatural-parity contribution to $R$ and shows rather good agreement with the data.

\subsection{The UPE-to-NPE asymmetry of the transverse cross section}

\begin{figure}
\centering
\hspace*{-.8cm}\includegraphics[width=.57\textwidth]{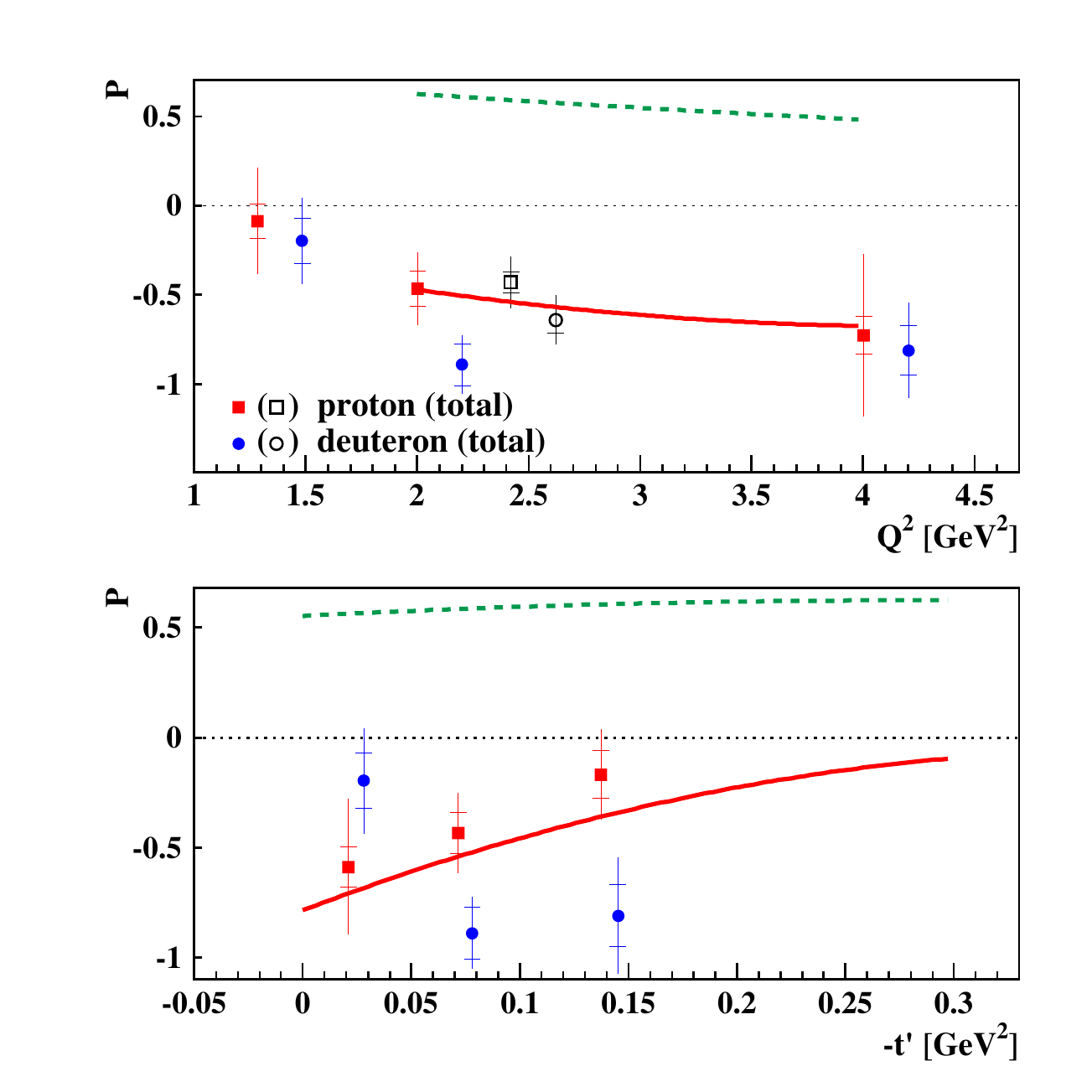}
\caption{ The $Q^{2}$  and $-t'$ dependences of the UPE-to-NPE asymmetry $P$ of the
transverse differential cross section for  exclusive $\omega$ 
electroproduction at HERMES. The open symbols represent the values over the
entire kinematic region. Otherwise as for Fig.~\ref{Aq2t}.}
\label{P_sigma}
\end{figure}

  The UPE-to-NPE asymmetry of the transverse differential cross section  is defined as~\cite{sigmat} 
\begin{align}
P=\frac{d\sigma^N_T-d\sigma^U_T}{d\sigma^N_T+d\sigma^U_T} &\equiv
\frac{d\sigma^N_T/d\sigma^U_T-1}{d\sigma^N_T/d\sigma^U_T+1} \nonumber \\
&=(1+\epsilon R)(2r^1_{1-1}-r^1_{00}),
\label{asymm} 
\end{align}
where $\sigma^N_T$ and $\sigma^U_T$ denote the part of the cross section due to NPE and UPE, 
respectively. 
Substituting  Eq.~(\ref{sigto}) in Eq.~(\ref{asymm}) leads to the  approximate relation 
\begin{eqnarray}
 P \approx \frac{2r^{1}_{1-1}-r^{1}_{00}}{1-r^{04}_{00}}.
\end{eqnarray}
The value of $P$ obtained in the entire kinematic region is 
$-0.42 \pm 0.06 \pm 0.08$ and  
$-0.64 \pm 0.07 \pm 0.12$ for proton and deuteron, respectively. 
This means that a large part of the transverse cross section is due to
UPE. In Fig.~\ref{P_sigma}, the $Q^{2}$  and $-t'$ dependences of the
UPE-to-NPE asymmetry of the transverse differential  cross section for exclusive $\omega$ production 
are presented. Again, the GK model calculation appears to fully
account for the unnatural-parity contribution and shows very good agreement with the
data both in shape and magnitude.

%%%%%%%%%%%%%%%%%%%%%%%%%%%%%%%%%%%%%%%%%%%%%%%%%%%%%%%%%
\subsection{Hierarchy of amplitudes}\label{sec:hierarchy}

In order to develop a hierarchy of amplitudes, in the following a number of 
relations between individual helicity amplitudes is considered. The resulting hierarchy is given in Eqs.~(\ref{hierarchy}) and (\ref{hierarchy-2}) 
below.

\subsubsection{$U_{10}$ versus $U_{11}$}

From Eqs.~(\ref{u1u}) and (\ref{u2u3simpl}), the relation  
\begin{eqnarray}
\nonumber
\frac{\sqrt{2(u_2^2+u_3^2)}}{u_1} \approx \frac{|U_{11} U^*_{10}|}{|U_{11}|^2+2 \epsilon |U_{10}|^2}\\
=\frac{|U_{10}/U_{11}|}{1+2 \epsilon |U_{10}/U_{11}|^2}
\label{u23u1}
\end{eqnarray} 
is obtained. Using the measured values of those SDMEs that determine $u_1$, $u_2$, and $u_3$, the following amplitude ratio is estimated:  
\begin{eqnarray}
\frac{|U_{10}|}{|U_{11}|} \approx \frac{\sqrt{2(u_2^2+u_3^2)}}{u_1} \approx 0.2. 
\label{u10u11}
\end{eqnarray}
In order to reach the best possible accuracy for such estimates, the mean values of SDMEs for the proton and deuteron are used and preference will be given to
 quantities that do not contain polarized SDMEs, which have  much less experimental accuracy 
than the unpolarized SDMEs.
The relatively large value
for the ratio $|U_{10}/U_{11}|$ is due to the large measured value of $u_3$.
However, as this value is compatible  with zero within about one standard deviation of the total 
uncertainty, the contribution of $u_3$ in Eq.~(\ref{u10u11}) can be neglected, which leads to the value of $0.06$ as lower bound on $|U_{10}/U_{11}|$. 

\subsubsection{$T_{11}$ versus $U_{11}$}

 With the above considerations, it follows from Eq.~(\ref{u1u})  that the contribution of $|U_{10}/U_{11}|^2$ is
only a few percent and hence will  be neglected everywhere.
Then, in  particular, the relation
\begin{eqnarray}
u_1 \approx 2|U_{11}|^2/\mathcal{N}
\label{appru1}
\end{eqnarray}
is valid with a precision of a few percent.
  
Equations (\ref{fnat}-\ref{asytpr}) show that the nucleon-helicity-flip amplitudes
$T_{1\pm\frac{1}{2}1\mp\frac{1}{2}}$
($T_{0\pm\frac{1}{2}0\mp\frac{1}{2}}$) are suppressed by a factor of about $\sqrt{-t'}/M$ compared to the amplitude $T_{11}$ ($T_{00}$) with 
diagonal
transitions ($\lambda'_N=\lambda_N$). Therefore, the second-order contributions of the amplitudes $T_{\lambda\pm\frac{1}{2}\lambda \mp\frac{1}{2}}$ 
for any $\lambda$ will be neglected compared to any bilinear product of $T_{00}$ and $T_{11}$. In this approximation, the relation
\begin{eqnarray}
\frac{2[\rm{Im} \{r^2_{1-1}\}-r^2_{1-1}]}{u_1}=1-\Bigl |\frac{T_{11}}{U_{11}} \Bigr |^2
\label{t11u11}
\end{eqnarray}
follows from Eqs.~(\ref{appl}) and (\ref{appru1}). Substituting numerical values for the SDMEs in Eq.~(\ref{t11u11}) leads to the estimate
$|T_{11}/U_{11}| \approx 0.6$. 

\subsubsection{$T_{00}$ versus $U_{11}$}

Using Eq.~(\ref{appru1}) and the  expression for $r^{04}_{00}$ from Refs.~\cite{Schill,DC-24} yields 
\begin{eqnarray}
\frac{2r^{04}_{00}}{u_1}=\frac{\widetilde{\sum}[\epsilon |T_{00}|^2+|T_{01}|^2+|U_{01}|^2]}{|U_{11}|^2}.
\label{r04u1}
\end{eqnarray}
Neglecting in the numerator of the right-hand side of Eq. (\ref{r04u1}) all positive terms except $|T_{00}|^2$, the inequality of 
interest is obtained:
\begin{eqnarray}
\frac{2r^{04}_{00}}{u_1}>\frac{\epsilon |T_{00}|^2}{|U_{11}|^2}.
\label{inr04u1}
\end{eqnarray}
Using for the estimate $\epsilon =0.8$ and values of SDMEs from Table~\ref{tab1} yields the result $|T_{00}/U_{11}| < 0.6$.

The same ratio can be estimated from other SDMEs. Using expressions for the SDMEs from~\cite{Schill,DC-24}, the following equations can be 
written:
\begin{eqnarray}
\nonumber
\rm{Re}\{r^5_{10}\}-\rm{Im}\{r^6_{10}\}=~~~~~~~~~~~~~~~~~~~~~~~~~~~~~~~~~\\
\frac{1}{\mathcal{N}\sqrt{2}}\widetilde{\sum}{\rm Re}[T_{11}T^*_{00}+T_{01}T^*_{10}-U_{01}U^*_{10}],
\label{r510-r610}\\
\nonumber
\rm{Im}\{r^7_{10}\}+\rm{Re}\{r^8_{10}\}=~~~~~~~~~~~~~~~~~~~~~~~~~~~~~~~~~\\
\frac{1}{\mathcal{N}\sqrt{2}}\widetilde{\sum}{\rm Im}[T_{11}T^*_{00}+T_{10}T^*_{01}-U_{10}U^*_{01}].
\label{r710-r810}
\end{eqnarray}
From  Eqs.~(\ref{fnat}-\ref{asytpr}), it follows that the terms $\widetilde{\sum}T_{01}T^*_{10}$ and $\widetilde{\sum}U_{01}U^*_{10}$ on the 
right-hand side of 
Eqs.~(\ref{r510-r610}, \ref{r710-r810}) are suppressed by a factor $(-t')/M^2$ compared to $T_{11}T^*_{00}$ and will be neglected.
The simplest consequence of Eqs.~(\ref{r510-r610}, \ref{r710-r810}) is the relation
\begin{eqnarray}
\nonumber
[\rm{Re}\{r^5_{10}\}-\rm{Im}\{r^6_{10}\}]^2+[\rm{Im}\{r^7_{10}\}+\rm{Re}\{r^8_{10}\}]^2= \\
\frac{1}{2\mathcal{N}^2}|T_{11}|^2|T_{00}|^2.
\label{a2b2}
\end{eqnarray}
Dividing this relation by $u_1^2/8$ and using Eq.~(\ref{appru1}), one gets the formula of interest: 
\begin{eqnarray}
\nonumber
\frac{[\rm{Re}\{r^5_{10}\}-\rm{Im}\{r^6_{10}\}]^2+[\rm{Im}\{r^7_{10}\}+\rm{Re}\{r^8_{10}\}]^2}{u^2_1/8}\approx\\
 \frac{|T_{11}|^2|T_{00}|^2}{|U_{11}|^4}.~~~
\label{r510-r610-r710-r810-u1}
\end{eqnarray}
Using numerical SDME values from Table~\ref{tab1} and $|T_{11}/U_{11}|=0.6$, the estimate $|T_{00}/U_{11}|\approx0.5$ is obtained, 
which is in agreement 
with the previous estimate. However, as the polarized SDMEs $\rm{Im}\{r^7_{10}\}$ and $\rm{Re}\{r^8_{10}\}$ have very large uncertainties, the 
latter result is less reliable than the former. Omitting the contribution of the polarized SDMEs in Eq.~(\ref{r510-r610-r710-r810-u1}) leads to the 
inequality
\begin{eqnarray}
\frac{8[\rm{Re}\{r^5_{10}\}-\rm{Im}\{r^6_{10}\}]^2}{u^2_1}<
 \frac{|T_{11}|^2|T_{00}|^2}{|U_{11}|^4},
\label{r510-r610-u1}
\end{eqnarray}
which provides the lower limit of $0.3$ for the same ratio $|T_{00}/U_{11}|$. This result combined with the former estimate leads to 
the boundaries $0.3 < |T_{00}/U_{11}| < 0.6$. \\

\subsubsection{$T_{00}$ versus $T_{01}$}

 In order to estimate the value of $|T_{01}|$, the quantity
\begin{eqnarray}
\frac{\sqrt{(r^5_{00})^2+(r^8_{00})^2}}{r^{04}_{00}}=\frac{\sqrt{2}|\widetilde{\sum}T_{01}T^*_{00}|}
{ \widetilde{\sum}[\epsilon|T_{00}|^2+|T_{01}|^2+|U_{01}|^2]}~~
\label{r5r8r04}
\end{eqnarray}
can be formed. 
Neglecting in the denominator of the right-hand side of Eq.~(\ref{r5r8r04}) all the terms except $\epsilon |T_{00}|^2$,  
the inequality 
\begin{eqnarray}
\frac{\sqrt{(r^5_{00})^2+(r^8_{00})^2}}{r^{04}_{00}}< \frac{\sqrt{2}|\widetilde{\sum}T_{01}T^*_{00}|}{\epsilon|T_{00}|^2}
\label{int01}
\end{eqnarray}
is obtained. The sum in the numerator of the right-hand side of Eq.~(\ref{int01}) is 
\begin{eqnarray}
\widetilde{\sum}T_{01}T^*_{00}=T_{0\frac{1}{2}1 \frac{1}{2}}T^*_{0\frac{1}{2}0\frac{1}{2}}
+T_{0-\frac{1}{2}1 \frac{1}{2}}T^*_{0-\frac{1}{2}0\frac{1}{2}}
\label{gensum}
\end{eqnarray}
according to Eq.~(\ref{sum-two}).
If the first product on the right-hand side of Eq.~(\ref{gensum}) dominates, then inequality (\ref{int01}) becomes simpler:
\begin{eqnarray}
\frac{\sqrt{(r^5_{00})^2+(r^8_{00})^2}}{r^{04}_{00}} <  \frac{\sqrt{2}}{\epsilon}\frac{|T_{01}|}{|T_{00}|}.
\label{estt01}
\end{eqnarray}
Numerically, this yields the estimate $|T_{01}/T_{00}| \simeq 0.3$. The dominant contribution to this number comes from the 
polarized SDME $r^8_{00}$
that is compatible with zero within about one standard deviation of the total uncertainty. Retaining only the contribution of the unpolarized SDME 
$r^5_{00}$
in Eq.~(\ref{estt01}) gives the following result: $|T_{01}/T_{00}|>0.1$. The experimental accuracy of the presented data is not sufficient 
to provide a reliable estimate for the upper bound to the ratio $|T_{01}/T_{00}|$. 
As shown in Appendix A, the upper limits for 
\begin{eqnarray}
 \mathcal{A} \equiv \frac{\widetilde{\sum}(|T_{01}|^2+|U_{01}|^2)}{|T_{00}|^2} 
\label{definA}
\end{eqnarray}
are $1.3 \pm 0.7$ for the proton and $1.1 \pm 1.2$ for the deuteron. 
In the below consideration the estimate based on Eq.~(\ref{estt01}), namely $|T_{01}/T_{00}| \simeq 0.3$, 
is assumed to be realistic. 

The numerator in the definition of $r^1_{00}$ is
$\widetilde{\sum}[|U_{01}|^2-|T_{01}|^2]$. The
values of $r^1_{00}$ are compatible with
zero within two standard deviations of the total experimental uncertainty,
hence $|U_{01}|$ cannot be much larger than $|T_{01}|$.\\ 
\indent Considering the SDME combinations  
($r^5_{11}-r^5_{1-1}$)
and ($\rm{Im}\{r^8_{1-1}\}-r^8_{11}$),  which are proportional
 to the real and imaginary parts of
$\widetilde{\sum}T_{10}(T_{11}-T_{1-1})^*$, respectively, it is possible in principle to
estimate the value
of $|T_{10}|$. Since these combinations are compatible with zero  within one
standard deviation of the total  uncertainty,  it can be  concluded  
that $|T_{10}|$ is negligibly small compared to the large  amplitude moduli
$|U_{11}|$, $|T_{11}|$, and $|T_{00}|$.

\subsubsection{Resulting hierarchy of amplitudes}

\noindent As a result, the following  hierarchy is obtained:
\begin{align}
& |U_{11}|^2>|T_{00}|^2 \sim |T_{11}|^2  \nonumber\\ 
\gg\ & |U_{10}|^2 \sim |T_{01}|^2 \sim |U_{01}|^2 \nonumber \\
\gg\ & |T_{10}|^2,|T_{1-1}|^2,|U_{1-1}|^2,
\label{hierarchy}
\end{align}   
where  negligibly small amplitudes are neglected.

However, there exists a possible alternative for the hierarchy
presented on the second line of Eq.~(\ref{hierarchy}), if  the helicity-flip amplitudes 
$T_{0-\frac{1}{2}1\frac{1}{2}}$
and $U_{0-\frac{1}{2}1\frac{1}{2}}$ are of the same order of magnitude as the
helicity-conserving amplitudes $T_{00}$ and $T_{11}$. Indeed, the 
sum $\widetilde{\sum}T_{01}T^*_{00}$ in Eq.~(\ref{int01}) is the sum of two
products, 
$T_{0\frac{1}{2}1\frac{1}{2}}T^*_{0\frac{1}{2}0\frac{1}{2}}$ and
$T_{0-\frac{1}{2}1\frac{1}{2}}T^*_{0-\frac{1}{2}0\frac{1}{2}}$,
according to Eq.~(\ref{gensum}). In order to obtain Eq.~(\ref{estt01})  from
Eq.~(\ref{int01}), the dominance of the first product was assumed. If instead   
the second product is assumed to be dominant,  Eq.~(\ref{estt01})  has to be replaced by 
\begin{align}
\nonumber
\frac{\sqrt{(r^5_{00})^2+(r^8_{00})^2}}{r^{04}_{00}} &\leq\frac{\sqrt{2}}{\epsilon}
\frac{|T_{0-\frac{1}{2}1\frac{1}{2}} T^{*}_{0-\frac{1}{2}0\frac{1}{2}}|}{|T_{00}|^2}\\
&=\frac{\sqrt{2}}{\epsilon}\frac{|T_{0-\frac{1}{2}1\frac{1}{2}}|}{|T_{00}|}
\frac{|T_{0-\frac{1}{2}0\frac{1}{2}}|}{|T_{00}|}.
\label{eq54}
\end{align}
The nucleon-helicity-flip amplitude $T_{0-\frac{1}{2}0\frac{1}{2}}$ is smaller than the
helicity-conserving amplitude $T_{00} \equiv T_{0\frac{1}{2}0\frac{1}{2}}$
 by a factor of about $\sqrt{-t'}/M \approx 0.3$ (see 
Eq.~(\ref{asytpr})). Substituting this factor for 
$|T_{0-\frac{1}{2}0\frac{1}{2}}/T_{00}|$,  using 
$\epsilon=0.8$ and the measured  SDME values,  the final estimate $|T_{0-\frac{1}{2}1\frac{1}{2}}| \simeq  |T_{00}|$ is obtained.
This result shows that the nucleon-helicity-flip amplitude $T_{0-\frac{1}{2}1\frac{1}{2}}$ could be of the same order of magnitude as $T_{00}$, 
while the values of $T_{01}$ and $U_{01}$ could be as given in the previous estimates.

As the SDME $r^1_{00}$, which is proportional to $\widetilde{\sum}[|U_{01}|^2-|T_{01}|^2]$,
was measured to be  compatible with zero, 
the value of $|U_{0-\frac{1}{2}1\frac{1}{2}}|$ should be about
the same as that  of $|T_{0-\frac{1}{2}1\frac{1}{2}}|$.
Then, the values of 
$|T_{0-\frac{1}{2}1\frac{1}{2}}|$, $|U_{0-\frac{1}{2}1\frac{1}{2}}|$, and $|T_{00}|$  
are of the same order of magnitude, so that 
the hierarchy of amplitudes becomes  
\begin{align}
|U_{11}|^2 &>|T_{00}|^2 \sim |T_{11}|^2 \sim |T_{0-\frac{1}{2}1\frac{1}{2}}|^2 \sim |U_{0-\frac{1}{2}1\frac{1}{2}}|^2  \nonumber\\ 
&\gg |U_{10}|^2 \sim |T_{01}|^2 \sim |U_{01}|^2\nonumber\\ 
&\gg |T_{10}|^2, |T_{1-1}|^2,|U_{1-1}|^2,
\label{hierarchy-2}
\end{align}
where again negligibly small amplitudes are neglected.
Note that the usually used Eq.~(\ref{sigto}) for $R$ is not applicable in
this case.
The estimation performed in Appendix A shows that the accuracy of the presented
data is not sufficient to decide  between hierarchies
(\ref{hierarchy}) and (\ref{hierarchy-2}).
The best way to get information on the amplitudes
$T_{0-\frac{1}{2}1\frac{1}{2}}$ and
$U_{0-\frac{1}{2}1\frac{1}{2}}$ is to study electroproduction of
$\omega$ mesons on transversely polarized protons, where
these amplitudes contribute linearly to the angular distribution.

\section{Summary}

 Exclusive $\omega $ electroproduction is studied at HERMES using a
longitudinally polarized lepton beam and unpolarized hydrogen and deuterium
targets in the kinematic region $Q^2>1.0$ GeV$^2$, $3.0$ GeV $< W < 6.3$ GeV, and $ - t' < 0.2$ GeV$^2$. 
The average kinematic values are $\langle Q^2  \rangle = 2.42$ GeV$^2$, $\langle W\rangle =
4.8$ GeV, and  $\langle -t' \rangle = 0.080$ GeV$^2$. Using an unbinned maximum
likelihood method, 15 unpolarized and, for the first time,  8
polarized spin density matrix elements are extracted. The kinematic dependences of all 23 SDMEs  are presented for proton
and deuteron data.  No significant differences between proton and deu\-te\-ron results are  seen.

 The SDMEs are presented in five classes
corresponding to different helicity transitions between the virtual photon and
the $\omega$ meson. While the values  of class-A and B SDMEs agree with the
hypothesis of $s$-channel helicity conservation,
the class-C SDME $r^{5}_{00}$ indicates a violation of this hypothesis.
 The values of those class-D SDMEs that  correspond
to the transition $\gamma^*_L\to \omega_T$  also indicate a small violation
 of the hypothesis of $s$-channel helicity conservation.
 
 Using the   SDMEs $r_{1-1}^{1}$ and $\mathrm{Im} \{r_{1-1}^2\}$, it is shown that for
 exclusive $\omega$-meson production
  the amplitude of the UPE transition $\gamma^{*}_{T} \to \omega_{T}$ is
 larger than the NPE amplitude for the same transition, i.e., $|U_{11}|^2 >|T_{11}|^2$.
The importance  of UPE transitions is also shown by a  combination  of SDMEs denoted  $u_1$.
This suggests  that at HERMES energies in exclusive $\omega$ electroproduction
the quark-exchange mechanism, or $\pi^{0}$, $a_1$... exchanges in Regge
phenomenology, plays a significant role.

The phase shift between those UPE amplitudes that describe transverse $\omega$
 production by transverse and longitudinal virtual photons, $U_{11}$ for 
 $\gamma^{*}_{T} \to \omega_{T}$ and  $U_{10}$ for $\gamma^{*}_{L} \to \omega_{T}$,
 respectively, as well as   the magnitude of the phase difference
 between the NPE amplitudes $T_{11}$ and $T_{00}$ is determined for the
 first time.
 
 The ratio $R$ between the differential longitudinal and transverse virtual-photon cross-sections  
 is  determined to be $R$ = 0.25 $\pm$ 0.03 $\pm$ 0.07 
for the $\omega$ meson, which is about four times smaller than
 in the case of  the $\rho^{0}$ meson. In contrast to the case of the $\rho^{0}$ meson,  $R$ 
 shows only a weak dependence on $Q^{2}$ for the $\omega $ meson.
 
 The UPE-to-NPE asymmetry of the transverse vir\-tu\-al-photon cross section 
is determined to be $P= -0.42 \pm 0.06 \pm 0.08$ 
and  $P= -0.64 \pm 0.07 \pm 0.12$  for the proton and deuteron data, respectively.

From the extracted SDMEs, two slightly different hierarchies of
helicity amplitudes can be derived, which remain indistinguishable for the 
 given experimental accuracy of the presented data.
Both hierarchies consistently mean that the UPE amplitude describing the $\gamma^*_T \rightarrow
\omega_T$ transition dominates over the two NPE amplitudes describing the $\gamma^*_L \rightarrow \omega_L$ and 
$\gamma^*_T \rightarrow \omega_T$ transitions, with the latter two being of similar magnitude.

Good agreement between the presented proton data and results of a
pQCD-inspired phenomenological mo\-del is found only when including pion-pole
contributions, which are of unnatural parity. The distinct $-t'$ dependence of the pion-pole contribution leads to a $-t'$ dependence of $R$. 
This invalidates for exclusive $\omega$ production at HERMES energies the common high-energy assumption of identifying $R$ 
with the ratio of the integrated longitudinal and transverse cross sections.

\begin{acknowledgement}
{\bf Acknowledgements~~}
We are grateful to Sergey Go\-los\-ko\-kov and Peter Kroll for fruitful  discussions on the comparison between our data and their model   
calculations.
We gratefully acknowledge the DESY management for its support and the staff
at DESY and the collaborating institutions for their significant effort.
This work was supported by 
the Ministry of Education and Science of Armenia;
the FWO-Flanders and IWT, Belgium;
the Natural Sciences and Engineering Research Council of Canada;
the National Natural Science Foundation of China;
the Alexander von Humboldt Stiftung,
the German Bundesministerium f\"ur Bildung und Forschung (BMBF), and
the Deu\-tsche Forschungsgemeinschaft (DFG);
the Italian Istituto Nazionale di Fisica Nucleare (INFN);
the MEXT, JSPS, and G-COE of Japan;
the Dutch Foundation for Fundamenteel Onderzoek der Materie (FOM);
the Russian Academy of Science and the Russian Federal Agency for 
Science and Innovations;
the Basque Foundation for Science (IKERBASQUE) and the UPV/EHU under program UFI 11/55;
the U.K.~Engineering and Physical Sciences Research Council, 
the Science and Technology Facilities Council,
and the Scottish Universities Physics Alliance;
as well as the U.S.~Department of Energy (DOE) and the National Science Foundation (NSF).
\end{acknowledgement}
%%%%%%%%%%%%%%%%%%%

\begin{appendix}
\section{Estimate of $T_{0-\frac{1}{2}1\frac{1}{2}}$ and $U_{0-\frac{1}{2}1\frac{1}{2}}$ values}
\label{app_a}

The normalization factor  $\mathcal{N}$  is given by (see, e.g.,~\cite{Schill,DC-24})
\begin{equation}
 \mathcal{N}=\mathcal{N}_T+\epsilon \mathcal{N}_L,
\label{nor1}
\end{equation}
with
\begin{align}
\mathcal{N}_T =\widetilde{\sum}( &|T_{11}|^2+|T_{01}|^2+|T_{-11}|^2 \nonumber\\
 &+|U_{11}|^2+|U_{01}|^2+|U_{-11}|^2),\label{nor2} \\
\mathcal{N} _L= \widetilde{\sum}( &|T_{00}|^2+2|T_{10}|^2+2|U_{10}|^2 ).\label{nor3}
\end{align}
Using Eqs.~(\ref{nor1}-\ref{nor3}) and the expression defining
$r^{04}_{00}$~\cite{Schill,DC-24},
\begin{equation}
r^{04}_{00}=\frac{1}{\mathcal{N}} \widetilde{\sum}( \epsilon|T_{00}|^2+|T_{01}|^2+|U_{01}|^2),
\label{defr04}
\end{equation}
the exact relation 
\begin{align}
1-r^{04}_{00}=\frac{1}{\mathcal{N}}\widetilde{\sum}[& |T_{11}|^2 + |U_{11}|^2 +|T_{1-1}|^2 +|U_{1-1}|^2 \nonumber \\
 & + 2 \epsilon(|T_{10}|^2+|U_{10}|^2)]
\label{1-r04}
\end{align}
is obtained.  
Neglecting, as usual, $\widetilde{\sum}[|T_{1-1}|^2+|U_{1-1}|^2+|T_{10}|^2+|U_{10}|^2]$
in this expression, we get the approximate relation 
\begin{eqnarray}
1-r^{04}_{00} \approx \frac{1}{\mathcal{N}}\widetilde{\sum}(|T_{11}|^2+|U_{11}|^2).  
\label{appr1-r04}
\end{eqnarray}
Neglecting also the small nucleon-helicity-flip amplitudes $T_{1-\frac{1}{2}1\frac{1}{2}}$ and $U_{1-\frac{1}{2}1\frac{1}{2}}$ in
 Eq.~(\ref{difff}) and then subtracting it from Eq.~(\ref{appr1-r04}), the relation   
\begin{eqnarray}
1-r^{04}_{00} +r^{1}_{1-1}- \rm{Im}\{r^2_{1-1}\} \approx \frac{2}{\mathcal{N}} |T_{11}|^2
\label{cc}
\end{eqnarray}
is obtained.
After neglecting in Eq.~(\ref{defr04}) only the nucleon-helicity-flip amplitude $T_{0-\frac{1}{2}0\frac{1}{2}}$,
it  can be  rewritten as 
\begin{equation}
r^{04}_{00} \approx \frac{1}{\mathcal{N}} \left[ \epsilon |T_{00}|^2+\widetilde{\sum}(|T_{01}|^2+|U_{01}|^2) \right].
\label{dd}
\end{equation}
Multiplying this equation by Eq.~(\ref{cc}) and dividing it by  Eq.~(\ref{a2b2}) with a factor
of four, the equation of interest reads
\begin{eqnarray}
\epsilon + \mathcal{A} \approx \nonumber ~~~~~~~~~~~~~~~~~~~~~~~~~~~~~~~~~~~~~~~~~~~~~~~~~~~~~~~~~~~\\ 
\frac{r^{04}_{00}(1-r^{04}_{00} +r^{1}_{1-1}- \rm{Im}\{r^2_{1-1}\})/4}
{[\rm{Re}\{r^5_{10}\}-\rm{Im}\{r^6_{10}\}]^2+[\rm{Im}\{r^7_{10}\}+\rm{Re}\{r^8_{10}\}]^2},~~~~
\label{cda2b2}
\end{eqnarray}
where the quantity $\mathcal{A}$ is defined in Eq.~(\ref{definA}).
 The value of $\mathcal{A}$  is close to zero, if  $|T_{0-\frac{1}{2}1\frac{1}{2}}|^2$
and $|U_{0-\frac{1}{2}1\frac{1}{2}}|^2$ are much smaller than $|T_{00}|^2$,
and it should be of the order of one if they are comparable to $|T_{00}|^2$. 
Since the uncertainties of the polarized SDMEs
$\rm{Im}\{r^7_{10}\}$ and $\rm{Re}\{r^8_{10}\}$ are large, the use of Eq.~(\ref{cda2b2}) for the present data is not very successful.
Indeed, using for numerical calculations $\epsilon=0.8$ and the values for the SDMEs in Eq.~(\ref{cda2b2}) from  Table~\ref{tab1} we get 
$\mathcal{A}=-0.56 \pm 0.20 
$ and $\mathcal{A}=0.50 \pm 1.8$ for the proton and deuteron data, respectively. 
In contrast, in $\rho^{0}$-meson production, the  
corresponding  values of $\mathcal{A}$~\cite{DC-24},  $-0.031 \pm 0.084$ and $-0.064 \pm 0.068$, exclude practically the possibility
that  the amplitudes  $T_{0-\frac{1}{2}1\frac{1}{2}}$ and
$U_{0-\frac{1}{2}1\frac{1}{2}}$ are  comparable to the dominant amplitudes $U_{11}$, $T_{00}$ and
$T_{11}$.\\ 
\indent If the contribution of $[\rm{Im}\{r^7_{10}\}+\rm{Re}\{r^8_{10}\}]$ in the denominator of the
right-hand side of Eq.~(\ref{cda2b2}) is neglected, the 
useful inequality 
\begin{eqnarray}
\mathcal{A} \leq
\frac{r^{04}_{00}(1-r^{04}_{00} +r^{1}_{1-1}- \rm{Im}\{r^2_{1-1}\})}
{4[\rm{Re}\{r^5_{10}\}-\rm{Im}\{r^6_{10}\}]^2} -\epsilon
\label{ineqcda2}
\end{eqnarray}
can be obtained.
The numerical estimates $\mathcal{A} \leq 1.3 \pm 0.7$ and $\mathcal{A} \leq 1.1 \pm 1.2$ for the proton and deuteron data, respectively, show that 
the possibility for the 
values of $|T_{0-\frac{1}{2}1\frac{1}{2}}|^2$ and $|U_{0-\frac{1}{2}1\frac{1}{2}}|^2$ to be of the same order of magnitude as $|T_{00}|^2$ is not 
excluded by the presented results on $\omega$ SDMEs. For comparison, when applying Eq.~(\ref{ineqcda2}) to the results on proton 
and deuteron data in exclusive $\rho^0$-meson production~\cite{DC-24}, 
one obtains $\mathcal{A} \leq 0.22 \pm 0.09$ and $\mathcal{A} \leq 0.28 \pm 0.09$, respectively. 
This shows that in this case the probability for the amplitudes 
$T_{0-\frac{1}{2}1\frac{1}{2}}$ and $U_{0-\frac{1}{2}1\frac{1}{2}}$ to be of the same order of magnitude as $T_{00}$ is small. 

\section{SDMEs for proton and deuteron}\label{sec:tables}

\begin{table*}[hbtc!] 
\renewcommand{\arraystretch}{1.2}
\centering
\caption{\label{tab1} The 23 unpolarized and polarized  $\omega$ SDMEs  from the
proton   and deuteron data.
The first uncertainty is statistical, the second  systematic.}
\begin{tabular}{|c|c|c|}
\hline 
 element & proton & deuteron \\ 
\hline
\begin{math}r^{04}_{00}\end{math}     & 0.168  $\pm$ 0.018 $\pm$ 0.036 & 0.160  $\pm$ 0.024 $\pm$ 0.038 \\ 
\begin{math}r^{1}_{1-1}\end{math}     & -0.175 $\pm$ 0.029 $\pm$ 0.039 & -0.215 $\pm$ 0.036 $\pm$ 0.047 \\ 
Im \begin{math}r^{2}_{1-1}\end{math}  & 0.171  $\pm$ 0.029 $\pm$ 0.023 & 0.248  $\pm$ 0.037 $\pm$ 0.039 \\ 
Re \begin{math}r^{5}_{10}\end{math}   & 0.037  $\pm$ 0.009 $\pm$ 0.012 & 0.045  $\pm$ 0.010 $\pm$ 0.014 \\ 
Im \begin{math}r^{6}_{10}\end{math}   & -0.061 $\pm$ 0.008 $\pm$ 0.012 & -0.043 $\pm$ 0.010 $\pm$ 0.009 \\ 
Im \begin{math}r^{7}_{10}\end{math}   & 0.109  $\pm$ 0.075 $\pm$ 0.021 & 0.021  $\pm$ 0.087 $\pm$ 0.004 \\ 
Re \begin{math}r^{8}_{10}\end{math}   & 0.169  $\pm$ 0.075 $\pm$ 0.035 & -0.083 $\pm$ 0.083 $\pm$ 0.017 \\ 
Re \begin{math}r^{04}_{10}\end{math}  & -0.010 $\pm$ 0.012 $\pm$ 0.002 & 0.020  $\pm$ 0.014 $\pm$ 0.005 \\ 
Re \begin{math}r^{1}_{10}\end{math}   & -0.014 $\pm$ 0.019 $\pm$ 0.005 & 0.016  $\pm$ 0.022 $\pm$ 0.009 \\ 
Im \begin{math}r^{2}_{10}\end{math}   & 0.039  $\pm$ 0.018 $\pm$ 0.007 & -0.003 $\pm$ 0.023 $\pm$ 0.002 \\ 
\begin{math}r^{5}_{00}\end{math}      & 0.042  $\pm$ 0.015 $\pm$ 0.012 & 0.036  $\pm$ 0.019 $\pm$ 0.014 \\ 
\begin{math}r^{1}_{00}\end{math}      & 0.006  $\pm$ 0.029 $\pm$ 0.008 & 0.107  $\pm$ 0.036 $\pm$ 0.023 \\ 
Im \begin{math}r^{3}_{10}\end{math}   & 0.059  $\pm$ 0.047 $\pm$ 0.012 & 0.038  $\pm$ 0.056 $\pm$ 0.008 \\ 
\begin{math}r^{8}_{00}\end{math}      & -0.142 $\pm$ 0.110 $\pm$ 0.029 & -0.017 $\pm$ 0.131 $\pm$ 0.004 \\ 
\begin{math}r^{5}_{11}\end{math}      & -0.059 $\pm$ 0.012 $\pm$ 0.022 & -0.025 $\pm$ 0.015 $\pm$ 0.015 \\ 
\begin{math}r^{5}_{1-1}\end{math}     & -0.043 $\pm$ 0.014 $\pm$ 0.006 & -0.021 $\pm$ 0.018 $\pm$ 0.001 \\ 
Im \begin{math}r^{6}_{1-1}\end{math}  & 0.036  $\pm$ 0.014 $\pm$ 0.008 & 0.056  $\pm$ 0.019 $\pm$ 0.013 \\ 
Im \begin{math}r^{7}_{1-1}\end{math}  & -0.092 $\pm$ 0.117 $\pm$ 0.018 & 0.113  $\pm$ 0.135 $\pm$ 0.028 \\ 
\begin{math}r^{8}_{11}\end{math}      & -0.079 $\pm$ 0.089 $\pm$ 0.017 & -0.097 $\pm$ 0.103 $\pm$ 0.020 \\ 
Im \begin{math}r^{8}_{1-1}\end{math}  & -0.060 $\pm$ 0.110 $\pm$ 0.012 & -0.150 $\pm$ 0.125 $\pm$ 0.034 \\ 
\begin{math}r^{04}_{1-1}\end{math}    & -0.004 $\pm$ 0.018 $\pm$ 0.004 & 0.060  $\pm$ 0.023 $\pm$ 0.016 \\ 
\begin{math}r^{1}_{11}\end{math}      & 0.014  $\pm$ 0.024 $\pm$ 0.004 & -0.037 $\pm$ 0.030 $\pm$ 0.007 \\ 
\begin{math}r^{3}_{1-1}\end{math}     & 0.023  $\pm$ 0.076 $\pm$ 0.010 & -0.122 $\pm$ 0.089 $\pm$ 0.025 \\ 
\hline
\end{tabular}
\end{table*}

\begin{table*}[hbtc!] 
\renewcommand{\arraystretch}{1.2}
\centering
\caption{\label{tab2} The 23 unpolarized and polarized $\omega$ SDMEs for the
proton data in $Q^2$ intervals: $1.00 - 1.57 - 2.55 - 10.00$ GeV$^2$. The first
uncertainty is statistical, the second  systematic.}
\begin{tabular}{|c|c|c|c|}
\hline 
element                               &  $\langle$Q$^{2}$$\rangle$ = 1.28 GeV$^{2}$ &  $\langle$Q$^{2}$$\rangle$ = 2.00 GeV$^{2}$ &  $\langle$Q$^{2}$$\rangle$ = 4.00 GeV$^{2}$  \\ 
\hline 
\begin{math}r^{04}_{00}\end{math}     & 0.164  $\pm$ 0.034 $\pm$ 0.022 & 0.166  $\pm$ 0.030 $\pm$ 0.044 & 0.179  $\pm$ 0.031 $\pm$ 0.036 \\ 
\begin{math}r^{1}_{1-1}\end{math}     & -0.032 $\pm$ 0.050 $\pm$ 0.032 & -0.175 $\pm$ 0.049 $\pm$ 0.037 & -0.314 $\pm$ 0.053 $\pm$ 0.090 \\ 
Im \begin{math}r^{2}_{1-1}\end{math}  & 0.172  $\pm$ 0.048 $\pm$ 0.027 & 0.133  $\pm$ 0.050 $\pm$ 0.043 & 0.163  $\pm$ 0.057 $\pm$ 0.029 \\ 
Re \begin{math}r^{5}_{10}\end{math}   & 0.038  $\pm$ 0.016 $\pm$ 0.018 & 0.022  $\pm$ 0.015 $\pm$ 0.010 & 0.053  $\pm$ 0.015 $\pm$ 0.022 \\ 
Im \begin{math}r^{6}_{10}\end{math}   & -0.062 $\pm$ 0.015 $\pm$ 0.012 & -0.069 $\pm$ 0.012 $\pm$ 0.014 & -0.046 $\pm$ 0.014 $\pm$ 0.013 \\ 
Im \begin{math}r^{7}_{10}\end{math}   & 0.163  $\pm$ 0.139 $\pm$ 0.030 & -0.006 $\pm$ 0.125 $\pm$ 0.009 & 0.170  $\pm$ 0.128 $\pm$ 0.042 \\ 
Re \begin{math}r^{8}_{10}\end{math}   & 0.088  $\pm$ 0.143 $\pm$ 0.021 & 0.078  $\pm$ 0.137 $\pm$ 0.028 & 0.280  $\pm$ 0.119 $\pm$ 0.067 \\ 
Re \begin{math}r^{04}_{10}\end{math}  & 0.005  $\pm$ 0.021 $\pm$ 0.004 & -0.060 $\pm$ 0.020 $\pm$ 0.011 & 0.016  $\pm$ 0.019 $\pm$ 0.022 \\ 
Re \begin{math}r^{1}_{10}\end{math}   & -0.005 $\pm$ 0.032 $\pm$ 0.013 & -0.090 $\pm$ 0.031 $\pm$ 0.012 & 0.073  $\pm$ 0.034 $\pm$ 0.016 \\ 
Im \begin{math}r^{2}_{10}\end{math}   & 0.012  $\pm$ 0.030 $\pm$ 0.012 & 0.042  $\pm$ 0.030 $\pm$ 0.003 & 0.036  $\pm$ 0.034 $\pm$ 0.016 \\ 
\begin{math}r^{5}_{00}\end{math}      & 0.031  $\pm$ 0.029 $\pm$ 0.001 & 0.029  $\pm$ 0.025 $\pm$ 0.012 & 0.068  $\pm$ 0.027 $\pm$ 0.016 \\ 
\begin{math}r^{1}_{00}\end{math}      & 0.009  $\pm$ 0.049 $\pm$ 0.011 & 0.039  $\pm$ 0.049 $\pm$ 0.013 & -0.032 $\pm$ 0.053 $\pm$ 0.015 \\ 
Im \begin{math}r^{3}_{10}\end{math}   & 0.044  $\pm$ 0.096 $\pm$ 0.008 & 0.047  $\pm$ 0.076 $\pm$ 0.009 & 0.073  $\pm$ 0.076 $\pm$ 0.018 \\ 
\begin{math}r^{8}_{00}\end{math}      & -0.147 $\pm$ 0.210 $\pm$ 0.039 & 0.035  $\pm$ 0.196 $\pm$ 0.026 & -0.197 $\pm$ 0.171 $\pm$ 0.045 \\ 
\begin{math}r^{5}_{11}\end{math}      & -0.074 $\pm$ 0.020 $\pm$ 0.021 & -0.050 $\pm$ 0.020 $\pm$ 0.012 & -0.070 $\pm$ 0.021 $\pm$ 0.029 \\ 
\begin{math}r^{5}_{1-1}\end{math}     & -0.047 $\pm$ 0.024 $\pm$ 0.007 & -0.078 $\pm$ 0.025 $\pm$ 0.021 & 0.008  $\pm$ 0.025 $\pm$ 0.009 \\ 
Im \begin{math}r^{6}_{1-1}\end{math}  & 0.070  $\pm$ 0.025 $\pm$ 0.013 & -0.015 $\pm$ 0.024 $\pm$ 0.017 & 0.043  $\pm$ 0.026 $\pm$ 0.026 \\ 
Im \begin{math}r^{7}_{1-1}\end{math}  & -0.326 $\pm$ 0.223 $\pm$ 0.058 & -0.161 $\pm$ 0.198 $\pm$ 0.030 & 0.046  $\pm$ 0.204 $\pm$ 0.023 \\ 
\begin{math}r^{8}_{11}\end{math}      & 0.276  $\pm$ 0.171 $\pm$ 0.049 & -0.120 $\pm$ 0.155 $\pm$ 0.021 & -0.312 $\pm$ 0.144 $\pm$ 0.080 \\ 
Im \begin{math}r^{8}_{1-1}\end{math}  & -0.507 $\pm$ 0.212 $\pm$ 0.093 & -0.026 $\pm$ 0.188 $\pm$ 0.005 & 0.185  $\pm$ 0.178 $\pm$ 0.063 \\ 
\begin{math}r^{04}_{1-1}\end{math}    & -0.004 $\pm$ 0.032 $\pm$ 0.000 & -0.023 $\pm$ 0.031 $\pm$ 0.003 & 0.008  $\pm$ 0.031 $\pm$ 0.014 \\ 
\begin{math}r^{1}_{11}\end{math}      & 0.063  $\pm$ 0.040 $\pm$ 0.015 & -0.037 $\pm$ 0.041 $\pm$ 0.012 & 0.003  $\pm$ 0.044 $\pm$ 0.012 \\ 
\begin{math}r^{3}_{1-1}\end{math}     & 0.074  $\pm$ 0.153 $\pm$ 0.013 & -0.110 $\pm$ 0.131 $\pm$ 0.021 & 0.088  $\pm$ 0.124 $\pm$ 0.024 \\ 
\hline
\end{tabular}
\end{table*}

\begin{table*}[hbtc!] 
\renewcommand{\arraystretch}{1.2}
\centering
\caption{\label{tab3} The 23 unpolarized and polarized $\omega$ SDMEs for the
proton data in $-t'$ intervals: $0.000 - 0.044 - 0.105 - 0.200$ GeV$^2$.
 The first uncertainty is statistical, the second  systematic.}
\begin{tabular}{|c|c|c|c|}
\hline 
 element & $\langle-t'\rangle$ = 0.021 GeV$^{2}$ & $\langle-t'\rangle$ = 0.072 GeV$^{2}$ & $\langle-t'\rangle$ = 0.147 GeV$^{2}$ \\ 
\hline
\begin{math}r^{04}_{00}\end{math}     & 0.136  $\pm$ 0.027 $\pm$ 0.036 & 0.197  $\pm$ 0.032 $\pm$ 0.027 & 0.212  $\pm$ 0.036 $\pm$ 0.032 \\ 
\begin{math}r^{1}_{1-1}\end{math}     & -0.239 $\pm$ 0.043 $\pm$ 0.023 & -0.141 $\pm$ 0.048 $\pm$ 0.043 & -0.120 $\pm$ 0.060 $\pm$ 0.048 \\ 
Im \begin{math}r^{2}_{1-1}\end{math}  & 0.220  $\pm$ 0.045 $\pm$ 0.033 & 0.138  $\pm$ 0.050 $\pm$ 0.015 & 0.111  $\pm$ 0.057 $\pm$ 0.012 \\ 
Re \begin{math}r^{5}_{10}\end{math}   & 0.015  $\pm$ 0.013 $\pm$ 0.008 & 0.032  $\pm$ 0.015 $\pm$ 0.010 & 0.081  $\pm$ 0.018 $\pm$ 0.025 \\ 
Im \begin{math}r^{6}_{10}\end{math}   & -0.051 $\pm$ 0.012 $\pm$ 0.012 & -0.077 $\pm$ 0.013 $\pm$ 0.013 & -0.058 $\pm$ 0.015 $\pm$ 0.018 \\ 
Im \begin{math}r^{7}_{10}\end{math}   & -0.143 $\pm$ 0.121 $\pm$ 0.037 & 0.340  $\pm$ 0.123 $\pm$ 0.071 & 0.277  $\pm$ 0.146 $\pm$ 0.073 \\ 
Re \begin{math}r^{8}_{10}\end{math}   & 0.151  $\pm$ 0.125 $\pm$ 0.039 & 0.232  $\pm$ 0.127 $\pm$ 0.044 & 0.151  $\pm$ 0.136 $\pm$ 0.039 \\ 
Re \begin{math}r^{04}_{10}\end{math}  & -0.022 $\pm$ 0.018 $\pm$ 0.004 & 0.010  $\pm$ 0.020 $\pm$ 0.006 & 0.006  $\pm$ 0.023 $\pm$ 0.002 \\ 
Re \begin{math}r^{1}_{10}\end{math}   & -0.020 $\pm$ 0.030 $\pm$ 0.007 & -0.013 $\pm$ 0.032 $\pm$ 0.001 & -0.029 $\pm$ 0.035 $\pm$ 0.011 \\ 
Im \begin{math}r^{2}_{10}\end{math}   & 0.017  $\pm$ 0.029 $\pm$ 0.008 & -0.003 $\pm$ 0.029 $\pm$ 0.005 & 0.125  $\pm$ 0.033 $\pm$ 0.023 \\ 
\begin{math}r^{5}_{00}\end{math}      & -0.016 $\pm$ 0.023 $\pm$ 0.029 & 0.059  $\pm$ 0.027 $\pm$ 0.011 & 0.100  $\pm$ 0.031 $\pm$ 0.012 \\ 
\begin{math}r^{1}_{00}\end{math}      & 0.032  $\pm$ 0.047 $\pm$ 0.033 & 0.067  $\pm$ 0.050 $\pm$ 0.024 & -0.106 $\pm$ 0.053 $\pm$ 0.067 \\ 
Im \begin{math}r^{3}_{10}\end{math}   & 0.063  $\pm$ 0.073 $\pm$ 0.010 & 0.076  $\pm$ 0.082 $\pm$ 0.018 & 0.121  $\pm$ 0.090 $\pm$ 0.036 \\ 
\begin{math}r^{8}_{00}\end{math}      & 0.155  $\pm$ 0.179 $\pm$ 0.033 & -0.138 $\pm$ 0.197 $\pm$ 0.026 & -0.442 $\pm$ 0.191 $\pm$ 0.115 \\ 
\begin{math}r^{5}_{11}\end{math}      & -0.059 $\pm$ 0.018 $\pm$ 0.012 & -0.051 $\pm$ 0.020 $\pm$ 0.015 & -0.068 $\pm$ 0.024 $\pm$ 0.048 \\ 
\begin{math}r^{5}_{1-1}\end{math}     & -0.034 $\pm$ 0.022 $\pm$ 0.002 & -0.060 $\pm$ 0.024 $\pm$ 0.007 & -0.052 $\pm$ 0.030 $\pm$ 0.011 \\ 
Im \begin{math}r^{6}_{1-1}\end{math}  & 0.010  $\pm$ 0.022 $\pm$ 0.000 & 0.090  $\pm$ 0.024 $\pm$ 0.020 & 0.020  $\pm$ 0.028 $\pm$ 0.009 \\ 
Im \begin{math}r^{7}_{1-1}\end{math}  & -0.027 $\pm$ 0.176 $\pm$ 0.004 & 0.244  $\pm$ 0.197 $\pm$ 0.046 & -0.601 $\pm$ 0.233 $\pm$ 0.165 \\ 
\begin{math}r^{8}_{11}\end{math}      & -0.136 $\pm$ 0.145 $\pm$ 0.023 & -0.155 $\pm$ 0.150 $\pm$ 0.029 & 0.038  $\pm$ 0.169 $\pm$ 0.010 \\ 
Im \begin{math}r^{8}_{1-1}\end{math}  & -0.182 $\pm$ 0.181 $\pm$ 0.046 & 0.085  $\pm$ 0.180 $\pm$ 0.017 & -0.055 $\pm$ 0.210 $\pm$ 0.025 \\ 
\begin{math}r^{04}_{1-1}\end{math}    & -0.006 $\pm$ 0.029 $\pm$ 0.003 & -0.007 $\pm$ 0.030 $\pm$ 0.006 & -0.023 $\pm$ 0.036 $\pm$ 0.008 \\ 
\begin{math}r^{1}_{11}\end{math}      & 0.009  $\pm$ 0.037 $\pm$ 0.005 & 0.023  $\pm$ 0.040 $\pm$ 0.006 & 0.033  $\pm$ 0.047 $\pm$ 0.029 \\ 
\begin{math}r^{3}_{1-1}\end{math}     & -0.016 $\pm$ 0.111 $\pm$ 0.006 & 0.160  $\pm$ 0.134 $\pm$ 0.036 & -0.154 $\pm$ 0.156 $\pm$ 0.054 \\ 
\hline
\end{tabular}
\end{table*} 

\begin{table*}[hbtc!] 
\renewcommand{\arraystretch}{1.2}
\centering
\caption{\label{tab4} The 23 unpolarized and polarized $\omega$ SDMEs for  the
deuteron data in $Q^2$ intervals: $1.00 - 1.57 - 2.55 - 10.00$ GeV$^2$. The first
uncertainty is statistical, the second  systematic.}
\begin{tabular}{|c|c|c|c|}
\hline 
element                               &  $\langle$Q$^{2}$$\rangle$ = 1.28 GeV$^{2}$ &  $\langle$Q$^{2}$$\rangle$ = 2.00 GeV$^{2}$ &  $\langle$Q$^{2}$$\rangle$ = 4.00 GeV$^{2}$  \\ 
\hline 
\begin{math}r^{04}_{00}\end{math}     & 0.148  $\pm$ 0.043 $\pm$ 0.025 & 0.132  $\pm$ 0.041 $\pm$ 0.053 & 0.186  $\pm$ 0.040 $\pm$ 0.034 \\ 
\begin{math}r^{1}_{1-1}\end{math}     & -0.045 $\pm$ 0.063 $\pm$ 0.030 & -0.347 $\pm$ 0.058 $\pm$ 0.075 & -0.258 $\pm$ 0.072 $\pm$ 0.070 \\ 
Im \begin{math}r^{2}_{1-1}\end{math}  & 0.232  $\pm$ 0.063 $\pm$ 0.045 & 0.216  $\pm$ 0.065 $\pm$ 0.063 & 0.313  $\pm$ 0.073 $\pm$ 0.056 \\ 
Re \begin{math}r^{5}_{10}\end{math}   & 0.059  $\pm$ 0.020 $\pm$ 0.021 & 0.056  $\pm$ 0.017 $\pm$ 0.015 & 0.025  $\pm$ 0.020 $\pm$ 0.014 \\ 
Im \begin{math}r^{6}_{10}\end{math}   & -0.034 $\pm$ 0.018 $\pm$ 0.006 & -0.039 $\pm$ 0.016 $\pm$ 0.009 & -0.055 $\pm$ 0.021 $\pm$ 0.015 \\ 
Im \begin{math}r^{7}_{10}\end{math}   & -0.174 $\pm$ 0.160 $\pm$ 0.032 & 0.225  $\pm$ 0.150 $\pm$ 0.044 & -0.068 $\pm$ 0.156 $\pm$ 0.015 \\ 
Re \begin{math}r^{8}_{10}\end{math}   & -0.026 $\pm$ 0.154 $\pm$ 0.005 & -0.197 $\pm$ 0.148 $\pm$ 0.039 & 0.020  $\pm$ 0.140 $\pm$ 0.004 \\ 
Re \begin{math}r^{04}_{10}\end{math}  & -0.004 $\pm$ 0.027 $\pm$ 0.007 & 0.020  $\pm$ 0.024 $\pm$ 0.011 & 0.040  $\pm$ 0.025 $\pm$ 0.012 \\ 
Re \begin{math}r^{1}_{10}\end{math}   & -0.039 $\pm$ 0.037 $\pm$ 0.019 & 0.052  $\pm$ 0.037 $\pm$ 0.015 & 0.025  $\pm$ 0.046 $\pm$ 0.008 \\ 
Im \begin{math}r^{2}_{10}\end{math}   & 0.014  $\pm$ 0.037 $\pm$ 0.013 & 0.003  $\pm$ 0.036 $\pm$ 0.012 & -0.028 $\pm$ 0.049 $\pm$ 0.004 \\ 
\begin{math}r^{5}_{00}\end{math}      & 0.074  $\pm$ 0.033 $\pm$ 0.007 & 0.050  $\pm$ 0.032 $\pm$ 0.012 & -0.006 $\pm$ 0.035 $\pm$ 0.031 \\ 
\begin{math}r^{1}_{00}\end{math}      & 0.079  $\pm$ 0.061 $\pm$ 0.028 & 0.077  $\pm$ 0.059 $\pm$ 0.012 & 0.143  $\pm$ 0.073 $\pm$ 0.048 \\ 
Im \begin{math}r^{3}_{10}\end{math}   & 0.124  $\pm$ 0.107 $\pm$ 0.031 & 0.009  $\pm$ 0.095 $\pm$ 0.002 & 0.016  $\pm$ 0.096 $\pm$ 0.004 \\ 
\begin{math}r^{8}_{00}\end{math}      & 0.186  $\pm$ 0.248 $\pm$ 0.041 & -0.024 $\pm$ 0.242 $\pm$ 0.005 & -0.088 $\pm$ 0.211 $\pm$ 0.019 \\ 
\begin{math}r^{5}_{11}\end{math}      & -0.027 $\pm$ 0.026 $\pm$ 0.013 & -0.054 $\pm$ 0.025 $\pm$ 0.018 & -0.001 $\pm$ 0.030 $\pm$ 0.011 \\ 
\begin{math}r^{5}_{1-1}\end{math}     & -0.040 $\pm$ 0.031 $\pm$ 0.005 & -0.049 $\pm$ 0.031 $\pm$ 0.010 & 0.021  $\pm$ 0.036 $\pm$ 0.009 \\ 
Im \begin{math}r^{6}_{1-1}\end{math}  & 0.062  $\pm$ 0.031 $\pm$ 0.016 & 0.050  $\pm$ 0.032 $\pm$ 0.004 & 0.057  $\pm$ 0.035 $\pm$ 0.021 \\ 
Im \begin{math}r^{7}_{1-1}\end{math}  & 0.399  $\pm$ 0.250 $\pm$ 0.079 & -0.053 $\pm$ 0.236 $\pm$ 0.011 & -0.003 $\pm$ 0.234 $\pm$ 0.001 \\ 
\begin{math}r^{8}_{11}\end{math}      & -0.332 $\pm$ 0.193 $\pm$ 0.059 & -0.103 $\pm$ 0.184 $\pm$ 0.020 & -0.022 $\pm$ 0.164 $\pm$ 0.005 \\ 
Im \begin{math}r^{8}_{1-1}\end{math}  & -0.260 $\pm$ 0.234 $\pm$ 0.075 & -0.051 $\pm$ 0.216 $\pm$ 0.033 & -0.129 $\pm$ 0.200 $\pm$ 0.029 \\ 
\begin{math}r^{04}_{1-1}\end{math}    & 0.043  $\pm$ 0.040 $\pm$ 0.013 & 0.005  $\pm$ 0.039 $\pm$ 0.008 & 0.150  $\pm$ 0.040 $\pm$ 0.040 \\ 
\begin{math}r^{1}_{11}\end{math}      & 0.009  $\pm$ 0.048 $\pm$ 0.003 & -0.027 $\pm$ 0.051 $\pm$ 0.011 & -0.104 $\pm$ 0.060 $\pm$ 0.012 \\ 
\begin{math}r^{3}_{1-1}\end{math}     & -0.006 $\pm$ 0.174 $\pm$ 0.001 & -0.337 $\pm$ 0.157 $\pm$ 0.071 & 0.021  $\pm$ 0.141 $\pm$ 0.005 \\ 
\hline
\end{tabular}
\end{table*}

\begin{table*}[hbtc!] 
\renewcommand{\arraystretch}{1.2}
\centering
\caption{\label{tab5} The 23 unpolarized and polarized $\omega$ SDMEs for
the 
deuteron data in $-t'$ intervals: $0.000 - 0.044 - 0.105 - 0.200$ GeV$^2$.
The first uncertainty is statistical, the second  systematic.} 
\begin{tabular}{|c|c|c|c|}
\hline 
 element & $\langle-t'\rangle$ = 0.021 GeV$^{2}$ & $\langle-t'\rangle$ = 0.071 GeV$^{2}$ & $\langle-t'\rangle$ = 0.147 GeV$^{2}$ \\ 
\hline
\begin{math}r^{04}_{00}\end{math}     & 0.153  $\pm$ 0.034 $\pm$ 0.031 & 0.147  $\pm$ 0.041 $\pm$ 0.036 & 0.215  $\pm$ 0.050 $\pm$ 0.028 \\ 
\begin{math}r^{1}_{1-1}\end{math}     & -0.167 $\pm$ 0.054 $\pm$ 0.029 & -0.298 $\pm$ 0.063 $\pm$ 0.074 & -0.238 $\pm$ 0.074 $\pm$ 0.083 \\ 
Im \begin{math}r^{2}_{1-1}\end{math}  & 0.281  $\pm$ 0.056 $\pm$ 0.044 & 0.198  $\pm$ 0.064 $\pm$ 0.036 & 0.309  $\pm$ 0.070 $\pm$ 0.067 \\ 
Re \begin{math}r^{5}_{10}\end{math}   & 0.030  $\pm$ 0.015 $\pm$ 0.010 & 0.043  $\pm$ 0.018 $\pm$ 0.012 & 0.070  $\pm$ 0.024 $\pm$ 0.024 \\ 
Im \begin{math}r^{6}_{10}\end{math}   & -0.050 $\pm$ 0.016 $\pm$ 0.008 & -0.045 $\pm$ 0.017 $\pm$ 0.010 & -0.030 $\pm$ 0.022 $\pm$ 0.011 \\ 
Im \begin{math}r^{7}_{10}\end{math}   & -0.067 $\pm$ 0.130 $\pm$ 0.010 & 0.041  $\pm$ 0.150 $\pm$ 0.008 & 0.201  $\pm$ 0.179 $\pm$ 0.055 \\ 
Re \begin{math}r^{8}_{10}\end{math}   & 0.062  $\pm$ 0.136 $\pm$ 0.015 & -0.406 $\pm$ 0.153 $\pm$ 0.078 & -0.011 $\pm$ 0.143 $\pm$ 0.003 \\ 
Re \begin{math}r^{04}_{10}\end{math}  & 0.032  $\pm$ 0.022 $\pm$ 0.004 & -0.020 $\pm$ 0.025 $\pm$ 0.006 & 0.050  $\pm$ 0.030 $\pm$ 0.014 \\ 
Re \begin{math}r^{1}_{10}\end{math}   & 0.028  $\pm$ 0.035 $\pm$ 0.002 & 0.007  $\pm$ 0.038 $\pm$ 0.009 & 0.001  $\pm$ 0.045 $\pm$ 0.001 \\ 
Im \begin{math}r^{2}_{10}\end{math}   & -0.060 $\pm$ 0.034 $\pm$ 0.012 & 0.082  $\pm$ 0.038 $\pm$ 0.022 & -0.020 $\pm$ 0.048 $\pm$ 0.016 \\ 
\begin{math}r^{5}_{00}\end{math}      & 0.007  $\pm$ 0.027 $\pm$ 0.021 & 0.036  $\pm$ 0.033 $\pm$ 0.018 & 0.089  $\pm$ 0.043 $\pm$ 0.012 \\ 
\begin{math}r^{1}_{00}\end{math}      & 0.092  $\pm$ 0.057 $\pm$ 0.043 & 0.117  $\pm$ 0.055 $\pm$ 0.039 & 0.145  $\pm$ 0.080 $\pm$ 0.005 \\ 
Im \begin{math}r^{3}_{10}\end{math}   & -0.009 $\pm$ 0.081 $\pm$ 0.001 & 0.160  $\pm$ 0.099 $\pm$ 0.033 & 0.059  $\pm$ 0.119 $\pm$ 0.016 \\ 
\begin{math}r^{8}_{00}\end{math}      & 0.029  $\pm$ 0.209 $\pm$ 0.004 & -0.302 $\pm$ 0.223 $\pm$ 0.063 & 0.211  $\pm$ 0.256 $\pm$ 0.058 \\ 
\begin{math}r^{5}_{11}\end{math}      & -0.030 $\pm$ 0.022 $\pm$ 0.008 & -0.032 $\pm$ 0.027 $\pm$ 0.011 & -0.022 $\pm$ 0.032 $\pm$ 0.038 \\ 
\begin{math}r^{5}_{1-1}\end{math}     & -0.029 $\pm$ 0.027 $\pm$ 0.000 & -0.025 $\pm$ 0.032 $\pm$ 0.004 & 0.014  $\pm$ 0.042 $\pm$ 0.013 \\ 
Im \begin{math}r^{6}_{1-1}\end{math}  & 0.077  $\pm$ 0.028 $\pm$ 0.022 & 0.063  $\pm$ 0.033 $\pm$ 0.014 & 0.008  $\pm$ 0.035 $\pm$ 0.009 \\ 
Im \begin{math}r^{7}_{1-1}\end{math}  & -0.157 $\pm$ 0.208 $\pm$ 0.023 & 0.411  $\pm$ 0.238 $\pm$ 0.085 & 0.087  $\pm$ 0.267 $\pm$ 0.024 \\ 
\begin{math}r^{8}_{11}\end{math}      & 0.005  $\pm$ 0.163 $\pm$ 0.001 & 0.018  $\pm$ 0.182 $\pm$ 0.007 & -0.325 $\pm$ 0.186 $\pm$ 0.089 \\ 
Im \begin{math}r^{8}_{1-1}\end{math}  & -0.165 $\pm$ 0.193 $\pm$ 0.024 & -0.100 $\pm$ 0.228 $\pm$ 0.040 & -0.172 $\pm$ 0.229 $\pm$ 0.047 \\ 
\begin{math}r^{04}_{1-1}\end{math}    & 0.021  $\pm$ 0.034 $\pm$ 0.001 & 0.052  $\pm$ 0.041 $\pm$ 0.022 & 0.140  $\pm$ 0.048 $\pm$ 0.052 \\ 
\begin{math}r^{1}_{11}\end{math}      & 0.009  $\pm$ 0.045 $\pm$ 0.005 & -0.013 $\pm$ 0.053 $\pm$ 0.005 & -0.145 $\pm$ 0.059 $\pm$ 0.038 \\ 
\begin{math}r^{3}_{1-1}\end{math}     & 0.030  $\pm$ 0.132 $\pm$ 0.011 & -0.083 $\pm$ 0.165 $\pm$ 0.029 & -0.247 $\pm$ 0.177 $\pm$ 0.068 \\ 
\hline
\end{tabular}
\end{table*}

\newpage
\begin{table*}[hbtc!] 
\renewcommand{\arraystretch}{1.2}
\centering
\caption{\label{tab6} The 23 unpolarized and polarized $\omega$ SDMEs in the Diehl representation~\cite{Diehl} for  proton  and deuteron data in the entire
kinematic region. The first uncertainty is statistical, the second  systematic.} 
\begin{tabular}{|c|c|c|}
\hline 
 element & proton & deuteron \\ 
\hline
\begin{math}u^{00}_{++} + \epsilon \cdot u^{00}_{00}\end{math}                        & 0.168  $\pm$ 0.018 $\pm$ 0.036 & 0.160  $\pm$ 0.024 $\pm$ 0.038 \\ 
Re \begin{math}u^{00}_{0+}\end{math}                                                  & -0.010 $\pm$ 0.012 $\pm$ 0.002 & 0.020  $\pm$ 0.014 $\pm$ 0.005 \\ 
\begin{math}u^{00}_{-+}\end{math}                                                     & -0.004 $\pm$ 0.018 $\pm$ 0.004 & 0.060  $\pm$ 0.023 $\pm$ 0.016 \\ 
Re \begin{math}(u^{0+}_{0+} - u^{-0}_{0+})\end{math}                                  & 0.014  $\pm$ 0.024 $\pm$ 0.004 & -0.037 $\pm$ 0.030 $\pm$ 0.007 \\ 
Re \begin{math}(u^{0+}_{++} - u^{-0}_{++} + 2\epsilon \cdot u^{0+}_{00})\end{math}    & 0.006  $\pm$ 0.029 $\pm$ 0.008 & 0.107  $\pm$ 0.036 $\pm$ 0.023 \\ 
Re \begin{math}u^{0+}_{-+}\end{math}                                                  & -0.014 $\pm$ 0.019 $\pm$ 0.005 & 0.016  $\pm$ 0.022 $\pm$ 0.009 \\ 
Re \begin{math}(u^{0-}_{0+} - u^{+0}_{0+})\end{math}                                  & -0.175 $\pm$ 0.029 $\pm$ 0.039 & -0.215 $\pm$ 0.036 $\pm$ 0.047 \\ 
Re \begin{math}u^{0+}_{-+}\end{math}                                                  & 0.039  $\pm$ 0.018 $\pm$ 0.007 & -0.003 $\pm$ 0.023 $\pm$ 0.002 \\ 
\begin{math}u^{-+}_{-+}\end{math}                                                     & 0.171  $\pm$ 0.029 $\pm$ 0.023 & 0.248  $\pm$ 0.037 $\pm$ 0.039 \\ 
Re \begin{math}(u^{++}_{0+} + u^{--}_{0+})\end{math}                                  & -0.059 $\pm$ 0.012 $\pm$ 0.022 & -0.025 $\pm$ 0.015 $\pm$ 0.015 \\ 
Re \begin{math}u^{-+}_{0+}\end{math}                                                  & 0.042  $\pm$ 0.015 $\pm$ 0.012 & 0.036  $\pm$ 0.019 $\pm$ 0.014 \\ 
Re \begin{math}(u^{-+}_{++} + \epsilon \cdot u^{-+}_{00})\end{math}                   & 0.037  $\pm$ 0.009 $\pm$ 0.012 & 0.045  $\pm$ 0.010 $\pm$ 0.014 \\ 
Re \begin{math}u^{++}_{-+}\end{math}                                                  & -0.043 $\pm$ 0.014 $\pm$ 0.006 & -0.021 $\pm$ 0.018 $\pm$ 0.001 \\ 
Re \begin{math}u^{+-}_{0+}\end{math}                                                  & -0.061 $\pm$ 0.008 $\pm$ 0.012 & -0.043 $\pm$ 0.010 $\pm$ 0.009 \\ 
\begin{math}u^{+-}_{-+}\end{math}                                                     & 0.036  $\pm$ 0.014 $\pm$ 0.008 & 0.056  $\pm$ 0.019 $\pm$ 0.013 \\ 
Im \begin{math}u^{00}_{0+}\end{math}                                                  & 0.059  $\pm$ 0.047 $\pm$ 0.012 & 0.038  $\pm$ 0.056 $\pm$ 0.008 \\ 
Im \begin{math}(u^{0+}_{0+} - u^{-0}_{0+})\end{math}                                  & 0.023  $\pm$ 0.076 $\pm$ 0.010 & -0.122 $\pm$ 0.089 $\pm$ 0.025 \\ 
Im \begin{math}(u^{0+}_{++} - u^{-0}_{++})\end{math}                                  & 0.109  $\pm$ 0.075 $\pm$ 0.021 & 0.021  $\pm$ 0.087 $\pm$ 0.004 \\ 
Im \begin{math}(u^{0-}_{0+} - u^{+0}_{0+})\end{math}                                  & -0.092 $\pm$ 0.117 $\pm$ 0.018 & 0.113  $\pm$ 0.135 $\pm$ 0.028 \\ 
Im \begin{math}(u^{++}_{0+} + u^{--}_{0+})\end{math}                                  & -0.079 $\pm$ 0.089 $\pm$ 0.017 & -0.097 $\pm$ 0.103 $\pm$ 0.020 \\ 
Im \begin{math}u^{-+}_{0+}\end{math}                                                  & -0.142 $\pm$ 0.110 $\pm$ 0.029 & -0.017 $\pm$ 0.131 $\pm$ 0.004 \\ 
Im \begin{math}u^{-+}_{++}\end{math}                                                  & 0.169  $\pm$ 0.075 $\pm$ 0.035 & -0.083 $\pm$ 0.083 $\pm$ 0.017 \\ 
Im \begin{math}u^{+-}_{0+}\end{math}                                                  & -0.060 $\pm$ 0.110 $\pm$ 0.012 & -0.150 $\pm$ 0.125 $\pm$ 0.034 \\ 
\hline
\end{tabular}
\end{table*}

\begin{table*}[hbtc!] 
\renewcommand{\arraystretch}{1.2}
\caption{\label{tab7} The definition of intervals and the mean values for kinematic variables.}
% for hydrogen and deuterium data}
\centering
\begin{tabular}{|c|c|c|c|c|} 
\hline
 bin & $\langle Q^{2} \rangle$ [GeV$^2$]  &$\langle-t' \rangle$ [GeV$^2$] & $\langle W \rangle$ [GeV]  & $\langle x_{B} \rangle$\\ 
\hline
\hline
``overall'' 							& 2.42  	& 0.080 	& 4.80 	& 0.097 \\
\hline 
1.00 GeV$^2 <Q^{2}< 1.57$ GeV$^2$ 	& 1.28 	& 0.082 	& 4.87 	& 0.059 \\  
1.57 GeV$^2 <Q^{2}< 2.55$ GeV$^2$ 	& 2.00 	& 0.079 	& 4.78 	& 0.085 \\
$Q^{2}>2.55$ GeV$^2$ 				& 4.00	& 0.078 	& 4.91 	& 0.147 \\
%\hline
%$Q^{2}$ $> 1.00$ GeV$^2$
%\hline
%bin & $\langle-t^{'}\rangle$ [GeV$^2$] \\
%\hline
%0.000 GeV$^2 <-t'< 0.200$ GeV$^2$ 	& 2.38	& 0.080 	& \\ 
\hline
0.000 GeV$^2 <-t'< 0.044$ GeV$^2$ 	& 2.38	& 0.021	& 4.73	& 0.097 \\
0.044 GeV$^2 <-t'< 0.105$ GeV$^2$ 	& 2.49	& 0.072	& 4.78	& 0.099 \\
0.105 GeV$^2 <-t'< 0.200$ GeV$^2$ 	& 2.39	& 0.147	& 4.85	& 0.095 \\
\hline
\end{tabular}
\end{table*}

%\end{appendix}

\newpage

\begin{table*}
\caption{\small The correlation matrix for the 23 SDMEs
obtained from the hydrogen target data. The column headings do not indicate the real and imaginary parts of any
SDMEs in order to keep the table compact.
}
\renewcommand{\arraystretch}{1.3}
\setlength{\tabcolsep}{3pt}
\hspace*{2.0cm}
\begin{sideways}
\begin{footnotesize}
\begin{tabular}{|l|c|c|c|c|c|c|c|c|c|c|c|c|c|c|c|c|c|c|c|c|c|c|c| }
\hline
SDME & $r^{04}_{00}$  & $r^{04}_{10}$ & $r^{04}_{1-1}$  & $r^1_{11}$ & $r^1_{00}$ & $r^1_{10}$ & $r^1_{1-1}$ & $r^2_{10}$ & $r^2_{1-1}$ & $r^5_{11}$ & $r^5_{00}$ & $r^5_{10}$ & $r^5_{1-1}$ & $r^6_{10}$ & $r^6_{1-1}$ & $r^3_{10}$ & $r^3_{1-1}$ & $r^7_{10}$ & $r^7_{1-1}$ & $r^8_{11}$ & $r^8_{00}$ & $r^8_{10}$ & $r^8_{1-1}$ \\
\hline
$r^{04}_{00}$   & 1.00&\\
Re $r^{04}_{10}$& 0.07& 1.00&\\
$r^{04}_{1-1}$  & 0.03&-0.06& 1.00&\\
$r^1_{11}$      & 0.02&-0.01&-0.37& 1.00&\\
$r^1_{00}$      &-0.07&-0.08& 0.06&-0.38& 1.00&\\
Re $r^1_{10}$   &-0.04& 0.17& 0.02& 0.00& 0.08& 1.00&\\
$r^1_{1-1}$     & 0.20& 0.01& 0.01&-0.02&-0.02&-0.03& 1.00&\\
Im $r^2_{10}$   & 0.09&-0.20& 0.08& 0.01& 0.00&-0.04& 0.04& 1.00&\\
Im $r^2_{1-1}$  &-0.23&-0.06&-0.03&-0.00& 0.01&-0.01&-0.13&-0.05& 1.00&\\
$r^5_{11}$      &-0.18& 0.03&-0.08&-0.16& 0.11&-0.07& 0.00& 0.07& 0.06&1.00&\\
$r^5_{00}$      & 0.48& 0.12& 0.01& 0.12&-0.37&-0.08& 0.08& 0.09&-0.10&-0.38&1.00&\\
Re $r^5_{10}$   & 0.11& 0.34&-0.07&-0.04&-0.13&-0.24& 0.10&-0.02&-0.13& 0.03& 0.15&1.00&\\
$r^5_{1-1}$     & 0.04&-0.07& 0.18& 0.01& 0.03& 0.07&-0.15& 0.13& 0.03& 0.23&-0.03&-0.06&1.00&\\
Im $r^6_{10}$   &-0.14&-0.05&-0.12&-0.04&-0.14&-0.15&-0.14&-0.10& 0.15&-0.01&-0.03& 0.14&-0.09&1.00&\\
Im $r^6_{1-1}$  & 0.02& 0.12&-0.03& 0.15&-0.00& 0.13& 0.01& 0.07&-0.10&-0.32& 0.05& 0.02&-0.07& 0.01&1.00&\\
Im $r^3_{10}$   & 0.00&-0.01& 0.04&-0.02&-0.01&-0.00& 0.02&-0.05& 0.00& 0.04&-0.00&-0.01& 0.03&-0.01&-0.06&1.00&\\
Im $r^3_{1-1}$  & 0.03&-0.01& 0.03&-0.00& 0.00& 0.03& 0.01& 0.01&-0.08&-0.04& 0.02&-0.02& 0.01&-0.05&-0.01&-0.06&1.00&\\
Im $r^7_{10}$   & 0.00&-0.01& 0.02& 0.00&-0.01& 0.00& 0.01&-0.01& 0.01& 0.03& 0.01&-0.01& 0.03& 0.03&-0.03& 0.34&-0.08&1.00&\\
Im $r^7_{1-1}$  & 0.03&-0.03& 0.01& 0.01&-0.00& 0.01& 0.02& 0.03& 0.00&-0.02& 0.02&-0.03& 0.02&-0.02& 0.05&-0.06& 0.32&-0.07&1.00&\\
$r^8_{11}$      & 0.02& 0.03&-0.01&-0.02& 0.03& 0.01&-0.04&-0.04& 0.05& 0.04&-0.02& 0.01& 0.01& 0.01&-0.01&-0.07& 0.07&-0.02&-0.26&1.00&\\
$r^8_{00}$      &-0.07& 0.04& 0.00& 0.03&-0.09& 0.03&-0.01& 0.01&-0.01&-0.03& 0.06& 0.01&-0.01& 0.00& 0.01&-0.11& 0.03&-0.08& 0.05&-0.38&1.00&\\
Re $r^8_{10}$   & 0.02&-0.02&-0.01& 0.02& 0.03&-0.02& 0.01&-0.00& 0.01&-0.01&-0.00& 0.01&-0.00&-0.01& 0.01&-0.10& 0.12& 0.13& 0.04& 0.00&-0.01&1.00&\\
$r^8_{1-1}$     & 0.00&-0.01&-0.03&-0.02& 0.00&-0.00&-0.00&-0.02&-0.02&-0.00& 0.00& 0.00& 0.01& 0.03& 0.02&-0.08&-0.01&-0.07& 0.11&-0.26& 0.03&-0.03& 1.00\\
\hline
SDME & $r^{04}_{00}$  & $r^{04}_{10}$ & $r^{04}_{1-1}$  & $r^1_{11}$ & $r^1_{00}$ & $r^1_{10}$ & $r^1_{1-1}$ & $r^2_{10}$ & $r^2_{1-1}$ & $r^5_{11}$ & $r^5_{00}$ & $r^5_{10}$ & $r^5_{1-1}$ & $r^6_{10}$ & $r^6_{1-1}$ & $r^3_{10}$ & $r^3_{1-1}$ & $r^7_{10}$ & $r^7_{1-1}$ & $r^8_{11}$ & $r^8_{00}$ & $r^8_{10}$ & $r^8_{1-1}$  \\
\hline
\end{tabular}
\end{footnotesize}
\end{sideways}
\label{tab8}
\end{table*}

%\newpage

\begin{table*}
\caption{\small The correlation matrix for the 23 SDMEs
obtained from the deuterium target data. The column headings do not indicate the real and imaginary parts of any
SDMEs in order to keep the table compact.
}
\renewcommand{\arraystretch}{1.3}
\setlength{\tabcolsep}{3pt}
\hspace*{2.0cm}
\begin{sideways}
\begin{footnotesize}
\begin{tabular}{|l|c|c|c|c|c|c|c|c|c|c|c|c|c|c|c|c|c|c|c|c|c|c|c| }
\hline
SDME & $r^{04}_{00}$  & $r^{04}_{10}$ & $r^{04}_{1-1}$  & $r^1_{11}$ & $r^1_{00}$ & $r^1_{10}$ & $r^1_{1-1}$ & $r^2_{10}$ & $r^2_{1-1}$ & $r^5_{11}$ & $r^5_{00}$ & $r^5_{10}$ & $r^5_{1-1}$ & $r^6_{10}$ & $r^6_{1-1}$ & $r^3_{10}$ & $r^3_{1-1}$ & $r^7_{10}$ & $r^7_{1-1}$ & $r^8_{11}$ & $r^8_{00}$ & $r^8_{10}$ & $r^8_{1-1}$  \\
\hline
$r^{04}_{00}$   & 1.00&\\
Re $r^{04}_{10}$& 0.11& 1.00&\\
$r^{04}_{1-1}$  &-0.03&-0.09& 1.00&\\
$r^1_{11}$      &-0.00&-0.01&-0.38& 1.00&\\
$r^1_{00}$      & 0.07&-0.06& 0.07&-0.40& 1.00&\\
Re $r^1_{10}$   &-0.02& 0.22& 0.09& 0.00& 0.12& 1.00&\\
$r^1_{1-1}$     & 0.21& 0.09&-0.03& 0.02& 0.01&-0.07& 1.00&\\
Im $r^2_{10}$   & 0.01&-0.23& 0.06& 0.04& 0.01&-0.04& 0.02& 1.00&\\
Im $r^2_{1-1}$  &-0.20&-0.06&-0.04& 0.01&-0.04&-0.05&-0.11&-0.05& 1.00&\\
$r^5_{11}$      &-0.21& 0.03&-0.02&-0.20& 0.13&-0.09& 0.02& 0.05&-0.05&1.00&\\
$r^5_{00}$      & 0.51& 0.13&-0.03& 0.15&-0.40&-0.09& 0.08& 0.10&-0.07&-0.40&1.00&\\
Re $r^5_{10}$   & 0.13& 0.38&-0.14&-0.07&-0.11&-0.23& 0.17&-0.00&-0.12& 0.05& 0.16&1.00&\\
$r^5_{1-1}$     &-0.01&-0.13& 0.26&-0.02& 0.04& 0.10&-0.16& 0.11&-0.02& 0.32&-0.06&-0.15&1.00&\\
Im $r^6_{10}$   &-0.09&-0.02&-0.09&-0.07&-0.06&-0.12&-0.09&-0.24& 0.14&-0.01&-0.00& 0.16&-0.08&1.00&\\
Im $r^6_{1-1}$  &-0.01& 0.08&-0.02& 0.20&-0.03& 0.13& 0.01& 0.09&-0.27&-0.32& 0.04&-0.02&-0.08&-0.08&1.00&\\
Im $r^3_{10}$   & 0.02& 0.02&-0.00& 0.01& 0.04& 0.05& 0.02& 0.04& 0.02&-0.02& 0.01& 0.00&-0.00&-0.05& 0.03&1.00&\\
Im $r^3_{1-1}$  &-0.01&-0.00& 0.01&-0.01&-0.02&-0.03& 0.01& 0.03& 0.04& 0.03&-0.00& 0.02&-0.00& 0.02&-0.05&-0.10&1.00&\\
Im $r^7_{10}$   & 0.03& 0.02&-0.00& 0.03&-0.03&-0.03& 0.03&-0.00& 0.01& 0.01&-0.04&-0.03& 0.00&-0.03&-0.02& 0.44&-0.09&1.00&\\
Im $r^7_{1-1}$  &-0.02& 0.01& 0.02&-0.02& 0.01& 0.01&-0.01& 0.01& 0.00&-0.03& 0.02& 0.03&-0.01&-0.01&-0.02&-0.08& 0.43&-0.05&1.00&\\
$r^8_{11}$      & 0.02&-0.03&-0.03&-0.03& 0.00&-0.03& 0.02& 0.01&-0.01& 0.04&-0.02& 0.00& 0.00& 0.01&-0.03&-0.05& 0.14&-0.05&-0.24&1.00&\\
$r^8_{00}$      &-0.03& 0.02&-0.01& 0.01&-0.00&-0.02&-0.01&-0.05& 0.02&-0.01&-0.00&-0.02&-0.00& 0.02&-0.02&-0.06&-0.02& 0.03& 0.02&-0.40&1.00&\\
Re $r^8_{10}$   & 0.02& 0.03& 0.01&-0.01&-0.03&-0.01&-0.00&-0.02& 0.01& 0.02&-0.01&-0.04& 0.04& 0.03&-0.02&-0.08& 0.08& 0.17& 0.00& 0.01& 0.06&1.00&\\
$r^8_{1-1}$     & 0.00& 0.02& 0.00& 0.01&-0.01& 0.00&-0.05&-0.02& 0.03&-0.01& 0.02& 0.03& 0.00& 0.02& 0.03&-0.04&-0.03&-0.05& 0.11&-0.21& 0.03&-0.00& 1.00\\
\hline
SDME & $r^{04}_{00}$  & $r^{04}_{10}$ & $r^{04}_{1-1}$  & $r^1_{11}$ & $r^1_{00}$ & $r^1_{10}$ & $r^1_{1-1}$ & $r^2_{10}$ & $r^2_{1-1}$ & $r^5_{11}$ & $r^5_{00}$ & $r^5_{10}$ & $r^5_{1-1}$ & $r^6_{10}$ & $r^6_{1-1}$ & $r^3_{10}$ & $r^3_{1-1}$ & $r^7_{10}$ & $r^7_{1-1}$ & $r^8_{11}$ & $r^8_{00}$ & $r^8_{10}$ & $r^8_{1-1}$  \\
\hline
\end{tabular}
\end{footnotesize}
\end{sideways}
\label{tab9}
\end{table*}

\end{appendix}
\end{document}